\documentclass[tradiabstract]{aa} 
\usepackage{graphicx}
\usepackage{txfonts}
\usepackage{natbib}
\usepackage{longtable}
\usepackage{lscape}
\usepackage{multirow}
\usepackage{color}
\usepackage[colorlinks=true, bookmarksopen, bookmarksdepth=2, allcolors=blue, breaklinks=true]{hyperref}

\bibpunct{(}{)}{;}{a}{}{,}

\newcommand{\kms}{km\,s$^{-1}$}

\newcommand{\degree}{$^{\circ}$}

\newcommand{\nodata}{...}

\begin{document}
\title{The Serpens filament: at the onset of slightly supercritical collapse}

\author{Y. Gong\inst{1,2} \and G.~X. Li\inst{3,4} \and R.~Q. Mao\inst{2} \and C. Henkel\inst{1,5,6} \and K. M.~Menten\inst{1} \and M. Fang\inst{7} \and M. Wang\inst{2} \and J.~X. Sun\inst{2}
}
\institute{Max-Planck Institute f\"ur Radioastronomy, Auf dem H\"ugel 69, 53121 Bonn, Germany \\e-mail: ygong@mpifr-bonn.mpg.de, gongyan2444@gmail.com
\and Purple Mountain Observatory \& Key Laboratory for Radio Astronomy, Chinese Academy of Sciences, 2 West Beijing Road, 210008 Nanjing, P.R. China
\and University Observatory Munich, Scheinerstrasse 1, D-81679 M\"unchen, Germany
\and South-Western Institute for Astronomy Research, Yunnan University, Kunming, Yunnan 650500, P.R. China
\and Astronomy Department, Faculty of Science, King Abdulaziz University, P.O. Box 80203, Jeddah 21589, Saudi Arabia
\and Xinjiang Astronomical Observatory, Chinese Academy of Sciences, 830011 Urumqi, China
\and Department of Astronomy, University of Arizona, 933 North Cherry Avenue, Tucson, AZ 85721, USA
}

   \date{}

\abstract{The Serpens filament, as one of the nearest infrared dark clouds, is regarded as a pristine filament at a very early evolutionary stage of star formation. In order to study its molecular content and dynamical state, we mapped this filament in seven species including C$^{18}$O, HCO$^{+}$, HNC, HCN, N$_{2}$H$^{+}$, CS, and CH$_{3}$OH. Among them, HCO$^{+}$, HNC, HCN, and CS show self-absorption, while C$^{18}$O is most sensitive to the filamentary structure. A kinematic analysis demonstrates that this filament forms a velocity-coherent (trans-)sonic structure, a large part of which is one of the most quiescent regions in the Serpens cloud. Widespread C$^{18}$O depletion is found throughout the Serpens filament. Based on the Herschel dust-derived H$_{2}$ column density map, the line mass of the filament is 36--41~M$_{\odot}$~pc$^{-1}$, and its full width at half maximum width is 0.17$\pm$0.01~pc, while its length is $\approx 1.6$~pc. The inner radial column density profile of this filament can be well fitted with a Plummer profile with an exponent of 2.2$\pm$0.1, a scale radius of $0.018\pm 0.003$ pc, and a central density of $(4.0\pm 0.8)\times 10^{4}$~cm$^{-3}$. The Serpens filament appears to be slightly supercritical. The widespread blue-skewed HNC and CS line profiles and HCN hyperfine line anomalies across this filament indicate radial infall in parts of the Serpens filament. C$^{18}$O velocity gradients also indicate accretion flows along the filament. The velocity and density structures suggest that such accretion flows are likely due to a longitudinal collapse parallel to the filament's long axis. Both the radial infall rate ($\sim$72~$M_{\odot}$~Myr$^{-1}$, inferred from HNC and CS blue-skewed profiles) and the longitudinal accretion rate ($\sim$10~$M_{\odot}$~Myr$^{-1}$, inferred from C$^{18}$O velocity gradients) along the Serpens filament are lower than all previously reported values in other filaments. This indicates that the Serpens filament lies at an early evolutionary stage when collapse has just begun, or that thermal and non-thermal support are effective in providing support against gravity.}

\maketitle

\section{Introduction}\label{sec.intro}
For many decades filamentary structures have already been known to exist in molecular clouds \citep[e.g.,][]{1979ApJS...41...87S,1987ApJ...312L..45B,1989ApJ...338..925L}. More recently, the prevalence of filamentary structures was emphasized by Herschel observations of nearby molecular clouds, suggesting that filaments do not only serve as a fundamental stage in star-forming processes, but could also play an important role in the evolution of molecular clouds \citep{2010A&A...518L.102A,2014prpl.conf...27A,2017CRGeo.349..187A}. Substantial observations of dust extinction, dust continuum emission, and molecular lines have demonstrated that filaments are ubiquitous in star-forming, translucent, and even diffuse molecular clouds \citep[e.g.,][]{2009ApJ...700.1609M,2010A&A...518L.102A,2010A&A...518L.104M,2015MNRAS.450.4043W,2016A&A...591A...5L,2017ApJ...838...49X,2017ApJ...835L..14G,2018arXiv180807499M}. However, the formation and evolution of filaments are still not fully understood. Identifying and characterising isolated filaments at the earliest evolutionary stage of star formation is of great importance to tackle this question. 

\begin{figure*}[!htbp]
\centering
\includegraphics[width = 0.48 \textwidth]{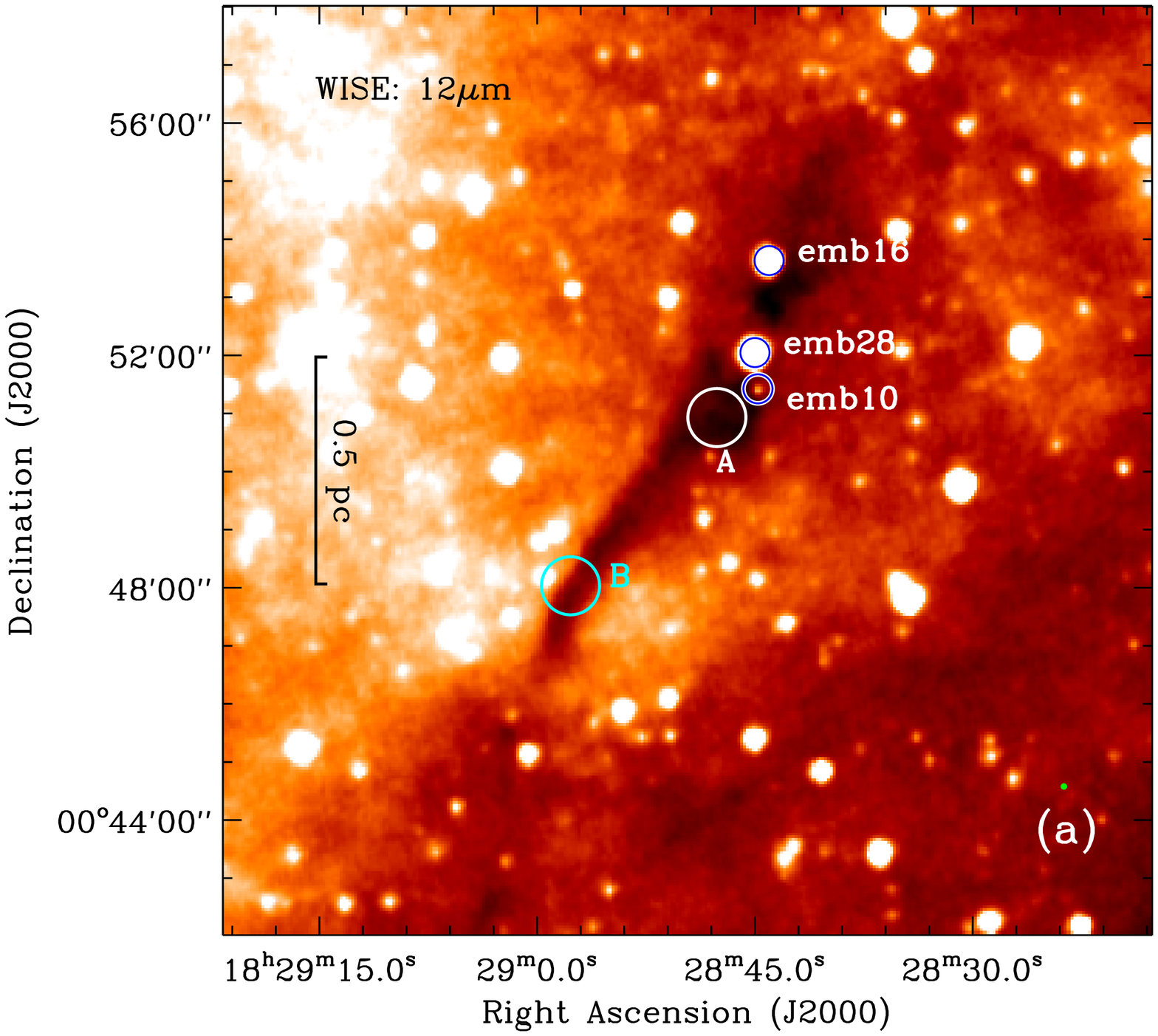}
\includegraphics[width = 0.48 \textwidth]{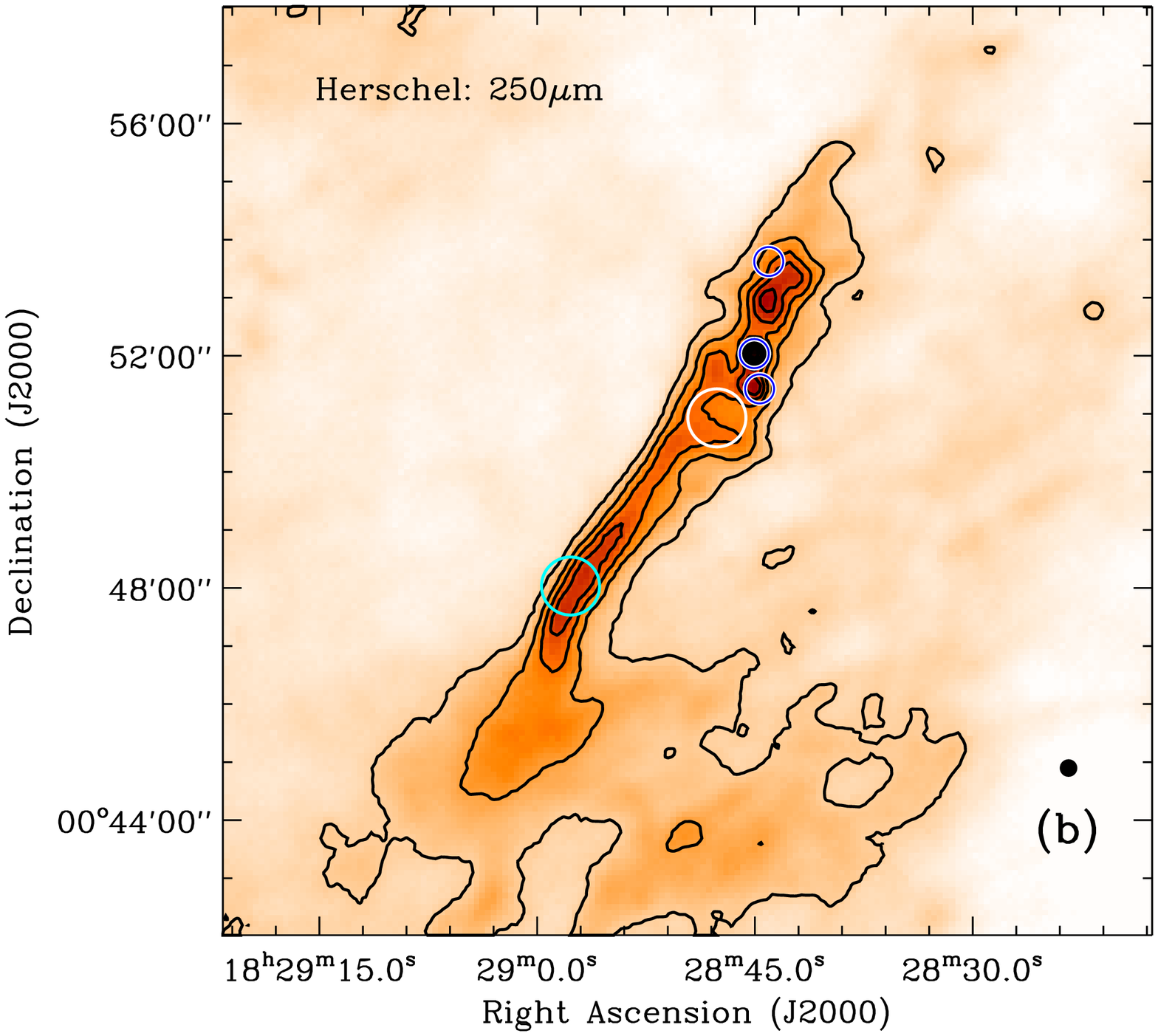}
\caption{{(a) The WISE 11.2~$\mu$m image of the Serpens filament which shows up in absorption. (b) The Herschel 250~$\mu$m image of the Serpens filament. The contours start at 0.5~Jy~beam$^{-1}$, and increase by 0.5 Jy~beam$^{-1}$. In both panels, the three blue circles mark the three embedded YSOs, emb10, emb16, and emb28 \citep{2009ApJ...692..973E}, while the white and cyan circles give the selected positions of Source A and Source B (see Sect.~\ref{sec.pmo} and Table~\ref{line}). The beam size is shown in the lower right of each panel.}\label{Fig:infra}}
\end{figure*}

The Serpens filament, named by \citet{2007ApJ...666..982E}, is also known as the ``starless cores'' region in \citet{2013ApJS..209...39B} and Filament 9 in \citet{2015A&A...584A.119R}. This filament resides in the Serpens cloud, the distance of which is about 440 pc as deduced from the trigonometric parallax measured with the Very Long Baseline Array (VLBA) \citep{2017ApJ...834..143O}. Figure~\ref{Fig:infra}a presents the distribution of $12\mu$m radiation in the region of interest here, observed by the Wide-Field Infrared Explorer (WISE). The filament appears in absorption against a diffuse background, making it an infrared dark cloud (IRDC). IRDCs are known to be the cold and dense part of molecular clouds \citep[e.g.,][]{1998ApJ...508..721C,2006A&A...450..569P,2011ApJS..193...10Z} and represent a very early evolutionary stage harboring no or only a limited amount of embedded protostars \citep[e.g.,][]{2005IAUS..227...23M,2007ARA&A..45..339B}. IRDCs, which predominantly have filamentary structures, can serve as ideal targets to study the initial conditions of dense molecular cloud regions prior to star formation. This IRDC was not included in previous IRDC catalogs, because of its location which is not fully covered by previous mid-infrared surveys, many of which have been restricted to the Galactic plane. In contrast to most IRDCs, this IRDC is characterized by a particularly low mass (see results given below) and its extreme proximity. The Serpens filament, similar to its well known sibling -- Serpens South \citep[e.g.,][]{2008ApJ...673L.151G}, is one of the nearest IRDCs, but has much less substructure (e.g., sub-filaments, fibres) than Serpens South. Furthermore, the Serpens filament was found to contain 7 dust cores \citep{2007ApJ...666..982E} and three class I young stellar objects (YSOs) \citep{2009ApJ...692..973E}, i.e., emb10 (also known as IRAS 18262+0050), emb16, and emb28, which are marked with blue circles in Fig.~\ref{Fig:infra}a. On the other hand, its southeastern part is free of YSOs, i.e., starless. \citet{2013ApJS..209...39B} argued that this region is going to be the next site of star formation in Serpens.


Based on the Herschel observations, the Serpens filament is estimated to have a size of $\sim$0.3~pc$\times$1.7 pc, and a position angle of 147\degr\,east of north \citep{2015A&A...584A.119R}. The highly rectilinear morphology of the Serpens filament is clearly seen in dust emission which matches the 12~$\mu$m absorption very well (see Fig.~\ref{Fig:infra}), but is invisible in either $^{12}$CO (2--1) or $^{13}$CO (2--1) maps \citep[see fig.~15 of][]{2013ApJS..209...39B}. More molecular line observations are crucial to study its molecular content and dynamical state.





\section{Observations and data reduction}\label{sec.obs}
\subsection{PMO-13.7 m observations}\label{sec.pmo}
We carried out multiple molecular line observations (project code: 17A004) toward the Serpens filament with the Purple Mountain Observatory 13.7 m (PMO-13.7 m) telescope from 2017 May 12 to 17. The 3$\times$3 beam sideband separation Superconducting Spectroscopic Array Receiver \citep{2012ITTST...2..593S} was employed as front-end, while a set of 18 fast Fourier transform spectrometers (FFTSs) was used as a back-end to analyze signals from both sidebands. During the observations, we used two different FFTS modes which provided instantaneous bandwidths of 1 GHz and 200 MHz, respectively. Each FFTS consists of 16384 channels, resulting in channel spacings of 61.0 kHz and 12.2 kHz (corresponding velocity spacings are given in Table~\ref{line}) for the two FFTS modes. The filament was observed in the On-The-Fly mode \citep{2018AcASn..59....3S} at a scanning rate of 50\arcsec\,per second and a dump time of 0.3 seconds. In order to minimize the overheads, the map was solely scanned along the long axis of this filament. The parameters of the observed transitions are given in Table~\ref{line}. Follow-up single-point On--Off observations with the 200 MHz FFTS mode were performed toward two peaks of the F=0--1 line of HCN (1--0), source A ($\alpha_{\rm J2000}$=18$^{\rm h}$28$^{\rm m}$47$\rlap{.}^{\rm s}62$, $\delta_{\rm J2000}$=00\degr50\arcmin55$\rlap{.}$\arcsec8) and source B ($\alpha_{\rm J2000}$=18$^{\rm h}$28$^{\rm m}$57$\rlap{.}^{\rm s}70$, $\delta_{\rm J2000}$=00\degr48\arcmin02$\rlap{.}$\arcsec0) (see Fig.~\ref{Fig:infra}a) to obtain higher quality HCN (1--0) line profiles. Furthermore, we also performed single-point On-Off observations toward source A to observe less abundant species and to confirm the observed self-absorption. For all the single-point On-Off observations, only spectra obtained with the central beam are analyzed in this work. The observations encompass a total of $\sim$34 observing hours.

The standard chopper-wheel method was used to calibrate the antenna temperature \citep{1976ApJS...30..247U}. The relationship $T_{\rm mb}= T_{\rm A}^{*}/B_{\rm eff}$ was used to convert antenna temperature, $T_{\rm A}^{*}$, to main-beam brightness temperature, $T_{\rm mb}$, where $B_{\rm eff}$ is the main beam efficiency. The half-power beam widths, $\theta_{\rm b}$, at different frequencies were found to be $\theta_{\rm b}=46.3^{\prime\prime}\times \frac{115.3}{\nu} $, where $\nu$ is the observed frequency in units of GHz. Two standard sources (W51D and S140) were used as spectral calibrators that were regularly observed with the same setups as for our scientific observations. A dispersion of $\sim$5\% in the peaks of observed lines toward these two standard sources indicate that the uncertainty on the temperature scale is better than the 10\% given in the telescope status report. Typical system temperatures were 121--142 K on a $T_{\rm A}^{*}$ scale. The pointing was found to be accurate to within $\sim$5\arcsec. Throughout this paper, velocities are given with respect to the local standard of rest (LSR).

Data reduction was performed with the GILDAS\footnote{\url{https://www.iram.fr/IRAMFR/GILDAS/}} software including CLASS and GREG \citep{2005sf2a.conf..721P}. Since the half-power beam widths are slightly different for each observed spectral line, all spectral maps have been convolved to an effective angular resolution of 1\arcmin\,(corresponding to a linear scale of 0.13 pc at a distance of 440 pc) for direct comparison between different tracers. In these maps, a single pixel size is 20\arcsec$\times$20\arcsec.

\subsection{Archival data}
We make also use of mid-infrared data from the WISE which mapped the full sky at 3.4, 4.6, 12, and 22~$\mu$m with angular resolutions of 6.1\arcsec, 6.4\arcsec, 6.5\arcsec, and 12.0\arcsec\,in the four bands \citep{2010AJ....140.1868W}. Additional ancillary data include level 2.5 processed {\it Herschel}\footnote{Herschel is an ESA space observatory with science instruments provided by European-led Principal Investigator consortia and with important participation from NASA.} images observed at 70/160~$\mu$m with the PACS instrument \citep{2010A&A...518L...2P} and at 250/350/500~$\mu$m with the SPIRE instrument \citep{2010A&A...518L...3G}. Their angular resolutions range from 6\arcsec\,to 37\arcsec. The H$_{2}$ column density map of this region is derived from the spectral energy distribution (SED) fitting from 160--500$\mu$m, and is obtained from the Herschel Gould Belt (GB) Survey\footnote{\url{http://www.herschel.fr/cea/gouldbelt/en/index.php}} \citep{2010A&A...518L.102A,2015A&A...584A..91K,2017MmSAI..88..783F}. The comparison with the extinction map from 2MASS on the same grid suggests that the derived H$_{2}$ column densities have an uncertainty of $\sim$10\% \citep{2015A&A...584A..91K}.




\section{Results}
\subsection{Molecular gas distributions}\label{sec.mor}
\begin{figure*}[!htbp]
\centering
\includegraphics[width = 0.95 \textwidth]{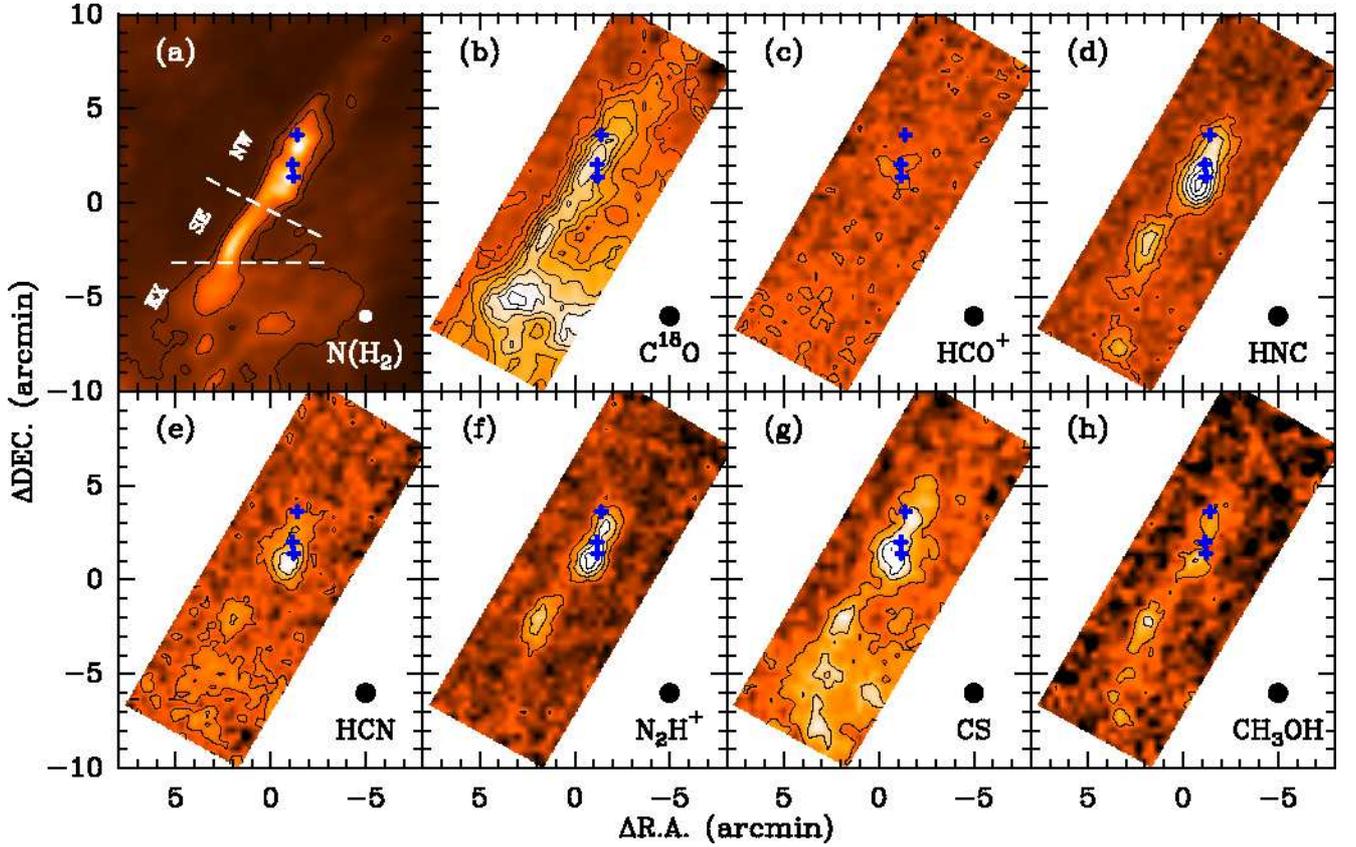}
\caption{{Spatial distributions of different molecular species, labeled in the lower right of each panel, over the Serpens filament. (a) The Herschel H$_{2}$ column density image. The contours are H$_{2}$ column densities of 4$\times 10^{21}$ cm$^{-2}$ and 6$\times 10^{21}$~cm$^{-2}$. This region is divided by the two white dashed lines into three individual regions which are labelled as NW (northwest), SE (southeast), and EX (extended). (b) The C$^{18}$O (1--0) intensity map integrated from 7.4 to 8.7~\kms. The contours start at 0.30~K~\kms\,(5$\sigma$), and increase by 0.18~K~\kms\,(3$\sigma$). (c) The HCO$^{+}$ (1--0) intensity map integrated from 6.5 to 9.5~\kms. The contours start at 0.27~K~\kms, and increase by 0.27~K~\kms. (d) The HNC (1--0) intensity map integrated from 7.0 to 9.0~\kms. The contours start at 0.24~K~\kms, and increase by 0.24~K~\kms. (e) The HCN (1--0) intensity map integrated from 0 to 1.6~\kms, from 7.0 to 8.6~\kms, and from 11.8 to 13.4~\kms relative to the frequency of the $F$=2--1 hyperfine structure component. The contours start at 0.27~K~\kms, and increase by 0.27~K~\kms. (f) The N$_{2}$H$^{+}$ (1--0) intensity map integrated from 6.5 to 9.5~\kms. The contours start at 0.30~K~\kms, and increase by 0.30~K~\kms. (g) The CS (2--1) intensity map integrated from 7.0 to 9.5~\kms. The contours start at 0.18~K~\kms, and increase by 0.18~K~\kms. (h) The CH$_{3}$OH (2$_{0}$--1$_{0}$) intensity map integrated from 7.5 to 8.8~\kms. The contours start at 0.12~K~\kms, and increase by 0.12~K~\kms. In all panels, the (0, 0) offset corresponds to $\alpha_{\rm J2000}$=18$^{\rm h}$28$^{\rm m}$49$\rlap{.}^{\rm s}$642, $\delta_{\rm J2000}$=00$^{\circ}$50$^{\prime}$01$\rlap{.}^{\prime \prime}$08, and the three blue crosses give the positions of the three embedded YSOs (see also Fig.~\ref{Fig:infra}). The beam size is shown in the lower right of each panel.}\label{Fig:molecules}}
\end{figure*}

Figure~\ref{Fig:molecules}a shows the Herschel H$_{2}$ column density image of the Serpens filament. In this panel, the two white dashed lines roughly divide the observed area into three individual regions that are labeled as NW (northwest), SE (southeast), and EX (extended), respectively. NW and SE represent the northwest and southeast part of the filamentary structure, while EX stands for the extended emission to the south and outside the filamentary structure. NW and SE have stronger 12~$\mu$m absorption than the EX region. Both 500~$\mu$m emission and absorption at 12~$\mu$m show that NW has a wider width than SE (see Figs.~\ref{Fig:infra} and \ref{Fig:molecules}a). The two YSOs emb10 and emb28 are seen projected near the center of NW, while emb16 is seen projected at the northeastern edge of NW. In NW, emb28 is located at the filament's narrowest place, from which it extends in two opposite directions. 

The Serpens filament has been mapped by our PMO-13.7 m observations in lines from 7 different species, the integrated intensity maps of which are presented in Fig.~\ref{Fig:molecules}b--\ref{Fig:molecules}h. These 7 species include C$^{18}$O, HCO$^{+}$, HNC, HCN, N$_{2}$H$^{+}$, CS, and CH$_{3}$OH. All but two of the detected lines have upper energy levels ($E_{\rm u}/k$) lower than 7 K (see Table~\ref{line}), and typical spectra toward source A are shown in Fig.~\ref{Fig:sp-A}. The observed properties and beam-averaged molecular column densities of source A are given in Table~\ref{Tab:a}, and more discussions about source A are presented in Appendix~\ref{app.b}. Among these detected lines, C$^{18}$O (1--0) is the brightest. It is worth noting that none of the three YSOs have associated peaks in the distribution of detected species. The distribution of different species will be discussed in the following.

C$^{18}$O. -- The whole filamentary structure is well detected in C$^{18}$O emission, which resembles the distribution of the Herschel H$_{2}$ column density image. However, such a structure is not found in either $^{12}$CO (2--1) or $^{13}$CO (2--1) maps \citep{2013ApJS..209...39B}. This is likely because there is significant $^{12}$CO and $^{13}$CO self-absorption in the filament and the bulk of optically thick CO/$^{13}$CO gas in the outer region blocks the view to the filamentary structure. Meanwhile, EX shows stronger C$^{18}$O (1--0) integrated intensities than NW and SE, which is opposite to that in the Herschel H$_{2}$ column density image. This is likely because the depletion of C$^{18}$O molecules in the innermost region of the filament is more significant than that in EX (see Sect.~\ref{sec.abun}).

HCO$^{+}$. -- HCO$^{+}$ (1--0) shows the weakest emission and is only marginally detected in NW and EX (see Fig.~\ref{Fig:molecules}c). However, H$^{13}$CO (1--0) is detected by our single-point observations toward source A. Its peak intensity (0.56~K) is much stronger than that of HCO$^{+}$ (1--0), suggesting that significant self-absorption is playing an important role.

HNC. -- In Fig.~\ref{Fig:molecules}d, the filament is clearly seen in HNC (1--0) with stronger emission in NW than in SE. Meanwhile, EX displays HNC (1--0) emission, which is not as extended as C$^{18}$O (1--0). Toward source A, HNC (1--0) shows self-absorption with a dip observed at the peak velocity of its $^{13}$C substitution HN$^{13}$C.

HCN. -- In Fig.~\ref{Fig:molecules}e, the integrated velocity range includes the three hyperfine structure (hfs) components of HCN (1--0) which are mostly weaker than HNC (1--0). HCN (1--0) is detected in NW and SE, and marginally detected in EX. The strongest emission is found in NW. However, the emission bridging NW and SE is not detected in HCN (1--0). This filament also shows widespread HCN (1--0) hyperfine line anomalies which will be discussed in Sect.~\ref{sec.hcn}.

N$_{2}$H$^{+}$. -- Figure~\ref{Fig:molecules}f gives the integrated intensity map of the three main hfs components of N$_{2}$H$^{+}$ (1--0). This transition is detected in NW and SE but not in EX. Both NW and SE are elongated, and the emission is stronger in NW than that in SE. Assuming that self-absorption is not present in N$_{2}$H$^{+}$ (1--0),  we can use the hfs line fitting subroutine in CLASS to determine the optical depth of N$_{2}$H$^{+}$ (1--0), which gives an optical depth of 0.32$\pm$0.15 for the main hfs line in source A (see Fig.~\ref{Fig:sp-A}). This implies that this line is optically thin.

CS. -- As demonstrated in Fig.~\ref{Fig:molecules}g, CS (2--1) is the second widespread tracer after C$^{18}$O (1--0) in the Serpens filament, but the CS (2--1) morphology is not as rectilinear as C$^{18}$O (1--0). Similar to HNC (1--0), CS (2--1) also displays self-absorption (see Fig.~\ref{Fig:sp-A}). On the other hand, CS (2--1) is more widespread than HNC (1--0), although their critical densities, where radiative decay rates match collisional de-excitation rates, are comparable \citep{2015PASP..127..299S}.

CH$_{3}$OH. -- CH$_{3}$OH is the only complex organic molecule observed by this study. CH$_{3}$OH ($J_{k}=$2$_{0}$--1$_{0}$ A) and CH$_{3}$OH (2$_{-1}$--1$_{-1}$ E) are detected. CH$_{3}$OH (2$_{0}$--1$_{0}$ E) and CH$_{3}$OH (2$_{1}$--1$_{1}$ E) are also covered by our observations, but remain undetected, i.e., $T_{\rm mb}<$0.21~K. CH$_{3}$OH (2$_{-1}$--1$_{-1}$ E) is also not shown in Fig.~\ref{Fig:molecules}. This is because its distribution is similar to that of CH$_{3}$OH (2$_{0}$--1$_{0}$ A) and its dynamical range is lower. Figure~\ref{Fig:molecules}h shows that CH$_{3}$OH emission appears in NW, SE, and EX. The CH$_{3}$OH emission peaks in SE, a starless region. This is reminiscent of CH$_{3}$OH in other starless cores, e.g., TMC-1 \citep{1988A&A...195..281F} and L1544 \citep{2014ApJ...795L...2V}. 

Overall, different molecular species show noticeable differences in their distributions. Among them, HCO$^{+}$, HNC, HCN, and CS are found to be affected by self-absorption. On the other hand, C$^{18}$O (1--0) is more sensitive for tracing the filamentary structure than the other tracers. Nearly all dense tracers (HNC, HCN, N$_{2}$H$^{+}$, and CH$_{3}$OH) are not detected in EX, indicating that EX is a less dense region than NW and SE.


\subsection{The properties of the Serpens filament derived from C$^{18}$O (1--0)}\label{sec.kine}
As shown in Sect.~\ref{sec.mor}, C$^{18}$O (1--0) data not only trace the distribution and velocity information of the filament, but also have the best signal-to-noise ratio among all the mapped lines. Furthermore, the difference between the systemic velocities derived by C$^{18}$O (1--0) and the other optically thin lines is less than 0.1~\kms \,(see Fig.~\ref{Fig:sp-A}), demonstrating that the velocity information derived from C$^{18}$O (1--0) will not be significantly affected by differences in molecular chemistry. Therefore, we simply apply Gaussian decomposition to the C$^{18}$O (1--0) data to study the kinematics of the Serpens filament. Before the fitting procedure, the rms noise level, $\sigma$, of each pixel is estimated from emission-free channels. We then create a mask that includes at least three adjacent channels with intensities higher than 3$\sigma$. Only spectra within this mask are used for the following analysis. Visual inspection of the spectra within this mask confirms that a single Gaussian component is a good approximation, so a single Gaussian component is assumed to obtain observed parameters after these spectra are binned to a channel width of 0.1~\kms\,to improve the signal-to-noise ratios of the resulting effective channels.

Figure~\ref{Fig:kine} presents the overall distribution of peak intensities, LSR velocities, and line widths derived from the single-component Gaussian decomposition of C$^{18}$O (1--0) data. In Fig.~\ref{Fig:kine}a, NW, SE, and EX are clearly seen, but in SE the line shows brighter peak intensities than in NW, which is opposite to what is found in its integrated intensity map (see Fig.~\ref{Fig:molecules}b). This is due to the line width variations. Meanwhile, the peak intensities of C$^{18}$O (1--0) range from 0.66~K to 2.60 K with a mean uncertainty of 0.19~K on a main beam brightness temperature scale. These peak intensities correspond to optical depths of 0.10--0.49 at an assumed C$^{18}$O (1--0) excitation temperature of 10 K which is a typical value of IRDCs \citep[e.g.,][]{2008ApJ...686..384D}. The low opacities imply that opacity broadening is negligible in the obtained line widths derived below \citep[e.g.][]{2016A&A...591A.104H}. Assuming local thermodynamic equilibrium (LTE) and a constant excitation temperature of 10 K and following equation (80) of \citet{2015PASP..127..266M}, the C$^{18}$O column densities are estimated to be (0.1--1.7)$\times 10^{15}$~cm$^{-2}$ in NW, (0.2--1.6)$\times 10^{15}$~cm$^{-2}$ in SE, and (0.2--2.3)$\times 10^{15}$~cm$^{-2}$ in EX. 

Figure~\ref{Fig:kine}b shows that the determined centroid LSR velocities fall in the range from 7.85 to 8.56~\kms\,with Gaussian fitting uncertainties of 0.01--0.06~\kms. The whole filament appears to be blue-shifted with respect to the ambient molecular gas ($\varv_{\rm lsr}\sim$8.4~\kms). The maximum velocity difference is about 0.7~\kms\,across the whole region. A continuous velocity distribution is seen across the Serpens filament, suggesting that this filament is velocity-coherent (see also Sect.~\ref{sec.pv}). At an offset of (3$\rlap{.}$\arcmin5, $-$7$\rlap{.}$\arcmin5) in Fig.~\ref{Fig:kine}b, there is a clump whose velocity is blue-shifted with respect to the bulk emission of EX. This blue-shifted clump is seen in the C$^{18}$O (1--0) and HNC (1--0) integrated intensity map (see Figs.~\ref{Fig:molecules}b and \ref{Fig:molecules}d). Velocity gradients both parallel and perpendicular to the filament's long axis are present in the Serpens filament, as indicated by black and red arrows in Fig.~\ref{Fig:kine}b, respectively. The velocity gradients parallel to the filament's long axis are indicated by four black arrows in Fig.~\ref{Fig:kine}b. The associated four regions are found to show the largest velocity gradients. The four parallel velocity gradients are estimated to be $\nabla \varv_{\|,1}=0.40\pm 0.05$~\kms~pc$^{-1}$, $\nabla \varv_{\|,2}=0.30\pm 0.09$~\kms~pc$^{-1}$, $\nabla \varv_{\|,3}=0.53\pm 0.06$~\kms~pc$^{-1}$, and $\nabla \varv_{\|,4}=0.55\pm 0.14$~\kms~pc$^{-1}$ over a length of about 3$\rlap{.}^{\prime}$9, 2$\rlap{.}^{\prime}$2, 3$\rlap{.}^{\prime}$9 and 2$\rlap{.}^{\prime}$5, respectively, suggesting that the velocity gradients in NW are slightly larger than those in SE. The velocity gradients perpendicular to the major axis around emb28 are $1.31\pm 0.78$~\kms~pc$^{-1}$ and $1.34\pm 0.62$~\kms~pc$^{-1}$ both over a length of about 1\arcmin, but keep in mind that the values have large uncertainties due to a small number of pixels.


Figure~\ref{Fig:kine}c shows that the derived line widths, $\Delta \varv$, range from 0.33 to 1.11~\kms. The mean error in the fitted line widths is 0.06~\kms. More than 70\% of the pixels have line widths of $<$0.7~\kms. In this plot, an enhanced line broadening ($\Delta \varv>$0.7~\kms) is clearly found around emb10 and emb28, toward which the four velocity gradients are converging (see Fig.~\ref{Fig:kine}b). Both thermal and non-thermal motions contribute to the observed line broadening, so we can extract the non-thermal velocity dispersion, $\sigma_{\rm NT}$, by subtracting the thermal velocity dispersion, $\sigma_{\rm th}$, from the observed line widths:
\begin{equation}\label{eq1}
\sigma_{\rm NT} = \sqrt{\frac{\Delta \varv^{2}}{8{\rm ln}\,2} - \frac{{\rm k}T_{\rm kin}}{m_{\rm i}}}
\end{equation}
where $\sigma_{\rm th} = \sqrt{\frac{{\rm k}T_{\rm kin}}{m_{\rm i}}}$ is the thermal velocity dispersion of C$^{18}$O, $m_{\rm i}$ is the mass weight of a C$^{18}$O molecule which is 30, k is the Boltzmann constant, and $T_{\rm kin}$ is the kinetic temperature. Following the method of \citet{2016A&A...587A..97H}, we express the derived $\sigma_{\rm NT}$ in terms of the H$_{2}$ sound speed, $c_{\rm s}$, i.e., the observed Mach number $\mathcal{M} = \sigma_{\rm NT}/c_{\rm s}$, where $c_{\rm s}$ is derived by adopting a mean molecular weight of 2.33 \citep{2008A&A...487..993K}. Although the dust temperature is found to be 12--15 K in this region according to the dust temperature map\footnote{\url{http://www.herschel.fr/cea/gouldbelt/en/index.php}}, gas kinetic temperature is found be lower than dust temperature in IRDCs \citep{2017A&A...606A.133S,2018RNAAS...2b..52W}, which is likely due to the background/foreground contributions of the dust emission. Thus, we assume a constant kinetic temperature of 10~K \citep{1989ApJS...71...89B} to derive the H$_{2}$ sound speed and thermal velocity dispersion, which gives $c_{\rm s}$=0.19~\kms\, and $\sigma_{\rm th}$=0.05~\kms. Using Equation~\ref{eq1}, we obtain the $\mathcal{M}$ distribution of the Serpens filament which is displayed in Fig.~\ref{Fig:kine}d. Except cloud edges and regions where the three YSOs are located, the filament is found to have a mean $\mathcal{M}$ of $<$1, which suggests that this filament is a velocity-coherent and (trans-)sonic filament. If the kinetic temperature were higher than 10~K, the non-thermal velocity dispersions would become even lower, so that it still would remain (trans-)sonic. Compared with other regions in the Serpens cloud \citep{2013ApJS..209...39B,2018ApJ...853..169D}, the Serpens filament has lower non-thermal velocity dispersions, which makes it one of the most quiescent regions in the Serpens cloud. The kinematic properties of this filament are analogous to velocity-coherent fibers in L1495/B213, NGC 1333, and the Orion integral filament \citep{2013A&A...554A..55H,2017A&A...606A.123H,2018A&A...610A..77H}, the Musca cloud \citep{2016A&A...587A..97H}, filaments in Barnard 5 \citep{2011ApJ...739L...2P}, dense filaments in OMC1 \citep{2018ApJ...861...77M}, filaments with H$_{2}$ column densities $\leqslant 8\times 10^{21}$ cm$^{-2}$ in nearby molecular clouds \citep{2013A&A...553A.119A}, and more distant filaments in IRDCs \citep{2017A&A...606A.133S,2018RNAAS...2b..52W}. Meanwhile, the Serpens filament, similar to the Musca cloud, appears to have much less substructure than the other aforementioned filaments, but this needs to be confirmed with higher angular resolution observations.

\subsection{C$^{18}$O and N$_{2}$H$^{+}$ fractional abundances relative to H$_{2}$}\label{sec.abun}
Among all mapped lines, C$^{18}$O (1--0) and N$_{2}$H$^{+}$ (1--0) are likely optically thin and have significant signal-to-noise ratios, so we only estimate their fractional abundances relative to H$_{2}$ in the Serpens filament. The column densities of C$^{18}$O and N$_{2}$H$^{+}$ are calculated by assuming optically thin emission and an excitation temperature of 10~K. Under the optical thin assumption, the different excitation temperatures within the range of 5--15~K only lead to deviations of $<$15\% in column densities. In order to match the angular resolution of molecular lines, the Herschel H$_{2}$ column density map was convolved to an angular resolution of 1\arcmin\, with a Gaussian kernel size of 47$\rlap{.}^{\prime\prime}2$, and then was linearly interpolated to the same grid as our molecular line data. The fractional abundances are directly derived from the ratios of molecular column densities and H$_{2}$ column densities. In the following analysis, only the C$^{18}$O and N$_{2}$H$^{+}$ data with signal-to-noise ratios higher than 5 are taken into account.

Figure~\ref{Fig:abundratio}a presents a pixel-by-pixel comparison between C$^{18}$O column densities and H$_{2}$ column densities within the H$_{2}$ column density range of (0.3--1.8)$\times 10^{22}$~cm$^{-2}$. We clearly see two different groups labeled by different colors where their column densities are roughly linearly correlated, respectively. The two groups are found to arise from EX and the Serpens filament including NW and SE (see Fig.~\ref{Fig:abund}a). Figure~\ref{Fig:abundratio}b demonstrates that the C$^{18}$O fractional abundances in the Serpens filament tend to be lower than those in EX. As shown in Figs.~\ref{Fig:abundratio}b and \ref{Fig:abund}a, the C$^{18}$O fractional abundances are found to be (0.7--1.5)$\times 10^{-7}$ with a median value of 1.1$\times 10^{-7}$ in EX, and (3.4--7.0)$\times 10^{-8}$ with a median value of 5.6$\times 10^{8}$ in the Serpens filament. The C$^{18}$O fractional abundances in the Serpens filament are lower than those in EX by roughly a factor of 2 (see Fig.~\ref{Fig:abund}a), which cannot be explained by the deviations caused by the different excitation temperatures. Rather, it is natural to assume that the gas-phase C$^{18}$O molecules have frozen out onto dust grain surfaces in cold dense regions \citep[e.g.,][]{2002ApJ...570L.101B,2007ARA&A..45..339B}. The low C$^{18}$O fractional abundances suggest the presence of widespread C$^{18}$O depletion throughout the Serpens filament. Meanwhile, such depletion is more significant in higher H$_{2}$ column density regions (see Fig.~\ref{Fig:abundratio}b), consistent with the predictions of previous chemical models \citep[e.g., fig. 3 of][]{2002ApJ...570L.101B}.

Figure~\ref{Fig:abundratio}c shows that the N$_{2}$H$^{+}$ column densities are linearly correlated with H$_{2}$ column densities with a high Pearson correlation coefficient of 0.86. Figure~\ref{Fig:abundratio}d indicates that the N$_{2}$H$^{+}$ fractional abundances tend to increase with increasing H$_{2}$ column densities but with a lower Pearson correlation coefficient of 0.51. The N$_{2}$H$^{+}$ fractional abundances also peak in dense regions (see Fig.~\ref{Fig:abund}b), better tracing dense cores. This trend is different from that observed for C$^{18}$O in Fig.~\ref{Fig:abundratio}b. This is mainly due to selective freeze-out in the sense that C$^{18}$O molecules are more easily depleted than N$_{2}$H$^{+}$ \citep{2007ARA&A..45..339B}. Meanwhile, N$_{2}$H$^{+}$ depletion has also been found in B68 \citep{2002ApJ...570L.101B}, but is not evident in the Serpens filament. This is mainly due to our lower linear resolution of 0.13 pc versus 0.01 pc in the case of B68. Within the H$_{2}$ column density range of (0.8--1.8)$\times 10^{22}$~cm$^{-2}$, the N$_{2}$H$^{+}$ abundances are found to be (0.8--2.3)$\times 10^{-10}$ with a median value of 1.3$\times 10^{-10}$. 

Figure~\ref{Fig:abundratio}e shows that fractional abundances of C$^{18}$O and N$_{2}$H$^{+}$ are anti-correlated. This is similar to what has been found in a sample of low-mass protostellar objects (see fig. 15 of \citet{2004A&A...416..603J}) and caused by the fact that N$_{2}$H$^{+}$ ions are mainly destroyed via the reaction N$_{2}$H$^{+}$+CO$\to$HCO$^{+}$+N$_{2}$ \citep[e.g.,][]{1991A&A...245..457M,2004A&A...416..603J}. Figure~\ref{Fig:abundratio}f shows that the N$_{2}$H$^{+}$/C$^{18}$O abundance ratios increase with increasing H$_{2}$ column densities, which is also evident in the N$_{2}$H$^{+}$/C$^{18}$O abundance ratio map (see Fig.~\ref{Fig:abund}c). This is due to the aforementioned selective freeze-out effect that causes C$^{18}$O molecules to become more significantly depleted in denser regions, whereas most N$_{2}$H$^{+}$ ions remain still in the gas phase. Meanwhile, the depletion of C$^{18}$O makes the removal mechanism for N$_{2}$H$^{+}$ more inefficient, reinforcing such a trend.

\subsection{Widespread HCN (1--0) hyperfine line anomalies}\label{sec.hcn}
Rotational transitions of HCN exhibit hfs which can be used to estimate their optical depths from the intensity ratios between individual components. Under local thermodynamic equilibrium (LTE) conditions, the ratios $R(I[F=1-1]/I[F=2-1])$ and $R(I[F=0-1]/I[F=2-1])$ (R$_{12}$ and R$_{02}$ hereafter) are expected to be $0.6<R_{12}<1.0$ and $0.2<R_{02}<1$ for HCN (1--0), depending on different optical depths \citep[e.g.,][]{2012MNRAS.420.1367L}. However, HCN hyperfine line anomalies, in which case the observed ratios $R_{12}$ and $R_{02}$ deviate from the LTE ranges, impair this usefulness of HCN hyperfine lines for optical depth and thus column density estimates.

Figure~\ref{Fig:hcn-anam} shows the integrated intensity maps of the three individual hfs lines. Surprisingly, the $F=0-1$ line is found to be stronger than the $F=2-1$ line, i.e., $R_{02}>1$, which demonstrates the presence of HCN (1--0) hyperfine line anomalies. On the other hand, the obtained $R_{12}$ values lie in the expected LTE range. For positions in NW, toward which both the $F=0-1$ and $F=2-1$ hfs lines are detected with at least 5$\sigma$, $R_{02}$ ranges from 0.6 to 1.7. The higher velocity resolution spectrum toward source A in NW reveals that all three individual hfs components show self-absorption (see Fig.~\ref{Fig:sp-A}). In SE, the integrated intensities of the $F=0-1$ line are higher than those of the $F=2-1$ line which is barely detected. This clearly shows that the entire SE region exhibits HCN (1--0) hyperfine line anomalies, even more pronounced than those in NW. In SE, we can constrain the lower limits ($3\sigma$) to $R_{02}$ which are found in the range of $\sim$1.0--2.0 for different pixels. The lower limit of $\sim$2.0 toward the $F=0-1$ peak of SE surpasses all reported R$_{02}$ values found in other sources \citep[e.g.,][]{2012MNRAS.420.1367L}. Such an extreme HCN (1--0) hyperfine line anomaly is confirmed by our deeper integration toward source B (see Fig.~\ref{Fig:sp-B}) which is found to have a R$_{02}$ value of $\sim$2.6.

\subsection{Widespread blue-skewed profiles}\label{sec.pv}
As shown in Fig.~\ref{Fig:sp-A}, three optically thick lines including HCN (1--0), HNC (1--0), and CS (2--1) are double-peaked with a stronger component on the blue-shifted side, while their corresponding rare isotopologues including H$^{13}$CN (1--0), HN$^{13}$C (1--0), and $^{13}$CS (2--1) peak at the dips of the self-absorbed lines. Such blue-skewed profiles are known to be an indication of inward motions \citep[e.g.,][]{1993ApJ...404..232Z,1996ApJ...465L.133M}. Figure~\ref{Fig:pv-abs} presents position-velocity (PV) diagrams across the crest of the filament in C$^{18}$O (1--0), HNC (1--0), and CS (2--1).  In Fig.~\ref{Fig:pv-abs}b, only a single velocity component is seen in C$^{18}$O (1--0), confirming that the Serpens filament is velocity-coherent. In Fig.~\ref{Fig:pv-abs}c-\ref{Fig:pv-abs}d, two main velocity components are found in HNC (1--0) and CS (2--1), while C$^{18}$O (1--0) lies between the two main velocity components. This structure, seen in HNC and CS, is more tentative in HCN (1--0) (see Fig.~\ref{Fig:pv2}a in Appendix~\ref{app.a}), but this is mainly due to this line's coarser spectral resolution (0.21~\kms). Generally, the blue-shifted component is stronger than the red-shifted component in HNC (1--0) and CS (2--1), indicating the presence of widespread blue-skewed profiles across the filament. This strongly suggests that the Serpens filament is undergoing radial gravitational collapse. Our observations demonstrate that the two velocity components identified in optically thick tracers can also arise from the same spectral component, so characterising filaments and their velocity structures in optically thin tracers is more reliable.







\section{Discussion}
\subsection{The density structure}
In order to study the radial column density profile of the Serpens filament, we use the H$_{2}$ column density maps from the Herschel Gould Belt Survey to constrain our physical model of filaments. We first extract the crest of the Serpens filament with the discrete persistent structure extractor (DisPerSe\footnote{http://www2.iap.fr/users/sousbie/web/html/indexd41d.html?}) algorithm \citep{2011MNRAS.414..350S}. Based on the discrete Morse theory, this algorithm can be used to extract filamentary structures in 2D and 3D datasets. The extracted skeletons are representive of the crest of the filaments. There are two important parameters, the persistence and robustness thresholds, in this algorithm. The persistence threshold describes the column density difference of a pair of critical points where the gradient is null. This value is set to be 4.0$\times 10^{20}$~cm$^{-2}$ to isolate the filamentary structure. The robustness threshold is defined as a local measure of how contrasted the critical points and filaments are with respect to their background. The robustness threshold is set to be 6$\times 10^{21}$ cm$^{-2}$. In order to make the structure as long as possible, we also assemble pieces of filaments that form an angle smaller than 70\degree. These parameters are adopted to identify the skeleton in the H$_{2}$ column density map without smoothing. Since we are only interested in the dense crest of the Serpens filament, the structures with column densities lower than 1$\times 10^{22}$~cm$^{-2}$ were trimmed to better visualize its crest. The resulting crest is shown as the blue line in Fig.~\ref{Fig:rprof}a. The spatial separation between every pixel and the crest is calculated as the radial distance. The width in NW is apparently broader than in SE, which is because NW is likely affected by the feedback of the three YSOs. SE is more pristine, so SE is used to derive the radial H$_{2}$ column density profile here. The radial column density profile of this filament is estimated within the white box in Fig.~\ref{Fig:rprof}a. The western part is contaminated by the emission of EX, leading to the asymmetry of the radial column density profile in Figs.~\ref{Fig:rprof}b and \ref{Fig:rprof}c. In order to avoid the errors caused by the contamination, we employ Gaussian functions to only fit the eastern part of SE. The fitted FWHM width is 0.19$\pm$0.01~pc, corresponding to a deconvolved width of 0.17$\pm$0.01 pc. Assuming a cylindrical geometry with a radius of 0.17~pc and a length of 1.6 pc, we estimate the average number density to be about 6.5$\times 10^{3}$~cm$^{-3}$.


Based on theoretical predictions of infinitely-long filaments \citep[e.g.,][]{1964ApJ...140.1056O,2011A&A...529L...6A,2016AA...586A..27K}, their radial column density profiles can be expressed in a Plummer-like function as
\begin{equation}\label{eq-r}
N(r) = A_{\rm p}\frac{n_{\rm c}\sqrt{8}r_{0}}{(1+(\frac{r}{\sqrt{8}r_{0}})^{2})^{(p-1)/2}}\,,
\end{equation}
where $A_{\rm p} = \frac{1}{\rm{cos}(\alpha)}\int_{\infty}^{\infty}\frac{\rm{d}u}{(1+u^{2})^{p/2}}$ where the inclination, $\alpha$, is set to 0 here, $n_{\rm c}$ is the number density at the center of the filament, $r_{0}$ is the scale radius. We note that the so-called characteristic radius, $R_{\rm flat}$, is also used in previous studies \citep[e.g.,][]{2011A&A...529L...6A}. According to their definitions, $R_{\rm flat}$ is simply $\sqrt{8}$ times $r_{0}$, i.e., $R_{\rm flat}=\sqrt{8}r_{0}$. During the fitting process, we first convolve Eq.~(\ref{eq-r}) with a Gaussian function with a FWHM of 37\arcsec\, ($\sim$0.08 pc) to take the contribution of the beam into account. In the fitting processes, the modeled $p$ varies from 1.0 to 4.0 with a step size of 0.1, the modeled $n_{\rm c}$ varies from 1$\times 10^{3}$ to 1$\times 10^{5}$ cm$^{-3}$ with a step size of 1$\times 10^{3}$ cm$^{-3}$, and the modeled $r_{0}$ varies from 0.001 to 0.030 pc with a step of 0.001 pc. The goodness of the fit is estimated from the chi-square ($\chi^{2}$) analysis, and we minimize $\chi^{2}$ between data points and models. Figure~\ref{Fig:rprof}c gives the best fits to the observed radial column density of the regions East and West of the crest lines (as defined by the rectangles in Fig.~\ref{Fig:rprof}a) within a radius range of $r<0.2$~pc, respectively. The best fit to the East leads to $p=$2.2$\pm$0.1, $n_{\rm c}=(4.0\pm 0.8) \times 10^{4}$~cm$^{-3}$, and $r_{0}=0.018\pm0.003$~pc, while the best fit to the West results in $p=$2.0$\pm$0.3, $n_{\rm c}=(2.5\pm 1.8) \times 10^{4}$~cm$^{-3}$, and $r_{0}=0.023\pm0.011$~pc. The given errors are the one-sigma uncertainties of the fit. Since the asymmetry of the radial column density profile is likely due to the contamination by EX, only the derived parameters of the East are used for the following discussions. When adopting different inclinations, $n_{c}$ changes by a factor of $1/{\rm cos(\alpha)}$. On the other hand, the $p$ value is dependent on the selected radius range. When the range extends out to 0.25 pc toward the East, this leads to a much higher $\chi^{2}$, e.g., a poorer fit. On the other hand, the derived $p$ value (2.2$\pm$0.1) of the East indicates that the profile is similar those found in IC 5146 \citep[$p\sim$2,][]{2011A&A...529L...6A} and other filaments in nearby molecular clouds (see Table~\ref{com}), but is shallower than that of an isolated isothermal filament in hydrostatic equilibrium \citep[$p$=4,][]{1964ApJ...140.1056O}. Comparing other filaments in nearby molecular clouds (see Table~\ref{com}), the physical properties of the Serpens filament are similar to those of Musca. At larger radial range ($r\gtrsim$0.2 pc), the observed radial column density profiles of both the East and West significantly deviate from the Plummer-like function. When fitting the tail within the radial range of 0.2--0.35 pc with a power law, the H$_{2}$ column density is found to decrease as $N_{\rm H2}\sim r^{-0.3}$ in the East profile and $N_{\rm H2}\sim r^{-0.5}$ in the West profile. At a radial range of $>$0.35 pc, the column densities of both the East and West stay nearly constant with mean values of 2.7$\times 10^{21}$~cm$^{-2}$ and 5.1$\times 10^{21}$~cm$^{-2}$, respectively.


Figure~\ref{Fig:maxprof}b shows the column density profile along the crest of the Serpens filament. There are only three maxima across the $\sim$1.5 pc long crest in this profile. In addition, this filament appears to have a separation at a projected scale of $\sim$0.85 pc indicated by the orange line in Fig.~\ref{Fig:maxprof}b, while a smaller separation at a projected scale of $\sim$0.26 pc is found in NW (indicated by the red line in Fig.~\ref{Fig:maxprof}b). Since little is known about the inclination, we should keep in mind that these scales can only be taken as lower limits. The larger separation of $\sim$0.85 pc is nearly 5 times the fitted FWHM width of the filament derived above. This is roughly consistent with the prediction that fragmentation is expected with separations of about four times the diameter of the filament in equilibrium or a dynamical system with the observed line mass (mass per unit length) close to the critical line mass \citep{1953ApJ...118..116C,1985MNRAS.214..379L,1992ApJ...388..392I}. On the other hand, the smaller separation of $\sim$0.26 pc in NW is close to the thermal Jeans length ($\sim$0.26 pc) of a spherical cloud \citep{1902RSPTA.199....1J} by assuming a number density of 6.5$\times 10^{3}$~cm$^{-3}$ averaged over the the entire length of the filament and a kinetic temperature of 10~K. Alternatively, the smaller separation is also possibly caused by the feedback of emb28 which is observed at the dip of the H$_{2}$ column density profile (see Fig.~\ref{Fig:maxprof}b).

\subsection{Gravitational collapse observed in a (trans-)sonic filament}
The stability of the Serpens filament can be assessed by comparing the observed line mass, $M_{\rm l}$, with the critical line mass, $M_{\rm l,crit}$, predicted by theoretical models. Based on the H$_{2}$ column density map from the Herschel Gould Belt Survey, we calculate the mass of the Serpens filament by integrating the column densities inside the contour of $N_{\rm H2}$ = \{6, 8, 10\}$\times 10^{21}$~cm$^{-2}$ (see Fig.~\ref{Fig:rprof}a for the contours). The resulting total mass is \{66, 50, 36\}~M$_{\odot}$ within a length of \{1.6, 1.3, 1.0\} pc, which gives a line mass, $M_{l}$, of \{41, 38, 36\}~$M_{\odot}$~pc$^{-1}$. Note that these values are about three times as high as 13~$M_{\odot}$~pc$^{-1}$ reported by \citet{2015A&A...584A.119R}, in which the number of bands, dust spectral indices, and color corrections adopted for SED fitting and the H$_{2}$ column density map is different from that of the Herschel GB team \citep{2015A&A...584A..91K,2017MmSAI..88..783F}. Detailed discussions about the differences are presented in Appendix~\ref{app.c}. The  H$_{2}$ column density map derived by the Herschel GB team is more appropriate and therefore used for this work. The critical line mass of an infinite filament in hydrostatic equilibrium is expected to be $m_{\rm l,crit} = \frac{2\sigma_{\rm eff}^{2}}{\rm G}$ \citep[e.g.,][]{1964ApJ...140.1056O,1992ApJ...388..392I}. When $\sigma_{\rm eff}$ is taken to be the sound speed, i.e. gravity is only balanced by thermal support, the critical line mass is 8--17~$M_{\odot}$~pc$^{-1}$ at a kinetic temperature of 5--10~K. As pointed out by previous studies \citep[e.g.,][]{2017A&A...606A.123H}, the calculation of $m_{\rm l,crit}$ should include the contribution from non-thermal motions in practice. When the equilibrium is achieved by the support from both the thermal and non-thermal motions, $\sigma_{\rm eff}$ should be taken to be the total velocity dispersion, $\sigma_{\rm t}$, which can be calculated from the equation below:
\begin{equation}\label{eq2}
\sigma_{\rm t} = \sqrt{\sigma_{\rm obs}^{2}+c_{\rm s}^{2}-\sigma_{\rm th}^{2}}\;,
\end{equation}
where $\sigma_{\rm obs}$ is the observed velocity dispersion. Because SE is more pristine, the properties of SE are more representive of the initial conditions. $\sigma_{\rm obs}$ is thus taken to be the median velocity dispersion (0.20~\kms) of SE. This leads to a critical line mass of 25--34~$M_{\odot}$~pc$^{-1}$ within a kinetic temperature range of 5--10~K. The observed line mass appears to be higher than the critical line mass, but does not exceed the critical value greatly, suggesting that the Serpens filament is slightly supercritical. 

As predicted previously by theoretical models \citep[e.g.,][]{1992ApJ...388..392I}, such slightly supercritical infinite filaments are able to collapse. Due to the gravitational focusing effect \citep{2004ApJ...616..288B,2011ApJ...740...88P,2016arXiv160305720L}, finite filaments are more prone to collapse at the ends of a filament even when such filaments are subcritical. Blue-skewed line profiles are known to be a signature of infall motions \citep[e.g.,][]{1993ApJ...404..232Z,1996ApJ...465L.133M,2016A&A...585A.149W}. Recently, \citet{2016MNRAS.459.2882M} pointed out that HCN (1--0) hyperfine line anomalies could also be ascribable to gravitational infall. The observed widespread blue-skewed line profiles and HCN (1--0) hyperfine line anomalies across the Serpens filament suggest that the filament is undergoing radial (i.e., perpendicular to the major axis) collapse. Therefore, our observations indicate that such radial collapse is present in a (trans-)sonic filament. Such collapse has been observed in other supersonic filaments \citep[e.g.,][]{2010A&A...520A..49S,2013ApJ...766..115K}, but evidence for radial collapse in a (trans-)sonic filament like the Serpens filament remains sparse.


\subsection{The radial collapse}\label{sec.infall}
Due to the greater gravitational potential in the central filament, ambient molecular material will collapse onto the center of the filament. As pointed out by previous studies \citep[e.g.,][]{2013ApJ...766..115K}, the infall rate, $\dot M_{\bot}$, can be estimated with
\begin{equation}\label{eq3}
\dot M_{\bot} = \frac{2\varv_{\rm i}M}{r{\rm cos}(\alpha)} \;,
\end{equation}
where $\varv_{\rm i}$ is the infall velocity, $M$ is the mass of the filament, $r$ is the radius of the filament, and $\alpha$ is the inclination which is used to correct projection effects. In order to estimate the infall velocity, we use the Hill5 model \citep{2005ApJ...620..800D} in the pyspeckit package \citep{2011ascl.soft09001G}. The Hill5 model employs an excitation temperature profile increasing linearly toward the center, rather than the two-slab model of \citet{1996ApJ...465L.133M}, so the Hill5 model is thought to provide a better fit of infall motions \citep{2005ApJ...620..800D}. In the Hill5 model, five physical parameters, including the opacity, $\tau_{0}$, the systemic LSR velocity, $\varv_{\rm s}$, the infall velocity, $\varv_{\rm i}$, the velocity dispersion, $\sigma_{0}$, and the peak excitation temperature, $T_{\rm e}$, can be fitted. As shown in Figs.~\ref{Fig:sp-A} and \ref{Fig:pv-abs}, both HNC (1--0) and CS (2--1) display widespread blue-skewed profiles, and are thus used as an input in the Hill5 model. Spectra in NW and SE are averaged to improve the signal-to-noise ratios, and the averaged spectra are fitted with the Hill5 model to derive the infall velocity. The fitted results are shown in Fig.~\ref{Fig:hill5} and given in Table~\ref{Tab:hill5}. These results suggest that both CS (2--1) and HNC (1--0) have high opacities ($\tau_{0} >1$). The fitted LSR velocities and velocity dispersions and the values derived from C$^{18}$O (1--0) agree within 0.1~\kms. The fitted infall velocities of NW and SE are found to be 0.05 to 0.13~\kms\,with an average value of 0.09~\kms. The fitted peak excitation temperatures are very low, i.e., $T_{\rm e}\sim 5$~K. Meanwhile, the HNC (1--0) and CS (2--1) spectra of source A are also fitted with the Hill5 model, and the results are shown in Fig.~\ref{Fig:sp-A}. The fitted parameters are consistent with the values derived with the average spectra.

Assuming that the whole filament is collapsing with a mean infall velocity of 0.09~\kms, we estimate its infall rate to be 72~$M_{\odot}$~Myr$^{-1}$ with Equation~\ref{eq3} when the inclination is set to be 0\degree. This infall rate is likely to be lower than what is found in Serpens South \citep[130~$M_{\odot}$~Myr$^{-1}$ under the same assumption,][]{2013ApJ...766..115K} and in the much more massive DR21 filament \citep[a few 10$^{3}$~$M_{\odot}$~Myr$^{-1}$,][]{2010A&A...520A..49S}. We also note that the derived infall rate onto the filament is a lower limit, because cos$(\alpha)$ is very likely to be $<1$ in reality. With this current infall rate, it needs about 0.9~Myr to double the mass of the Serpens filament. This timescale is comparable to the age ($\sim$1~Myr) to form class I YSOs \citep{2009ApJS..181..321E} which are found in the filament. On the other hand, this timescale is roughly twice the free-fall time ($\tau_{\rm ff} =\sqrt{\frac{3\pi}{32G\rho}}=3.8\times 10^{5}$ yr) at an average number density of 6.5$\times 10^{3}$~cm$^{-3}$. The low rate could be due to its relatively early evolutionary stage. This is also supported by the derived infall velocities which are less than 10\% of the free-fall velocity ($\varv_{\rm ff}= \sqrt{2GM/r}\sim 1.8$~\kms). As pointed out by \citet{2014prpl.conf..149T}, such a low infall velocity indicates that collapse has just begun. Alternatively, both thermal and non-thermal support are effective in providing support against gravity, leading to the low rate.

The mass (36--66~M$_{\odot}$) of the Serpens filament is similar to that (52~M$_{\odot}$) of the Serpens South filament \citep{2013ApJ...766..115K}, but the radial column density profile is different for the two filaments. The Serpens filament has 10 times lower central density and 3 times higher scale radius than the Serpens South filament. As collapse goes on, the central density will become higher and the scale radius will become smaller \citep[e.g.,][]{1992ApJ...388..392I}. This suggests that the Serpens filament may lie at an earlier evolutionary stage, prior to the Serpens South filament. This is reinforced by the fact that the Serpens filament contains less YSOs.





\subsection{The longitudinal collapse}\label{Sec.long}
Accretion flows along filaments are important for the growth of dense cores and protostars. A few observational studies indicate that there are accretion flows along the filaments' long axis to feed nascent stars or the dense cores \citep[e.g.,][]{2013ApJ...766..115K,2014A&A...561A..83P,2015ApJ...804...37L,2018ApJ...852...12Y,2018ApJ...855....9L}. As described in Sect.~\ref{sec.kine} (see also Fig.~\ref{Fig:kine}b), velocity gradients, $\nabla \varv_{\|}$, are also observed along the filament's long axis, which could be due to such accretion flows. The accretion rate along the filament, $\dot M_{\|}$, can be estimated with the following equation \citep{2013ApJ...766..115K},
\begin{equation}\label{eq4}
\dot M_{\|} = \frac{\nabla \varv_{\|}M}{{\rm tan}(\alpha)} \;.
\end{equation}
As mentioned in Sect.~\ref{sec.kine}, $\nabla \varv_{\|,1}$, $\nabla \varv_{\|,2}$,  $\nabla \varv_{\|,3}$, and $\nabla \varv_{\|,4}$ are $0.40\pm0.05$, $0.30\pm0.09$, $0.53\pm 0.06$, and $0.55 \pm 0.14$~\kms~pc$^{-1}$. The regions showing the four velocity gradients are found to have masses of $\sim$20~$M_{\odot}$, $\sim$10~$M_{\odot}$, $\sim$18~$M_{\odot}$, and $\sim$16~$M_{\odot}$ according to the H$_{2}$ column densities derived from dust emission (see Fig.~\ref{Fig:rprof}a). Assuming an inclination of 45\degree, we estimate the accretion rates to be $\sim$8~$M_{\odot}$~Myr$^{-1}$, $\sim$3~$M_{\odot}$~Myr$^{-1}$, $\sim$10~$M_{\odot}$~Myr$^{-1}$, and $\sim$9~$M_{\odot}$~Myr$^{-1}$. These values are lower than all reported accretion rates ($\gtrsim$50~$M_{\odot}$~Myr$^{-1}$) along filaments under the same assumption of an inclination of 45\degree\,\citep{2013ApJ...766..115K,2018ApJ...852...12Y,2018ApJ...855....9L}. If these low values are due to the earlier evolutionary stage of the Serpens filament, this may indicate that gas accretion can be accelerated during its evolution. The accretion rate per unit length, $\dot M_{\|}/L$, is found to be $\sim$20~$M_{\odot}$~Myr$^{-1}$~pc$^{-1}$, which is also lower than that ($\sim$90~$M_{\odot}$~Myr$^{-1}$~pc$^{-1}$) of Serpens South (see Table~\ref{com}). The low values of $\dot M_{\|}$ and $\dot M_{\|}/L$ may be the reason why star formation in the Serpens filament is not as active as that in Serpens South.

As shown in Figs.~\ref{Fig:kine}b and \ref{Fig:maxprof}c, the two velocity gradients $\nabla \varv_{\|,1}$ and $\nabla \varv_{\|,2}$ converge to source B in SE, while the other two velocity gradients $\nabla \varv_{\|,3}$ and $\nabla \varv_{\|,4}$ converge to emb28 in NW. Furthermore, the two velocity converging positions roughly coincide with the enhancements of H$_{2}$ column densities in Fig.~\ref{Fig:maxprof}b. In addition, the velocities change as nearly a linear function of offsets from the two converging positions (indicated by the arrows in Fig.~\ref{Fig:maxprof}c), which is roughly consistent with the velocity profiles expected in homologous free-fall collapse of filaments (see fig.~5 of \citet{2007A&A...464..983P}, fig.~5 of \citet{2014A&A...561A..83P}, and fig.~1 of \citet{2005ApJ...623..280M}). This suggests that such accretion flows are likely due to the longitudinal collapse of the Serpens filament. Toward the velocity converging place in NW, the line widths are apparently higher (see Fig.~\ref{Fig:maxprof}d), which may also result from the collision of the longitudinal collapsing gas indicated by arrows in Fig.~\ref{Fig:kine}b or the feedback of the YSOs. Together with the velocity gradients perpendicular to the filament's long axis and the widespread blue-skewed profiles (see Sect.~\ref{sec.kine} and \ref{sec.infall}), this implies that radial and longitudinal collapse occur hand in hand in the Serpens filament.





\section{Summary and conclusion}
We have mapped the Serpens filament (one of the nearest infrared dark clouds) in 7 different species (C$^{18}$O, HCO$^{+}$, HNC, HCN, N$_{2}$H$^{+}$, CS, and CH$_{3}$OH) near $\lambda=$3 mm with the PMO-13.7 m telescope, and have studied its physical properties and dynamical state. The main results include:

\begin{itemize}
\item[1.] Among the 7 species, HCO$^{+}$, HNC, HCN, CS are found to show self-absorption, while C$^{18}$O is most sensitive to the filamentary structure. Based on the C$^{18}$O data, the Serpens filament is found to be a velocity-coherent (trans-)sonic filament which is one of the most quiescent regions in the Serpens cloud. Widespread C$^{18}$O depletion is found throughout the Serpens filament where the C$^{18}$O fractional abundances are (3.4--7.0)$\times 10^{-8}$, lower than those in ambient regions by a factor of $\sim$2. N$_{2}$H$^{+}$ has fractional abundances of (0.8--2.3)$\times 10^{-10}$. The N$_{2}$H$^{+}$/C$^{18}$O abundance ratios are found to increase with increasing H$_{2}$ column densities, suggesting that N$_{2}$H$^{+}$ is less depleted than C$^{18}$O in dense regions. Furthermore, blue-skewed HNC and CS line profiles and HCN hyperfine line anomalies are widespread across this filament.

\item[2.] The filament has a total gas mass of $\sim$66~$M_{\odot}$ within a length of $\sim$1.6~pc, and its FWHM width is 0.17$\pm$0.01~pc. The radial column density profile is asymmetric in the Serpens filament. Within a radius range of $<$0.2 pc, the radial H$_{2}$ column density profile is well described as a Plummer-like profile. Our fit leads to an exponent ($p$) of 2.2$\pm$0.1, scale radius ($r_{0}$) of $0.018\pm 0.003$ pc, and central density ($n_{\rm c}$) of $(4.0\pm 0.8)\times 10^{4}$~cm$^{-3}$ for the East profile which is similar to Musca but shallower than that of an infinite hydrostatic filament. In the outer part ($r>$0.2 pc), the H$_{2}$ column densities fall as a power-law profile to a nearly constant column density. The column density profile across the filament crest presents two characteristic fragmentation scales of $\sim$0.85 pc and $\sim$0.26~pc. The former roughly agrees with the filament fragmentation scale, while the later could be due to the thermal Jeans instability or the feedback of young stellar objects.

\item[3.] The Serpens filament is found to be slightly supercritical with an observed line mass (mass per unit length) of 36--41~$M_{\odot}$~pc$^{-1}$. The widespread blue-skewed HNC and CS line profiles and HCN hyperfine line anomalies across this filament imply that radial collapse is present in a (trans-)sonic filament. This is consistent with the theoretical prediction that slightly supercritical filaments are prone to collapse.

\item[4.] Based on the HNC and CS blue-skewed profiles, the radial infall velocities of the Serpens filament are found to be less than 10\% of the free-fall velocity, and the resulting infall rate of radial collapse is found to be $\sim$72~$M_{\odot}$~Myr$^{-1}$, which implies that it needs about 0.9 Myr (roughly twice the free-fall time) to double the mass of the Serpens filament. The timescale is also comparable to the age ($\sim$1~Myr) to form class I young stellar objects. C$^{18}$O velocity gradients along the filament's long axis are indicative of gas accretion within this filament. This leads to an accretion rate of $\sim$10~$M_{\odot}$~Myr$^{-1}$ along this filament. Its velocity and density structures support that such accretion flows are likely due to the filament's longitudinal collapse. Both the infall rate and the accretion rate along the Serpens filament are lower than all previously reported values in other filaments. This indicates that the Serpens filament lies at an early evolutionary stage when collapse has just begun, or that thermal and non-thermal support are effective in providing support against gravity.
\end{itemize}


\section*{ACKNOWLEDGMENTS}\label{sec.ack}
We appreciate the assistance of the PMO-13.7 m staff during the observations. We are grateful to the FFTS team who made the special 200 MHz FFTS mode available to the community. Y. Gong also acknowledges the organizers of the treasure hunt around Yardangs in the Tsaidam Desert which triggers the idea of this paper. Stefano Pezzuto is acknowledged for providing information on the H$_{2}$ column density map. Shaobo Zhang is acknowledged for providing his AICer\footnote{https://github.com/shbzhang/aicer} plotting code. Xing Lu is acknowledged for introducing the pyspeckit package. This work was supported by the National Key Research \& Development Program of China under grant 2017YFA0402702, the National Natural Science Foundation of China (NSFC) under grant 11127903, and the Chinese Academy of Sciences under grant QYZDJ-SSW-SLH047. J.X.S. was supported by NSFC under grant U1531103. This publication makes use of data products from the Wide-field Infrared Survey Explorer, which is a joint project of the University of California, Los Angeles, and the Jet Propulsion Laboratory/California Institute of Technology, funded by the National Aeronautics and Space Administration. This research has made use of data from the Herschel Gould Belt survey (HGBS) project (http://gouldbelt-herschel.cea.fr). This research made use of NASA's Astrophysics Data System. This research made use of Astropy, a community-developed core Python package for Astronomy \citep{2013A&A...558A..33A}. The authors thank the anonymous referee for useful comments on the manuscript. 

\begin{table*}[!hbt]
\caption{Observed parameters of the molecular lines used in the work.}\label{line}
\scriptsize
\centering
\begin{tabular}{cccccccc}
\hline \hline
Line                         & Frequency                 & $E_{\rm u}/k$ & $B_{\rm eff}$ &$\delta \varv$ &  $\delta \nu$  & $\sigma$  & obs.Mode  \\ 
                             & (GHz)                     &  (K)         &             &  (\kms)      &   (kHz)        & (K)        &          \\ 
(1)                          & (2)                       &  (3)        &  (4)         &  (5)         &   (6)          & (7)        & (8)      \\
\hline
\multicolumn{8}{c}{The Serpens filament}\\                                                                               
\hline
C$^{18}$O (1--0)              &  109.78217              & 5.0        & 55.5\%   &  0.04        & 12.2 & 0.27 &  mapping   \\ 
HCO$^{+}$ (1--0)              &  89.18852               & 4.3        & 56.8\%   &  0.21        & 61   & 0.07 &  mapping   \\ 
HCN (1--0)                   &  88.63185               & 4.3        & 56.8\%   &  0.21        & 61   & 0.07 &  mapping   \\ 
N$_{2}$H$^{+}$ (1--0)         &  93173.76               & 4.5        & 61.9\%   &  0.04        & 12.2 & 0.20 &  mapping    \\ 
CS (2--1)                    & 97.98095                 & 7.1       & 60.6\%   &  0.04        & 12.2 & 0.13 &  mapping   \\ 
HNC (1--0)                   & 90.66357                 & 4.4       & 57.0\%   &  0.04         & 12.2 & 0.16 &  mapping   \\ 
CH$_{3}$OH ($J_{K}$=$2_{0}$--$1_{0}$ A$^{+}$) & 96.74138   & 7.0       & 63.4\%   &  0.19        & 61   & 0.07 &  mapping   \\ 
CH$_{3}$OH ($J_{K}$=$2_{-1}$--$1_{-1}$ E)     & 96.73936   & 12.5      & 63.4\%   &  0.19        & 61   & 0.07  &  mapping   \\ 
\hline                    
\multicolumn{8}{c}{Source A: ($\alpha_{\rm J2000}$=18$^{\rm h}$28$^{\rm m}$47$\rlap{.}^{\rm s}62$, $\delta_{\rm J2000}$=00\degr50\arcmin55$\rlap{.}$\arcsec8)}\\ 
\hline
HCN (1--0)                   &  88.63185                 & 4.3       & 56.8\%   &  0.04        &  12.2 & 0.06 & pointing   \\                 
H$^{13}$CO$^{+}$ (1--0)        & 86.75429                 & 4.2       & 56.7\% &  0.21          &  61  & 0.02 &  pointing   \\ 
H$^{13}$CN (1--0)             & 86.34018                 & 4.1       & 56.7\% &  0.21          &  61  & 0.02 &  pointing   \\ 
HN$^{13}$C (1--0)             & 87.09085                 & 4.2       & 56.7\% &  0.21          &  61  & 0.02 &  pointing   \\
$^{13}$CS  (2--1)             & 92.49426                 & 6.7       & 61.9\% &  0.20          &  61  & 0.02 &  pointing   \\ 
CCS ($J_{N}$=$8_{7}$--$7_{6}$) & 93.87011                 & 19.9      & 61.9\% &  0.19          &  61  & 0.04 &  pointing   \\
\hline                    
\multicolumn{8}{c}{Source B: ($\alpha_{\rm J2000}$=18$^{\rm h}$28$^{\rm m}$57$\rlap{.}^{\rm s}70$, $\delta_{\rm J2000}$=00\degr48\arcmin02$\rlap{.}$\arcsec0)}\\ 
\hline
HCN (1--0)                   &  88.63185                 & 4.3     & 56.8\%   &  0.04        &  12.2 & 0.06 & pointing   \\      
\hline
\end{tabular}
\tablefoot{(1) The observed transition. (2) The rest frequency of the observed transition. (3) The upper energy of the observed transition. (4) The main beam efficiency. (5) The channel width in units of \kms. (6) The channel width in units of kHz. (7) The rms noise level. (8) The obsevering mode.}
\normalsize
\end{table*}

\begin{table*}[!hbt]
\caption{Observed and physical properties of source A.}\label{Tab:a}
\scriptsize
\centering
\begin{tabular}{cccccccc}
\hline \hline
Line                          & $\varv_{\rm LSR}$  & $\Delta \varv$ & $\int T_{\rm mb}{\rm d}\varv$ &$N_{\rm mol}$                    \\ 
                              &  (\kms)          &  (\kms)        & (K~\kms)                    &  (cm$^{-2}$)                   \\ 
                              &  (1)             &  (2)           & (3)                         &  (4)                          \\ 
\hline
C$^{18}$O (1--0)               & 8.11$\pm$0.01   & 0.53$\pm$0.01       & 1.14$\pm$0.06  &  (5.9$\pm$0.3)$\times 10^{14}$           \\ 
N$_{2}$H$^{+}$ (1--0)          & 8.06$\pm$0.12   & 0.45$\pm$0.12       & 1.29$\pm$0.09   &  (3.1$\pm$0.2)$\times 10^{12}$           \\        
HCO$^{+}$ (1--0)              &  \nodata        & \nodata             &  $<$0.32        &  \nodata                                \\
H$^{13}$CO$^{+}$ (1--0)        & 8.20$\pm$0.21   & 0.70$\pm$0.20       & 0.42$\pm$0.01   & (2.5$\pm$0.1)$\times 10^{11}$            \\ 
HCN (1--0)                   &   \nodata        & \nodata             & 0.31$\pm$0.02   & $>$1.1$\times 10^{12}$                  \\ 
H$^{13}$CN (1--0)             &  8.04$\pm$0.21   & 0.50$\pm$0.21       & 0.11$\pm$0.01   & (4.0$\pm$0.4)$\times 10^{11}$           \\ 
HNC (1--0)                   & \nodata          & \nodata             & 1.33$\pm$0.05   & $>$2.4$\times 10^{12}$                  \\    
HN$^{13}$C (1--0)             & 8.09$\pm$0.21    & 0.77$\pm$0.21       & 0.47$\pm$0.01   & (1.1$\pm$0.2)$\times 10^{12}$           \\
CS (2--1)                    & \nodata          & \nodata             & 0.57$\pm$0.04   & $>$2.6$\times 10^{12}$                  \\ 
$^{13}$CS  (2--1)             & 8.18$\pm$0.20    & 0.69$\pm$0.20       & 0.05$\pm$0.01   & (2.5$\pm$0.5)$\times 10^{11}$           \\ 
CCS ($J_{N}$=$8_{7}$--$7_{6}$)               & 8.47$\pm$0.19   & 0.52$\pm$0.19   & 0.13$\pm$0.02 &  (2.0$\pm$0.3)$\times 10^{12}$  \\ 
CH$_{3}$OH ($J_{K}$=$2_{0}$--$1_{0}$ A$^{+}$) & 8.28$\pm$0.19   & 0.57$\pm$0.19   & 0.26$\pm$0.03 &  (1.5$\pm$0.2)$\times 10^{13}$  \\ 
CH$_{3}$OH ($J_{K}$=$2_{-1}$--$1_{-1}$ E)     & 8.23$\pm$0.19   & 0.54$\pm$0.19   & 0.15$\pm$0.03 &  (8.8$\pm$0.2)$\times 10^{12}$  \\     
\hline
\end{tabular}
\tablefoot{(1) The LSR velocity. (2) The FWHM line width. (3) The integrated intensity. For N$_{2}$H$^{+}$ (1--0), the integrated velocity range is 5--11~\kms. For HCO$^{+}$ (1--0), HCN (1--0), HNC (1--0), and CS (2--1), the integrated velocity range is 7--9~\kms. 3$\sigma$ is taken as the upper limit of the HCO$^{+}$ (1--0) integrated intensity. (4) The beam-averaged molecular column density is calculated by assuming optically thin emission and an excitation temperature of 10~K. Since HCN (1--0), HNC (1--0), and CS (2--1) suffer from self-absorption, their derived column densities can only be used as lower limits.}
\normalsize
\end{table*}

\begin{table*}[!hbt]
\caption{A comparison between the Serpens filament and other filaments in nearby clouds.}\label{com}
\scriptsize
\centering
\begin{tabular}{cccccccccccc}
\hline \hline
                             &                         &                           &                       &  \multicolumn{3}{c}{The Plummer profile}  &                        &                       &  \\
\cline{5-7}
Filaments                    &  length                 &  width\tablefootmark{(1)} & $M_{\rm l}$             &  $p$  &  $n_{\rm c}$  & $r_{0}$  & $\dot M_{\bot}$         & $\dot M_{\|}$         & $\dot M_{\|}/L$& ref. \\ 
                             & (pc)                   &  (pc)                     &  ($M_{\odot}$~pc$^{-1}$) &       & (cm$^{-4}$)  & (pc)     &  ($M_{\odot}$~Myr$^{-1}$) & ($M_{\odot}$~Myr$^{-1}$)& ($M_{\odot}$~Myr$^{-1}$~pc$^{-1}$)&      \\ 
\hline
The Serpens filament         & 1.6$\pm$0.1            & 0.17$\pm$0.01             & 41$\pm$5               &  2.2$\pm$0.1 & (4.0$\pm$0.8)$\times 10^{4}$ & 0.018$\pm$0.003 & $\sim$72      &   $\sim$10 & $\sim$20  & This paper \\
Serpens South\tablefootmark{(2)} & 0.53                   & 0.13                      & 97                     &  1.9$\pm$0.2 & (3.6$\pm$2.2)$\times 10^{5}$ & 0.004$\pm$0.002 &$\sim$ 208     & $\sim$45 & $\sim$90 & \citet{2013ApJ...766..115K} \\
Musca                        & 6.5                    & 0.07                      & 21--31               &  2.6$\pm$0.3 & (5--10)$\times 10^{4}$ & 0.016$\pm$0.005 &       &   &  &  \citet{2016AA...586A..27K}\\
TMC-1                        & 0.6                    & 0.12                      & 90                   &  2.27        & 7.0$\times 10^{4}$ & 0.015         &       &   &  &  \citet{2012AA...544A..50M} \\
filaments in IC5146          &                        & 0.06--0.18                & 1--152               & 1.3--2.4     &                   & 0.004--0.028   &       &   &  & \citet{2011AA...529L...6A} \\
F2 in OMC1                   & 0.45                   & 0.02$\pm$0.008            &                      & 5.1$\pm$1    &                   & 0.004$\pm$0.001&       &   &  & \citet{2018ApJ...861...77M} \\
\hline
\end{tabular}
\tablefoot{(1) The FWHM width. (2) Scaled to a distance of 415 pc.}
\normalsize
\end{table*}

\begin{table*}[!hbt]
\caption{The fitted parameters of the Hill5 model for the average spectra and source A.}\label{Tab:hill5}
\scriptsize
\centering
\begin{tabular}{cccccc}
\hline \hline
Line         & $\tau_{0}$     & $\varv_{\rm s}$ & $\varv_{\rm i}$   & $\delta_{0}$     &$T_{\rm e}$                    \\ 
             &               &  (\kms)        & (\kms)         & (K~\kms)         &  (K)                        \\ 
(1)          &  (2)          &  (3)           & (4)            &  (5)             &  (6)                        \\ 
\hline
\multicolumn{6}{c}{The average spectra centered at an offset of ($-$40\arcsec, 40\arcsec) in Fig.~\ref{Fig:molecules}}\\
\hline
HNC (1--0)   & 3.77$\pm$2.55 & 8.09$\pm$0.10  & 0.13$\pm$0.15  & 0.29$\pm$0.07    &  5.29$\pm$0.79              \\    
CS (2--1)    & 3.31$\pm$4.38  & 8.05$\pm$0.14  & 0.10$\pm$0.22  & 0.20$\pm$0.10    &  4.36$\pm$0.93              \\ 
\hline
\multicolumn{6}{c}{The average spectra centered at an offset of (100\arcsec, $-$120\arcsec) in Fig.~\ref{Fig:molecules}}\\
\hline
HNC (1--0)  & 4.72$\pm$4.94 & 8.30$\pm$0.14   & 0.09$\pm$0.19  & 0.28$\pm$0.10    &  4.49$\pm$0.85              \\
CS (2--1)   & 2.11$\pm$4.71  & 8.21$\pm$0.15  & 0.05$\pm$0.29  & 0.17$\pm$0.13    &  4.24$\pm$0.84              \\
\hline
\multicolumn{6}{c}{Source A}\\
\hline
HNC (1--0)  & 3.83$\pm$2.41& 8.11$\pm$0.09    & 0.10$\pm$0.13  & 0.29$\pm$0.06    &  5.56$\pm$0.79              \\       
CS (2--1)   & 3.90$\pm$4.69& 8.05$\pm$0.11    & 0.05$\pm$0.16  & 0.21$\pm$0.09    &  4.56$\pm$0.92              \\ 
\hline
\end{tabular}
\tablefoot{(1) The used transition. (2) The fitted opacity. (3) The fitted systemic LSR velocity. (4) The fitted infall velocity. (5) The fitted velocity dispersion. (6) The fitted peak excitation temperature.}
\normalsize
\end{table*}

\begin{figure*}[!htbp]
\centering
\includegraphics[width = 0.4 \textwidth]{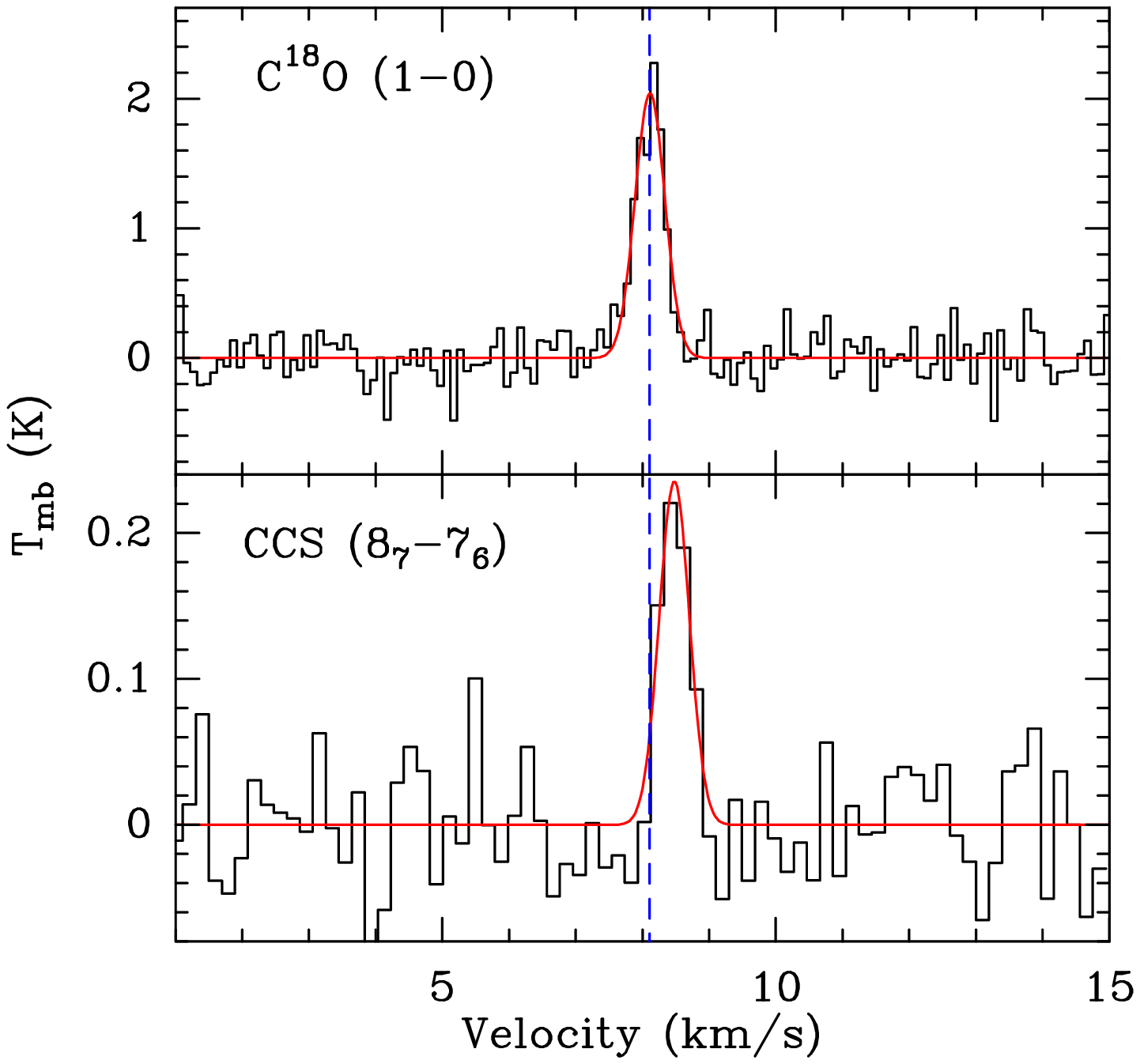}
\includegraphics[width = 0.4 \textwidth]{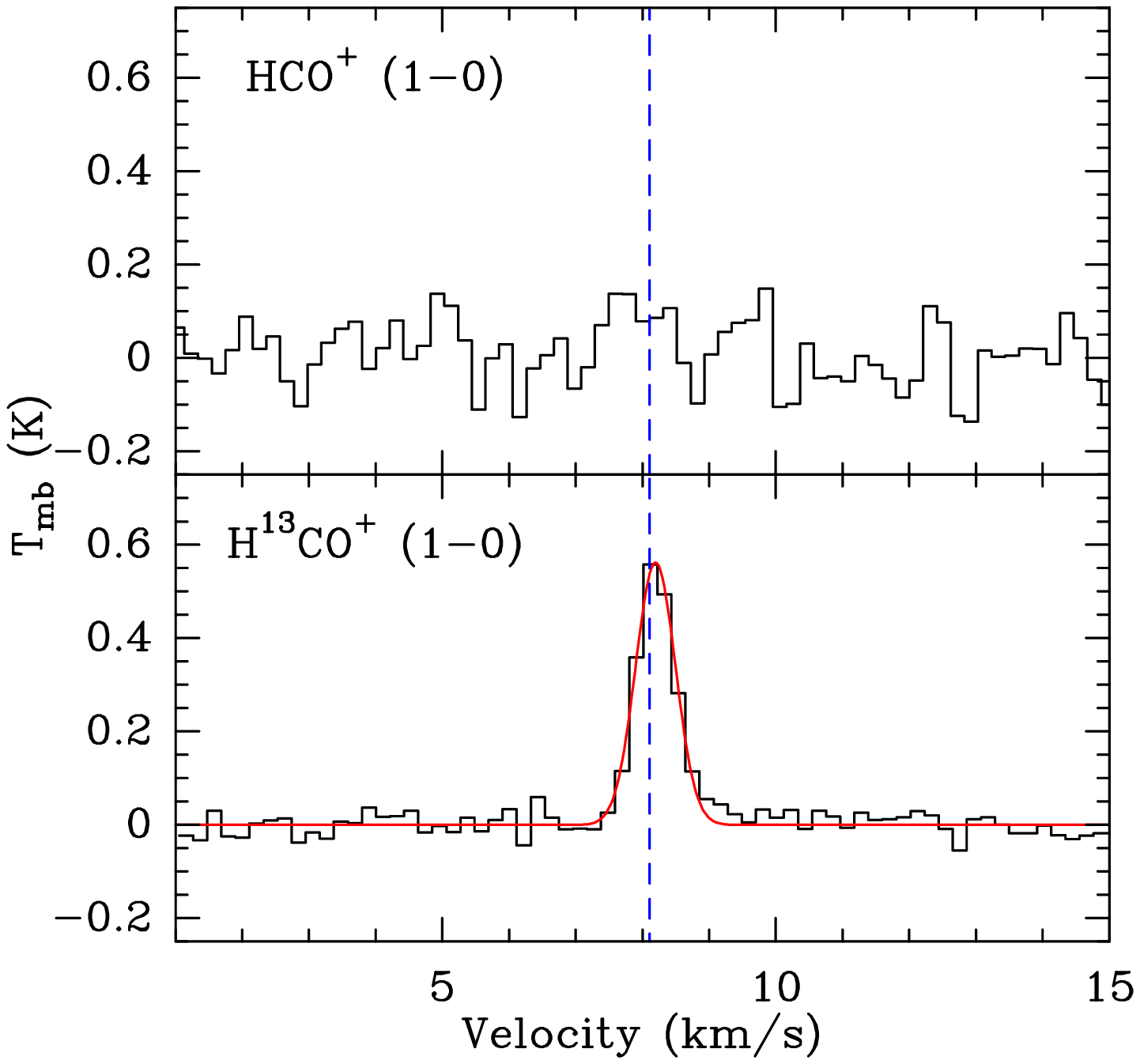}
\includegraphics[width = 0.4 \textwidth]{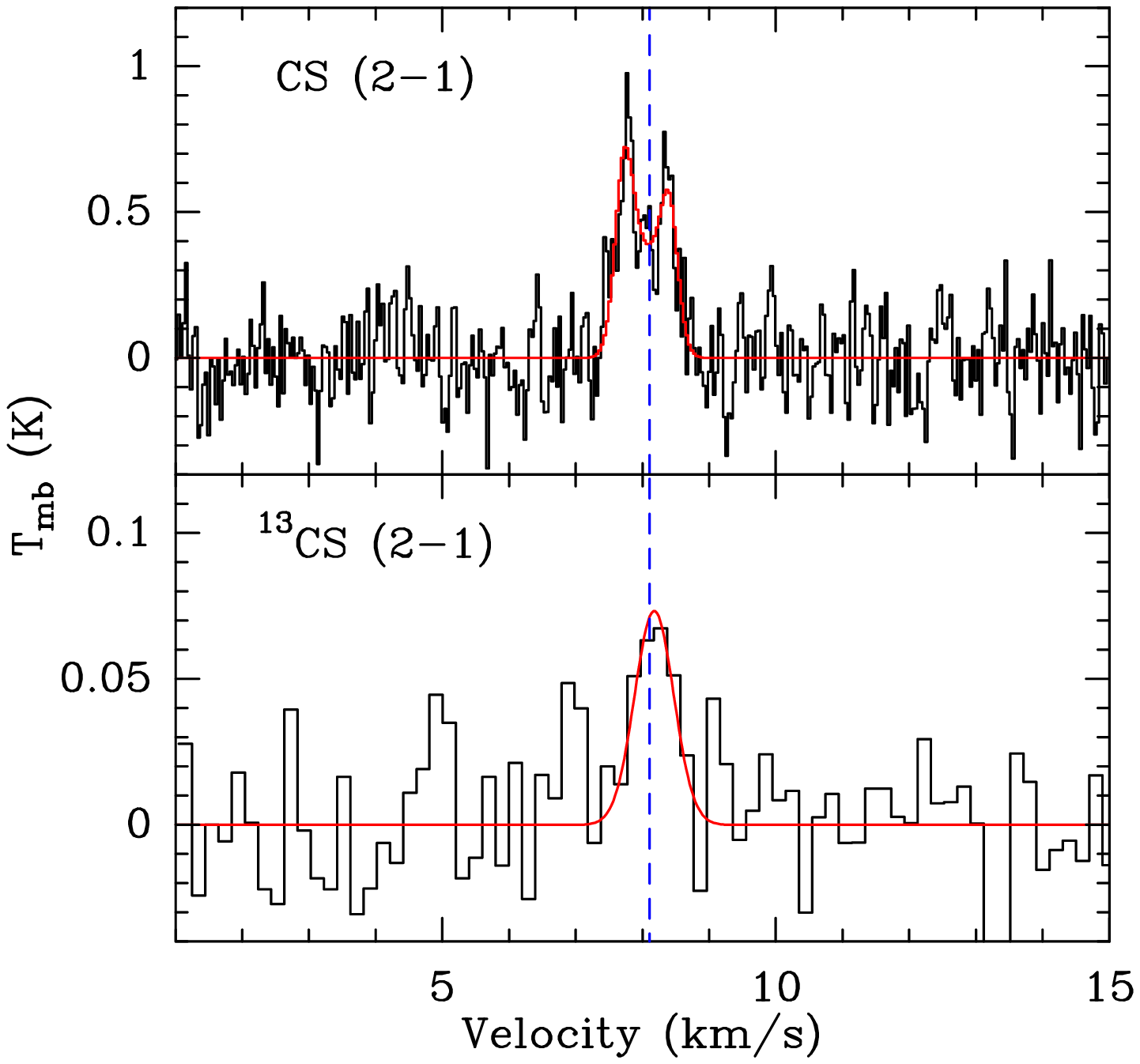}
\includegraphics[width = 0.4 \textwidth]{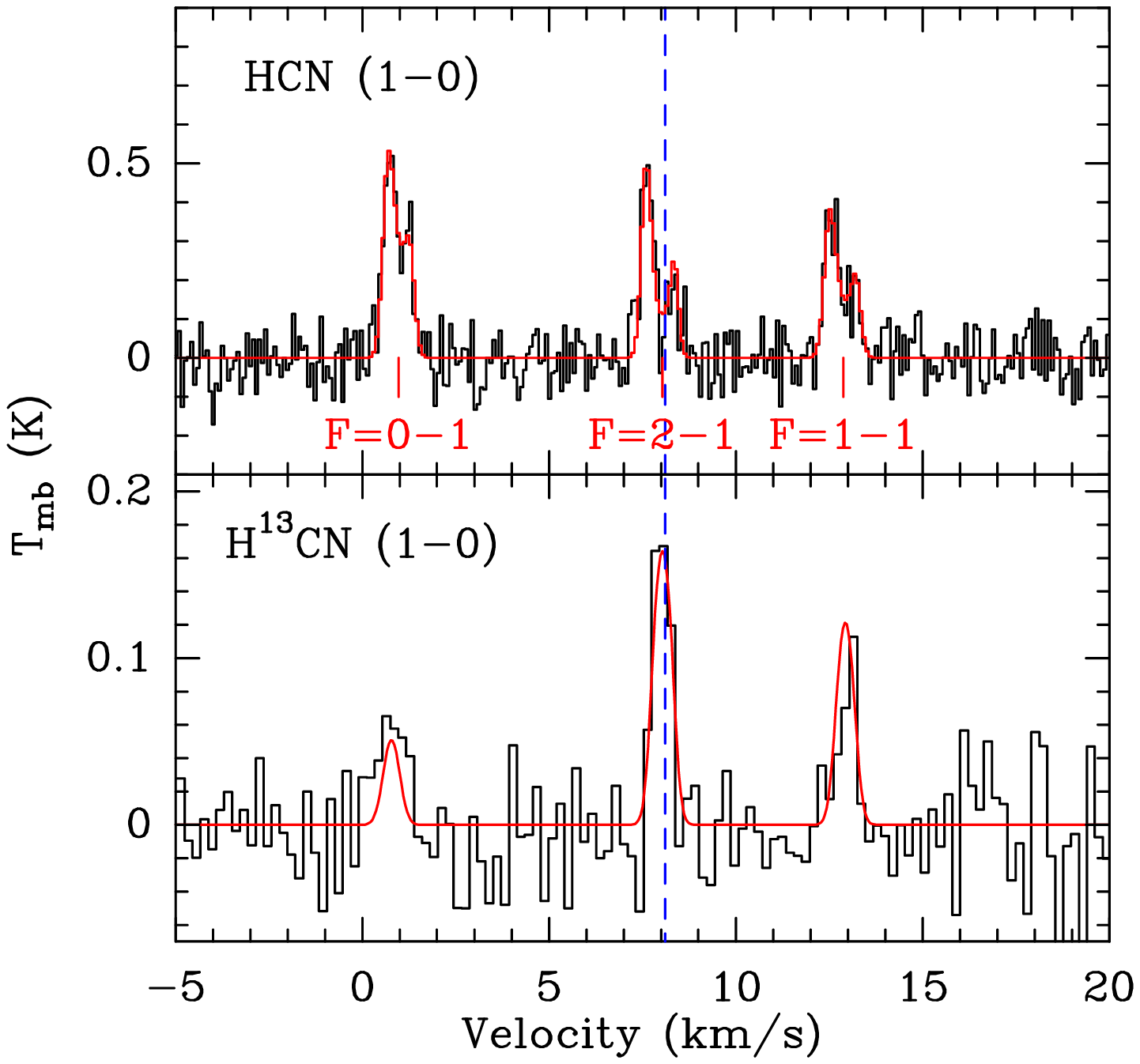}
\includegraphics[width = 0.4 \textwidth]{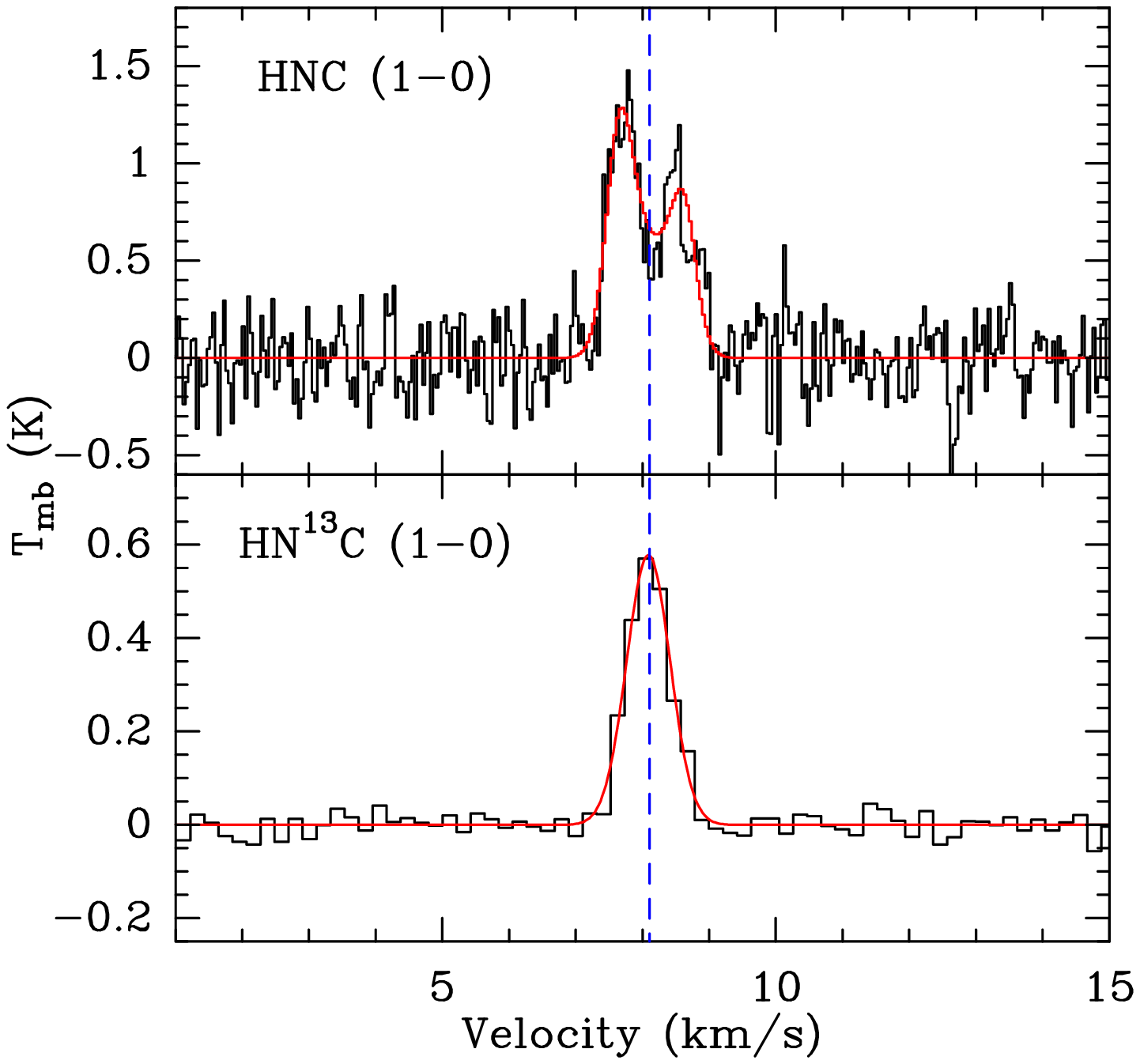}
\includegraphics[width = 0.4 \textwidth]{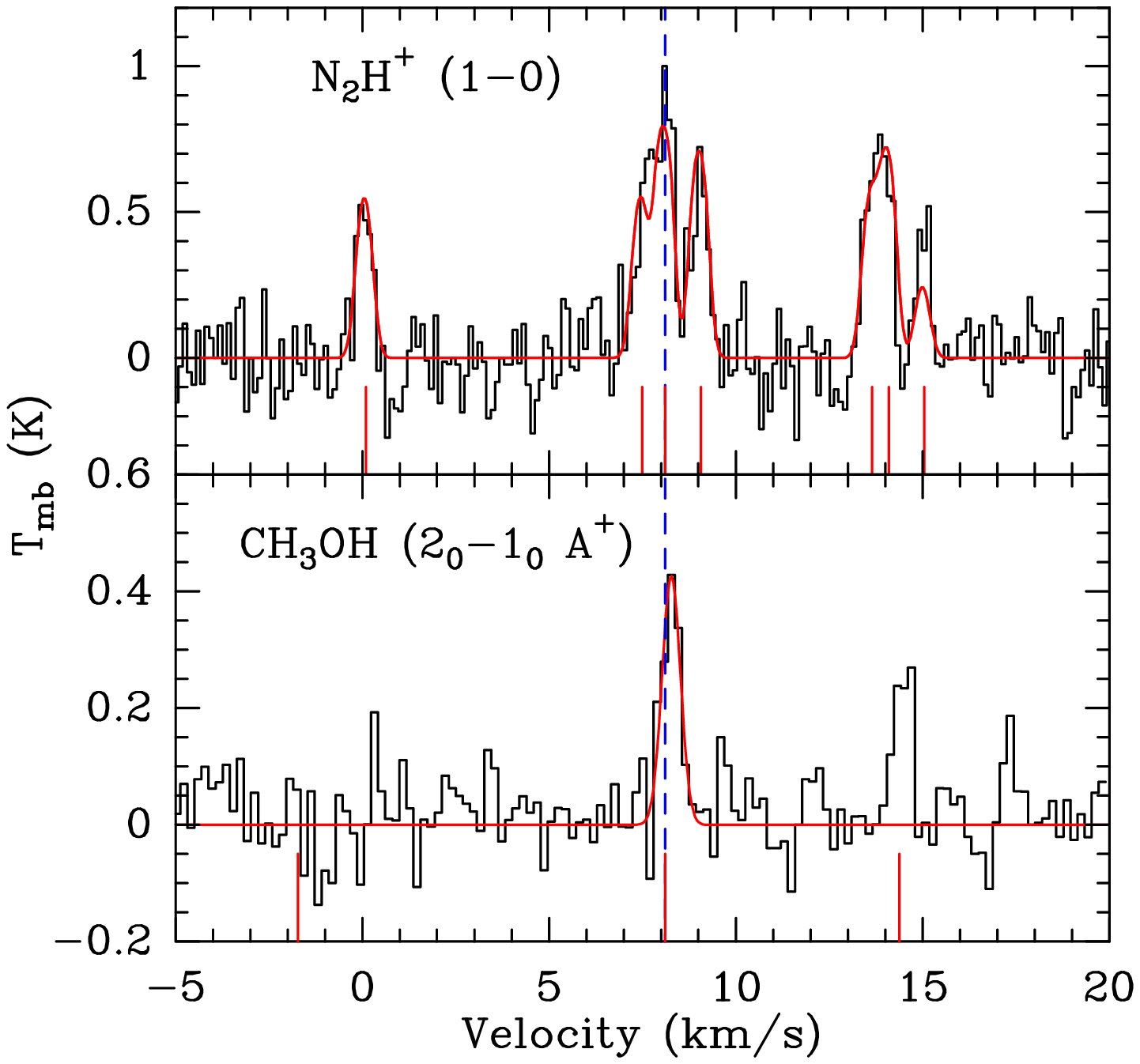}
\caption{{Single-pixel spectra (black lines) of all detected molecules toward Source A. Species are given in the upper left of each panel. In all panels, the vertical blue dashed lines represent the systemic velocity obtained from C$^{18}$O (1--0). Single-peaked spectra are modeled with a Gaussian fit. The blue-skewed line profiles of CS (2--1), HCN (1--0), and HNC (1--0) are fitted with the Hill5 model \citep{2005ApJ...620..800D}. For H$^{13}$CN (1--0) and N$_{2}$H$^{+}$ (1--0), the spectra are fitted with the hyperfine structure fitting subroutine in GILDAS. All fitted results are shown in red lines. The velocities of the N$_{2}$H$^{+}$ (1--0) hyperfine structure components are indicated by vertical red lines. In the CH$_{3}$OH panel where the velocity is given with respect to the rest frequency (96.74138 GHz) of CH$_{3}$OH (2$_{0}-1_{0}$ A$^{+}$), the CH$_{3}$OH triplet around 96.74 GHz is also indicated by three vertical red lines.}\label{Fig:sp-A}}
\end{figure*}

\begin{figure*}[!htbp]
\centering
\includegraphics[width = 0.95 \textwidth]{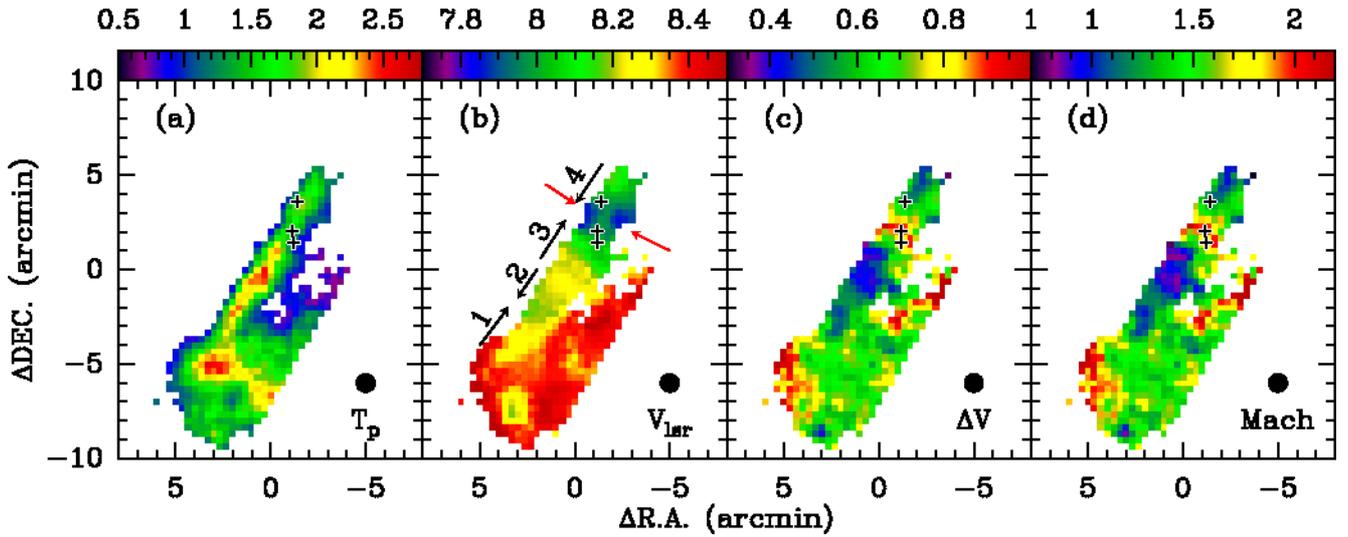}
\caption{{Maps of peak intensities (Fig.~\ref{Fig:kine}a), LSR velocities (Fig.~\ref{Fig:kine}b), line widths (Fig.~\ref{Fig:kine}c), and Mach numbers (Fig.~\ref{Fig:kine}d) derived from single-component Gaussian fits to our C$^{18}$O (1--0) data. The color bars represent main beam brightness temperatures in units of K in Fig.~\ref{Fig:kine}a, LSR velocities in units of \kms\,in Fig.~\ref{Fig:kine}b, and line widths in units of \kms\,in Fig.~\ref{Fig:kine}c. Velocity gradients along the filament and perpendicular to the filament are indicated by the black and red arrows in Fig.~\ref{Fig:kine}b, respectively. In all panels, the (0, 0) offset corresponds to $\alpha_{\rm J2000}$=18$^{\rm h}$28$^{\rm m}$49$\rlap{.}^{\rm s}$642, $\delta_{\rm J2000}$=00$^{\circ}$50$^{\prime}$01$\rlap{.}^{\prime \prime}$08, and the three black crosses give the positions of the three embedded YSOs (see Fig.~\ref{Fig:infra}). The mean uncertainties in derived peak intensities, LSR velocities, line widths, and Mach numbers are 0.19~K, 0.02~\kms, 0.06~\kms, and 0.1, respectively. The beam size is shown in the lower right of each panel.}\label{Fig:kine}}
\end{figure*}

\begin{figure*}[!htbp]
\centering
\includegraphics[width = 0.95 \textwidth]{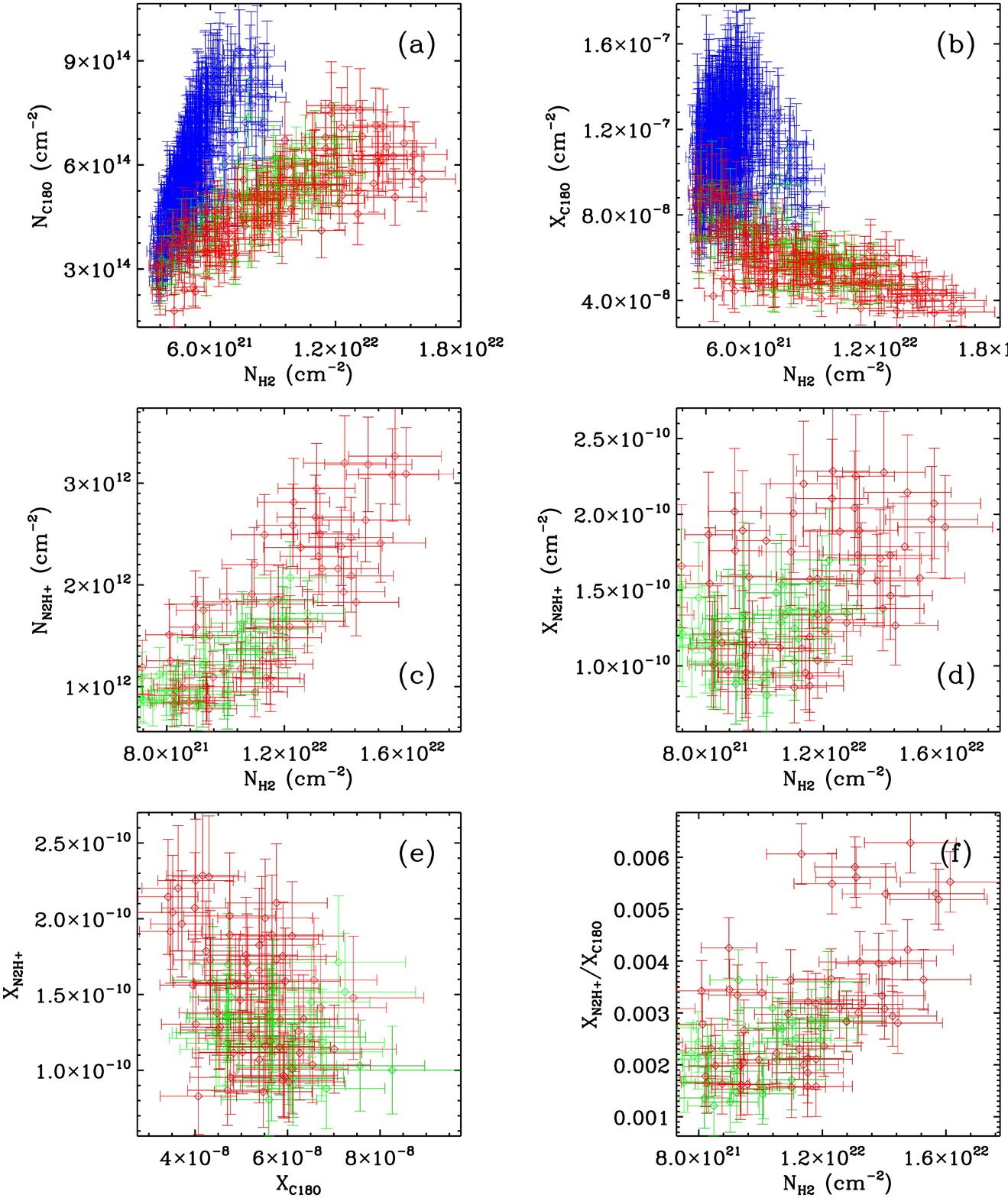}
\caption{{(a) C$^{18}$O column densities as a function of H$_{2}$ column densities. (b) C$^{18}$O fractional abundances as a function of H$_{2}$ column densities. (c) Similar to Fig.~\ref{Fig:abundratio}a but for N$_{2}$H$^{+}$ column densities. (d) Similar to Fig.~\ref{Fig:abundratio}b but for N$_{2}$H$^{+}$ fractional abundances. (e) C$^{18}$O fractional abundances as a function of N$_{2}$H$^{+}$ fractional abundances. (f) The N$_{2}$H$^{+}$/C$^{18}$O abundance ratios as a function of H$_{2}$ column densities. In all panels, the values of NW, SE, and EX are marked with red, green and blue symbols, respectively. In Figs.~\ref{Fig:abundratio}c--\ref{Fig:abundratio}f, only data from NW and SE, but not from EX, are presented.}\label{Fig:abundratio}}
\end{figure*}

\begin{figure*}[!htbp]
\centering
\includegraphics[width = 0.9 \textwidth]{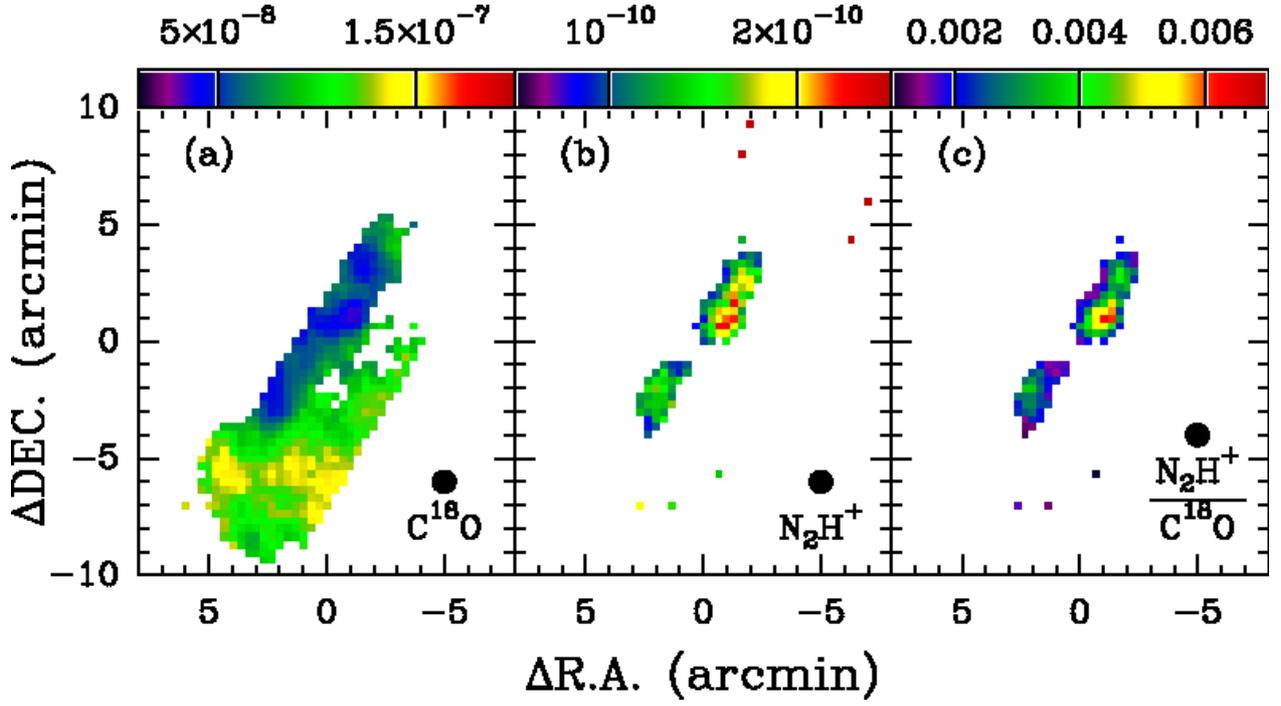}
\caption{{The distributions of C$^{18}$O (Fig.~\ref{Fig:abund}a) fractional abundances, N$_{2}$H$^{+}$ fractional abundances (Fig.~\ref{Fig:abund}b), and N$_{2}$H$^{+}$/C$^{18}$O abundance ratios (Fig.~\ref{Fig:abund}c). The (0, 0) offset corresponds to $\alpha_{\rm J2000}$=18$^{\rm h}$28$^{\rm m}$49$\rlap{.}^{\rm s}$642, $\delta_{\rm J2000}$=00$^{\circ}$50$^{\prime}$01$\rlap{.}^{\prime \prime}$08. The beam size is shown in the lower right of each panel.}\label{Fig:abund}}
\end{figure*}

\begin{figure*}[!htbp]
\centering
\includegraphics[width = 0.9 \textwidth]{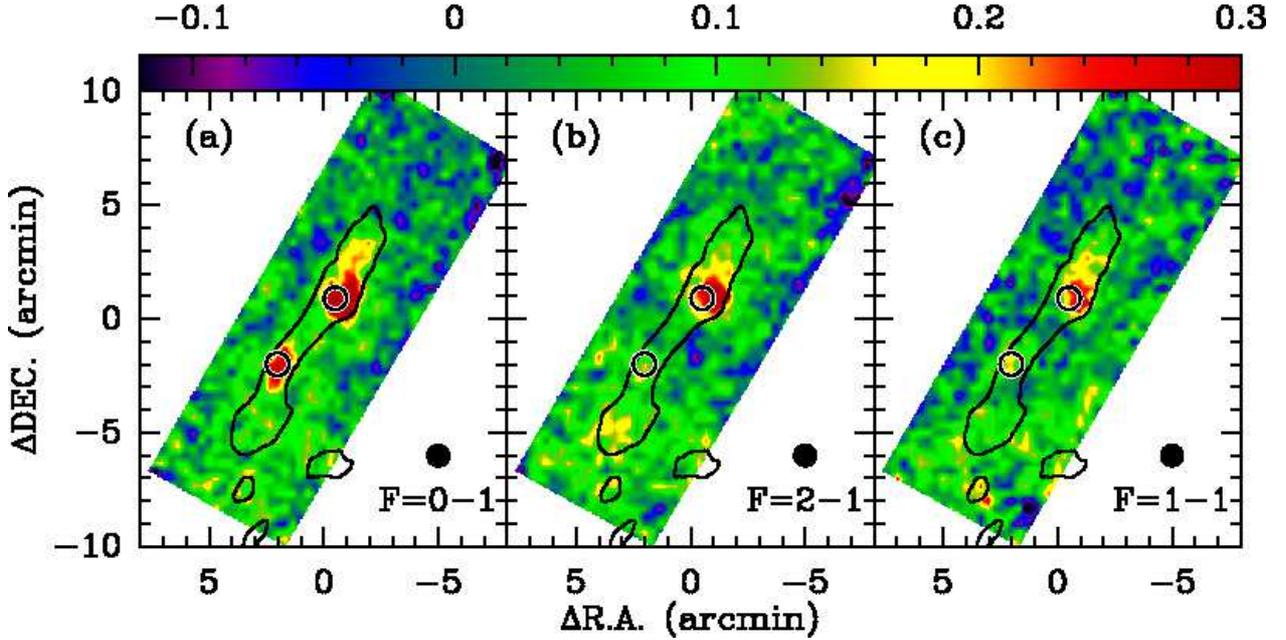}
\caption{{HCN (1--0) intensity maps of the Serpens filament integrated from 0.0 to 1.6~\kms\,(Fig.~\ref{Fig:hcn-anam}a), integrated from 7.0 to 8.6~\kms\,(Fig.~\ref{Fig:hcn-anam}b), and integrated from 11.7 to 13.6~\kms\,(Fig.~\ref{Fig:hcn-anam}c) with respect to the frequency of the $F=$2--1 hyperfine structure component. The three ranges correspond to the $F=0-1$, $F=2-1$, and $F=1-1$ hyperfine structure components, respectively. The color bar represents the integrated intensities in units of K~\kms. In all panels, the overlaid black contours denote H$_{2}$ column densities of 6$\times 10^{21}$~cm$^{-2}$, while the two open circles give the observed positions of sources A and B. In all panels, the (0, 0) offset corresponds to $\alpha_{\rm J2000}$=18$^{\rm h}$28$^{\rm m}$49$\rlap{.}^{\rm s}$642, $\delta_{\rm J2000}$=00$^{\circ}$50$^{\prime}$01$\rlap{.}^{\prime \prime}$08. The beam size is shown in the lower right of each panel.}\label{Fig:hcn-anam}}
\end{figure*}

\begin{figure*}[!htbp]
\centering
\includegraphics[width = 0.45 \textwidth]{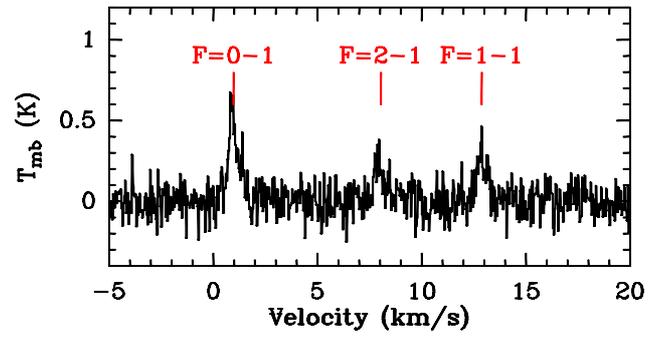}
\caption{{The observed HCN (1--0) hyperfine line anomaly (black lines) toward Source B. The three hyperfine structure components are indicated by vertical red lines.}\label{Fig:sp-B}}
\end{figure*}

\begin{figure*}[!htbp]
\centering
\includegraphics[width = 0.95 \textwidth]{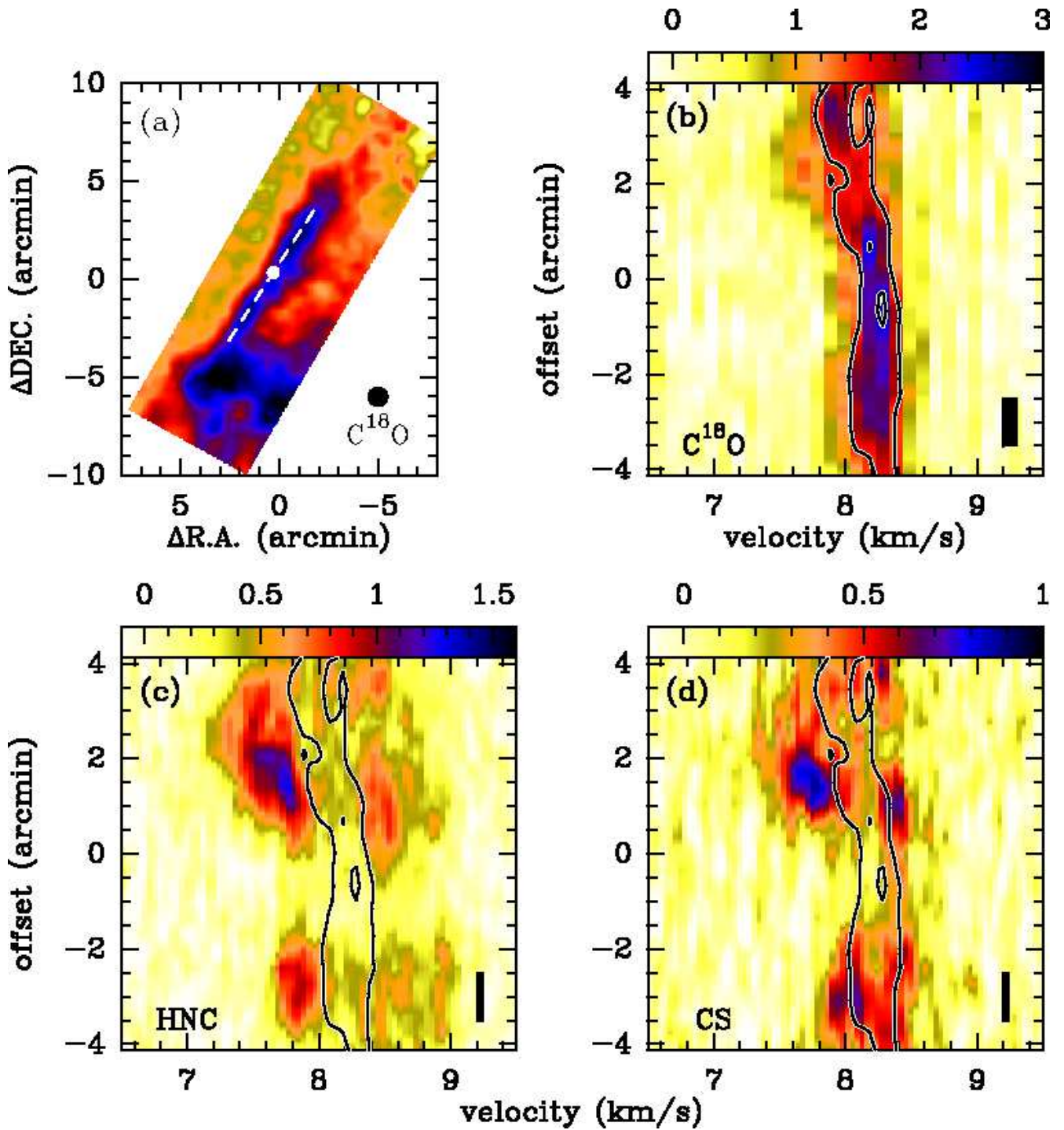}
\caption{{(a) The C$^{18}$O (1--0) integrated intensity map similar to Fig.~\ref{Fig:molecules}b but overlaid with a PV cut indicated by the white dashed line. (b) The PV diagram of C$^{18}$O (1--0) along the PV cut in Fig.~\ref{Fig:pv-abs}a. The channel width of C$^{18}$O (1--0) spectra has been binned to 0.1~\kms\,to achieve a higher dynamical range. (c) Similar to Fig.~\ref{Fig:pv-abs}b but for HNC (1--0). (d) Similar to Fig.~\ref{Fig:pv-abs}b but for CS (2--1). In Fig.~\ref{Fig:pv-abs}b--d, the color bars represent main beam brightness temperatures in units of K, and the black contours represent the C$^{18}$O (1--0) emission starting at 1.5~K (5$\sigma$) with increments of 0.9~K (3$\sigma$). The offsets are given with respect to the center of the PV cut ($\alpha_{\rm J2000}=$18$^{\rm h}$28$^{\rm m}$50$\rlap{.}^{\rm s}865$, $\delta_{\rm J2000}=$00\degr50\arcmin21$\rlap{.}$\arcsec799, also indicated by the white filled circle in Fig.~\ref{Fig:pv-abs}a), and increase from southeast to northwest. The resolution element is shown in the lower right of each panel.}\label{Fig:pv-abs}}
\end{figure*}

\begin{figure*}[!htbp]
\centering
\includegraphics[height = 0.4 \textwidth]{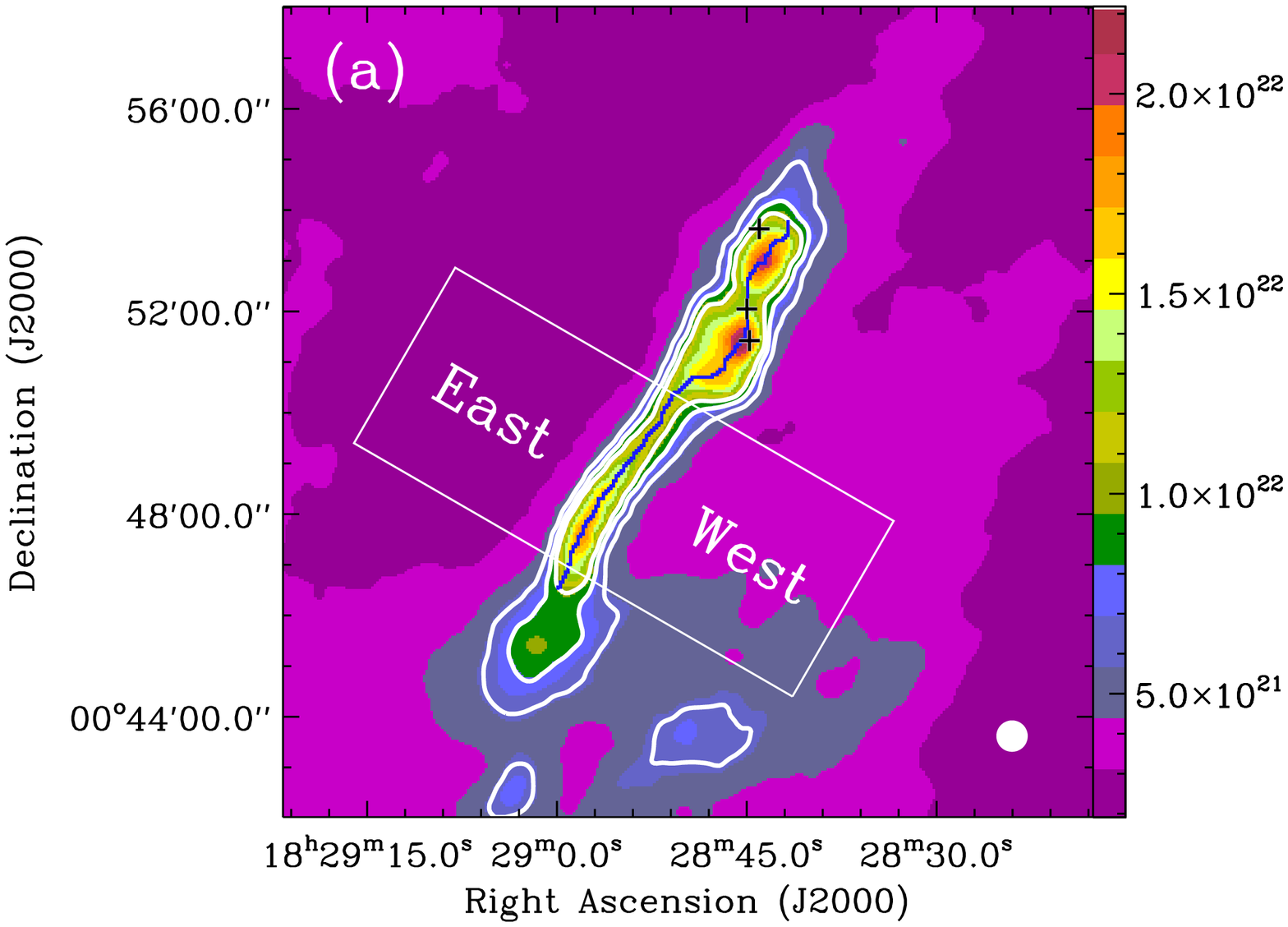}\\
\includegraphics[width = 0.4 \textwidth]{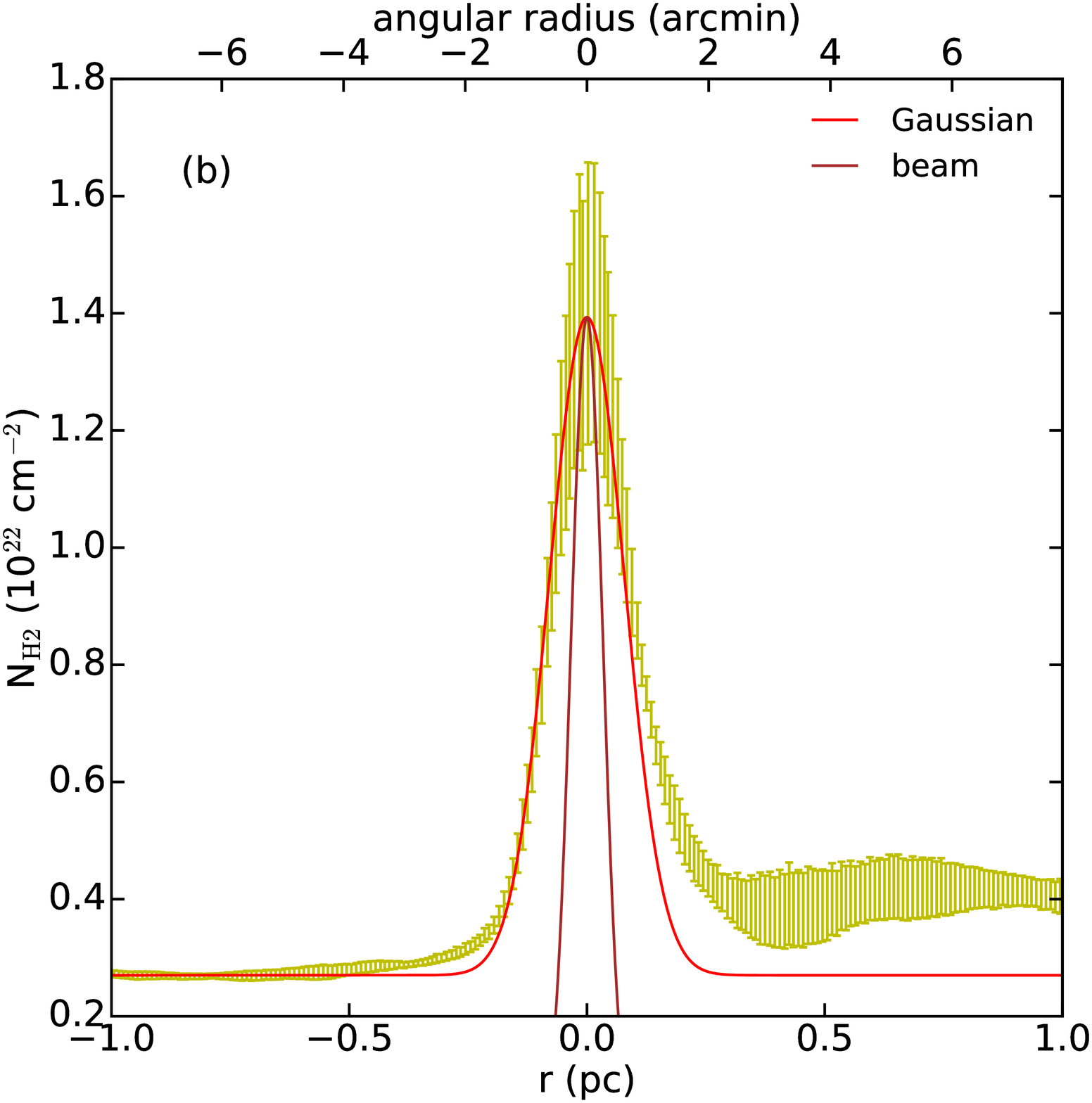}
\includegraphics[width = 0.4 \textwidth]{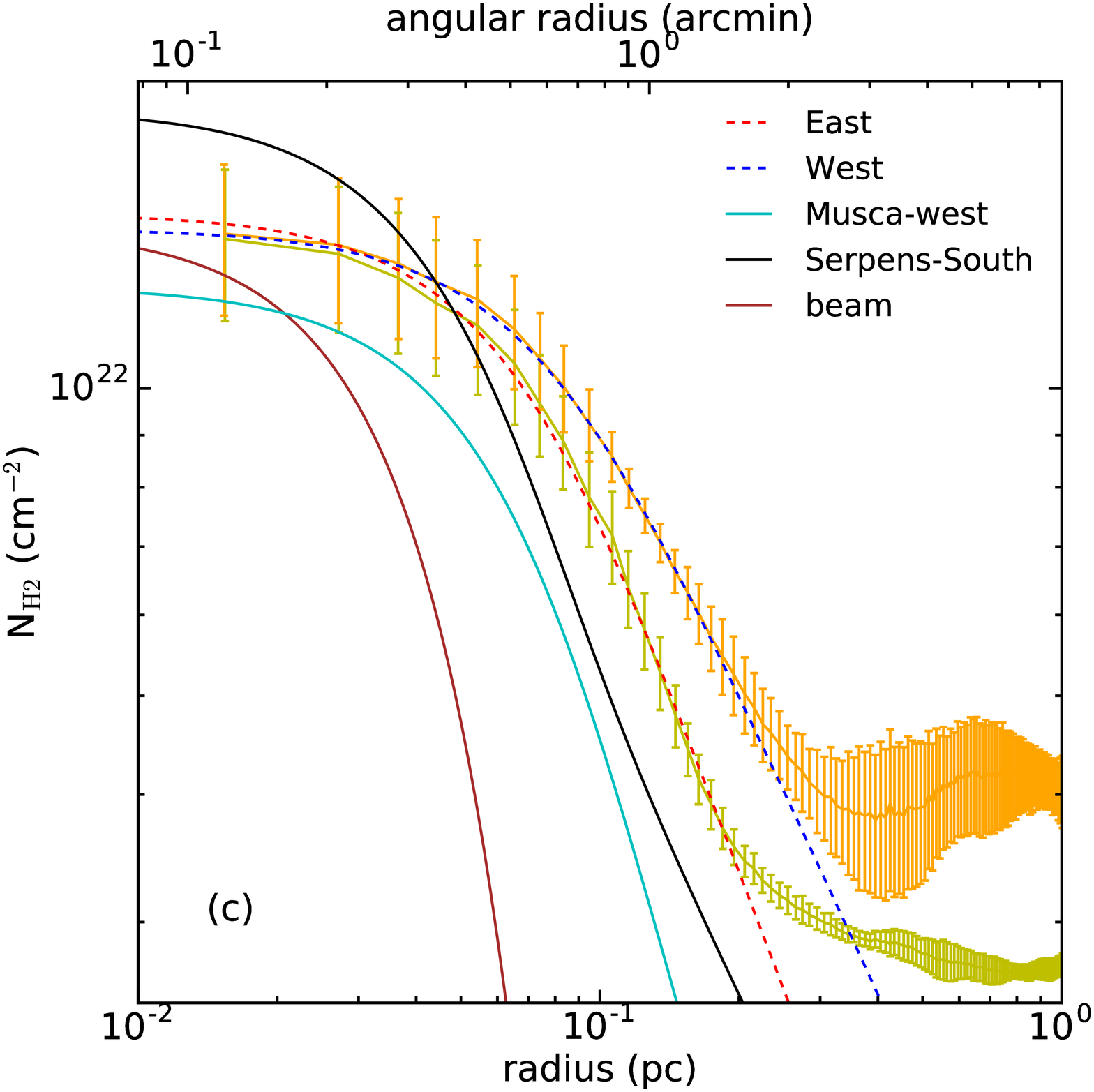}
\caption{{(a) The H$_{2}$ column density image of the Serpens filament obtained from the Herschel Gould belt survey. The contours correspond to column densities of 6$\times 10^{21}$~cm$^{-2}$, 8$\times 10^{21}$~cm$^{-2}$, and 1$\times 10^{22}$~cm$^{-2}$. The extracted crest of the Serpens filament is marked with the blue line. The emission within the white box is used to derive the radial profile of H$_{2}$ column density. The three black crosses give the positions of the three embedded YSOs. The beam size is shown in the lower right of this panel. (b) The mean radial profile of H$_{2}$ column densities within the white box in Fig.~\ref{Fig:rprof}a. The offsets increase from east to west. A Gaussian fit is indicated by the red line. (c) The mean radial H$_{2}$ column density profile on the west (yellow) and east (orange) sides of the crest. The red and blue dashed lines represent Plummer-like fits to the east and west (see Fig.~\ref{Fig:rprof}a) within a radius range of $<$0.2 pc, while the cyan and black lines represent the beam-convolved radial profile of Musca-West \citep{2016AA...586A..27K} and the Serpens South filament \citep{2013ApJ...766..115K}. In both Figs.~\ref{Fig:rprof}a and \ref{Fig:rprof}b. The beam size is shown by the brown line.}\label{Fig:rprof}}
\end{figure*}

\begin{figure*}[!htbp]
\centering
\includegraphics[height = 0.45 \textwidth]{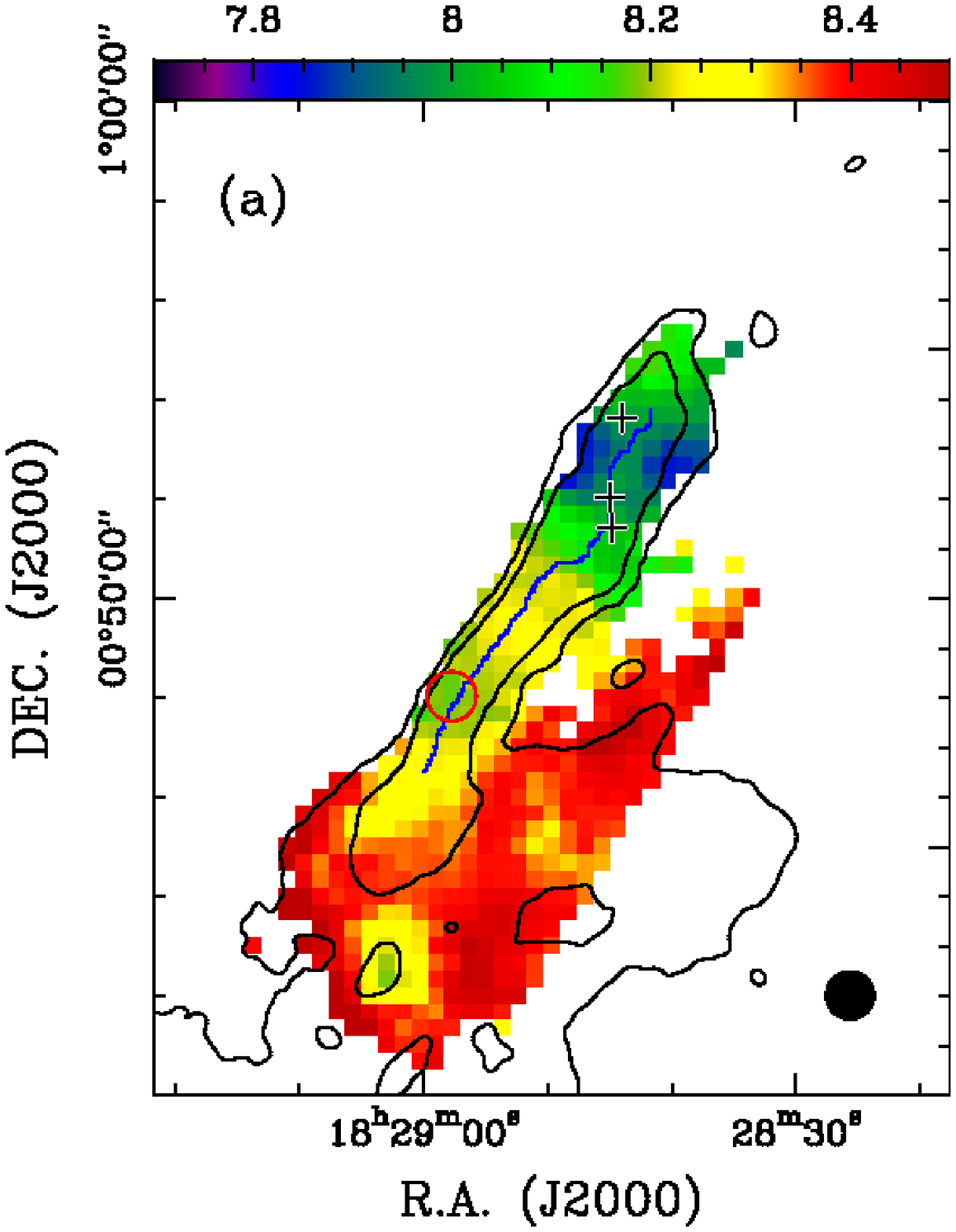}
\hspace*{1cm}
\includegraphics[height = 0.45 \textwidth]{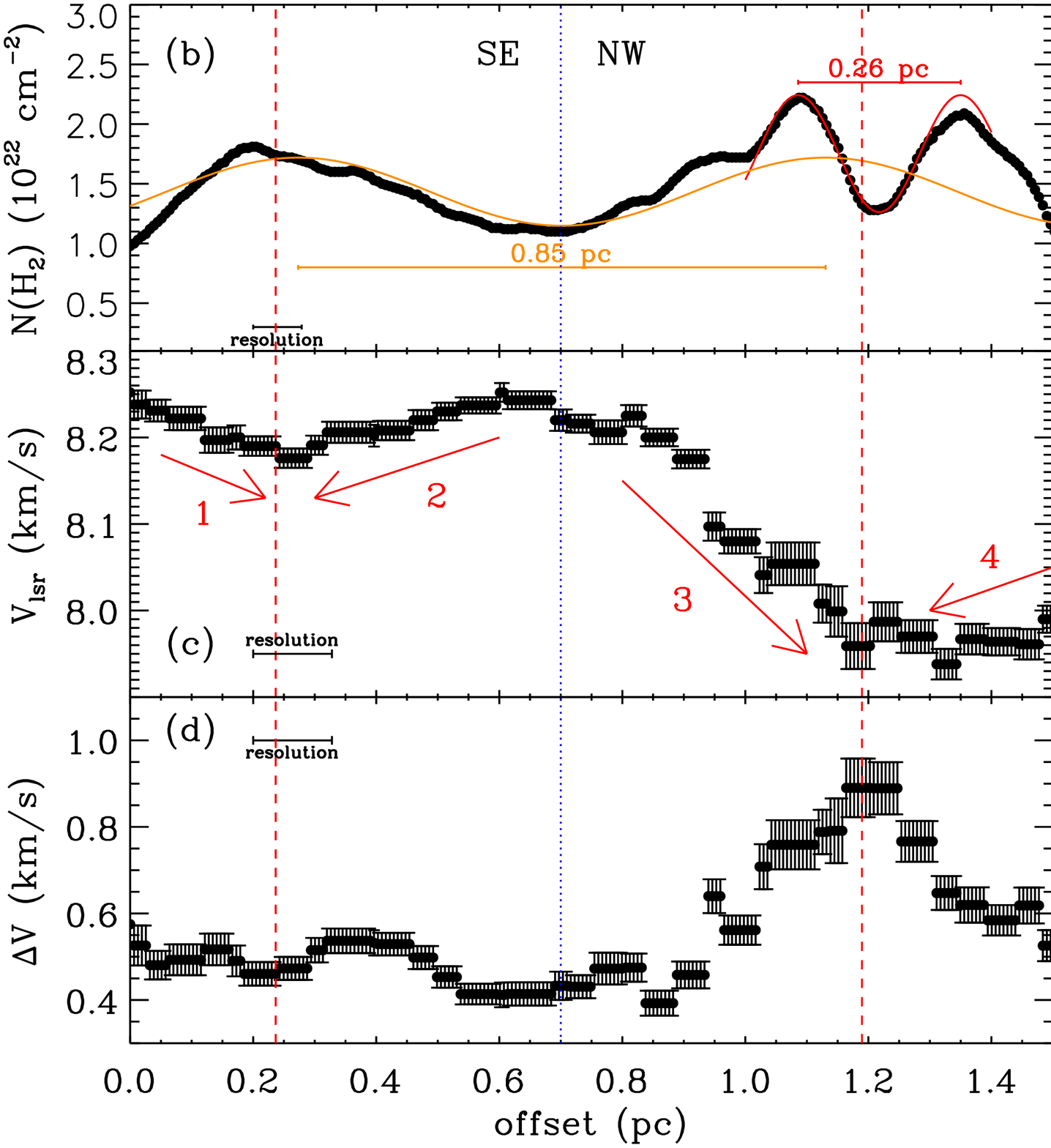}
\caption{{(a) The LSR velocity map similar to Fig.~\ref{Fig:kine}b but overlaid with the H$_{2}$ column density contours which represent 4$\times 10^{21}$~cm$^{-2}$ and 6$\times 10^{21}$~cm$^{-2}$. Source B is marked with the red circle, and the three black crosses give the positions of the three embedded YSOs, i.e., em10, em28, and em16. (b) The column density profile along the crest of the Serpens filament indicated by the blue line in Fig.~\ref{Fig:maxprof}a. The spatial resolution is $\sim$0.08 pc. The offset increases from south to north. The red line represents a sinusoidal fit to the profile within the offset range of 1.0--1.4 pc. A sinusoidal orange line is created to better visualize the variation at a large scale. (c) Similar to Fig.~\ref{Fig:maxprof}b but for the C$^{18}$O (1--0) velocity. The four red arrows correspond to the velocity gradients in Fig.~\ref{Fig:kine}b, indicated there by black arrows. (d) Similar to Fig.~\ref{Fig:maxprof}c but for the C$^{18}$O (1--0) line widths. In Figs.~\ref{Fig:maxprof}b--\ref{Fig:maxprof}d, the red dashed lines mark the positions of Source B and the YSO emb28.}\label{Fig:maxprof}}
\end{figure*}

\begin{figure*}[!htbp]
\centering
\includegraphics[width = 0.45 \textwidth]{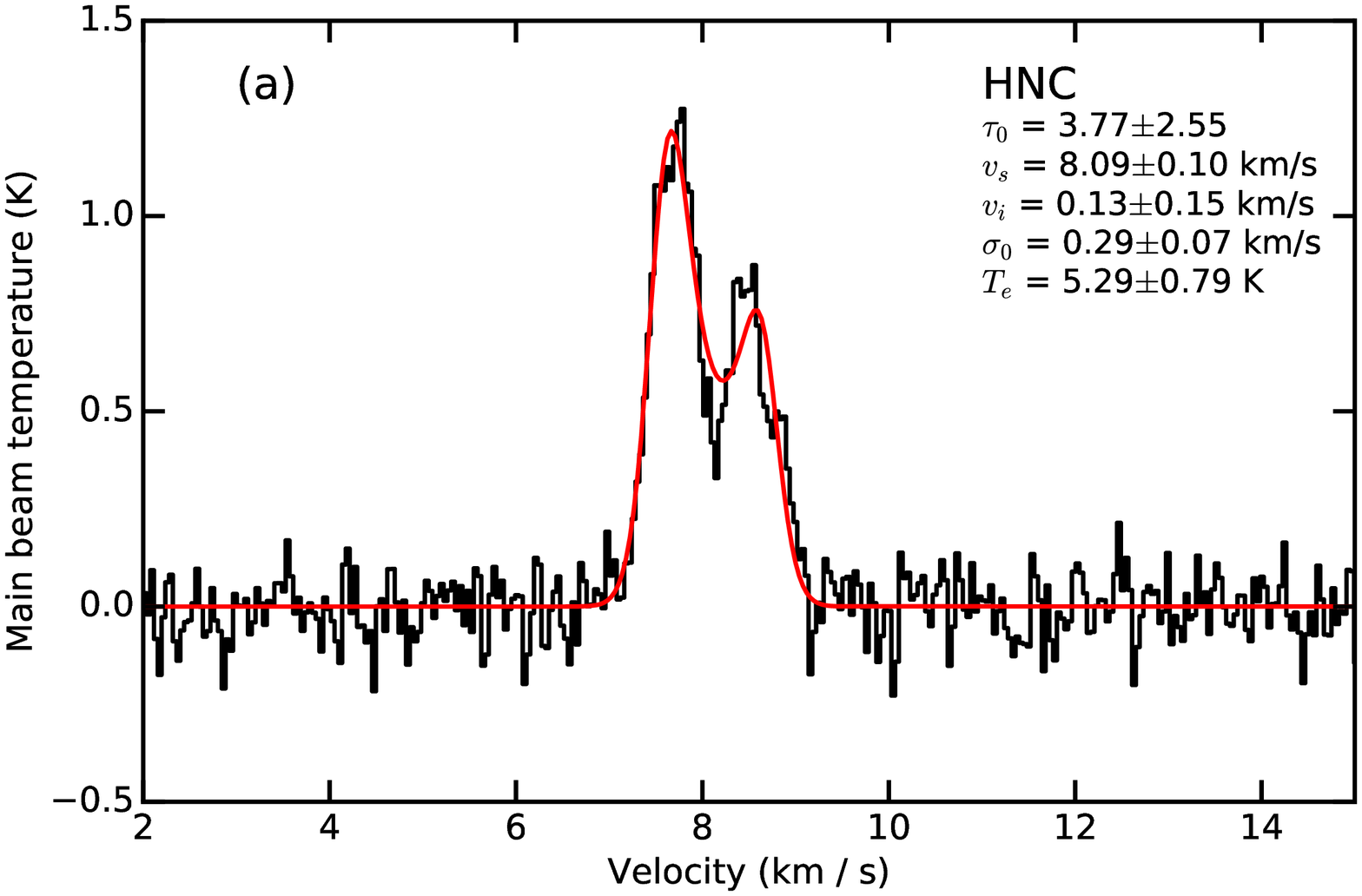}
\includegraphics[width = 0.45 \textwidth]{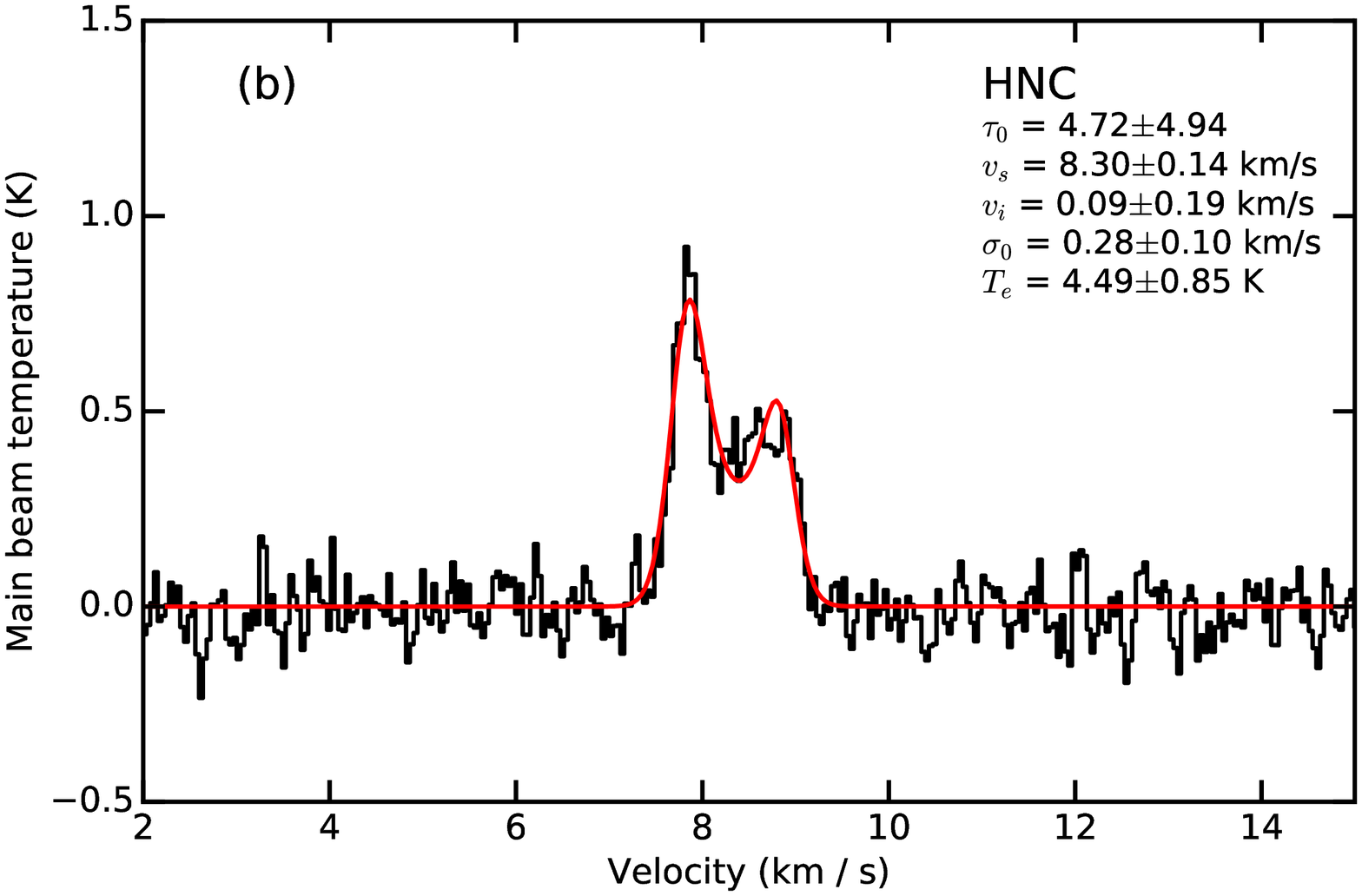}
\includegraphics[width = 0.45 \textwidth]{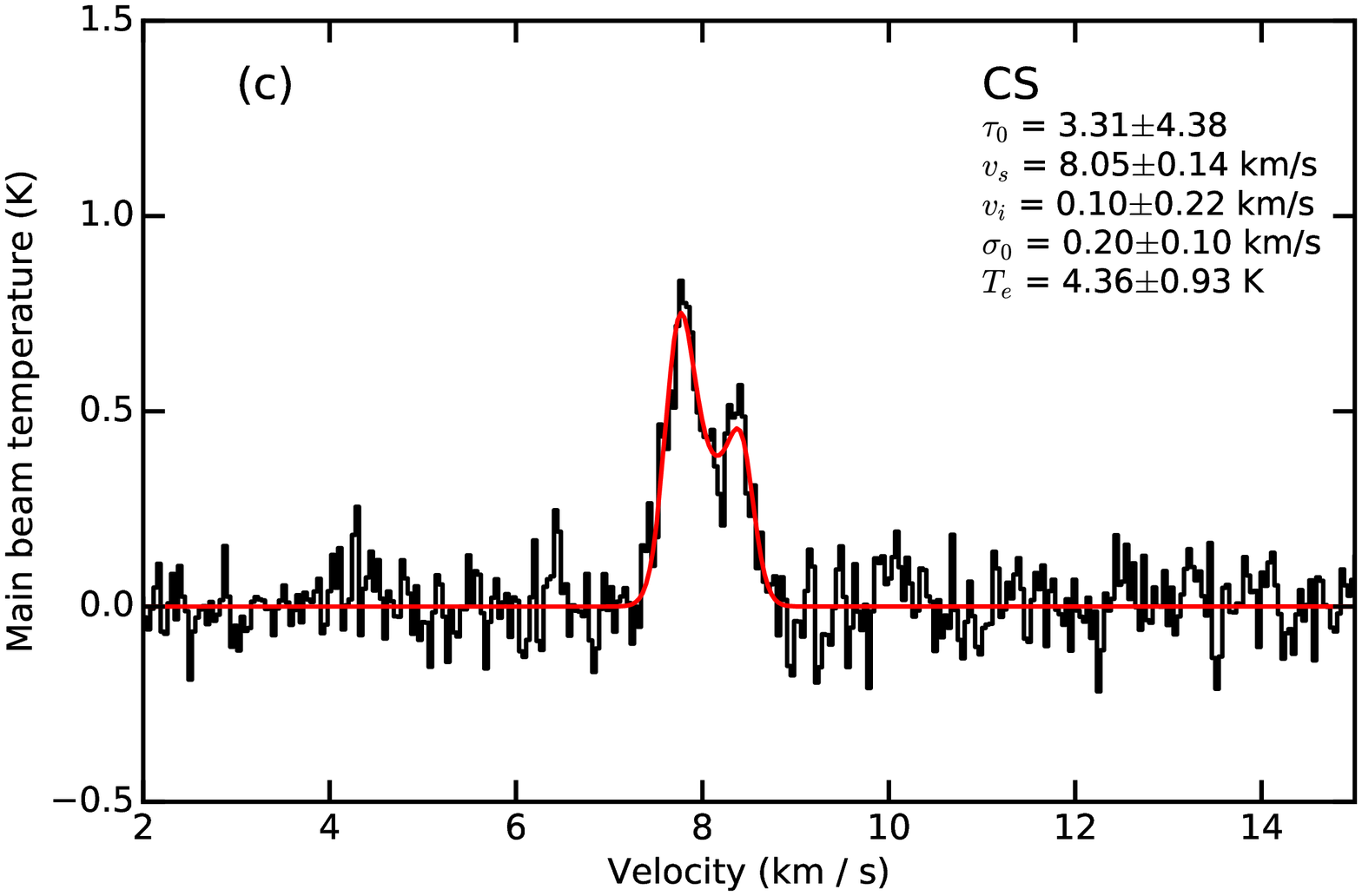}
\includegraphics[width = 0.45 \textwidth]{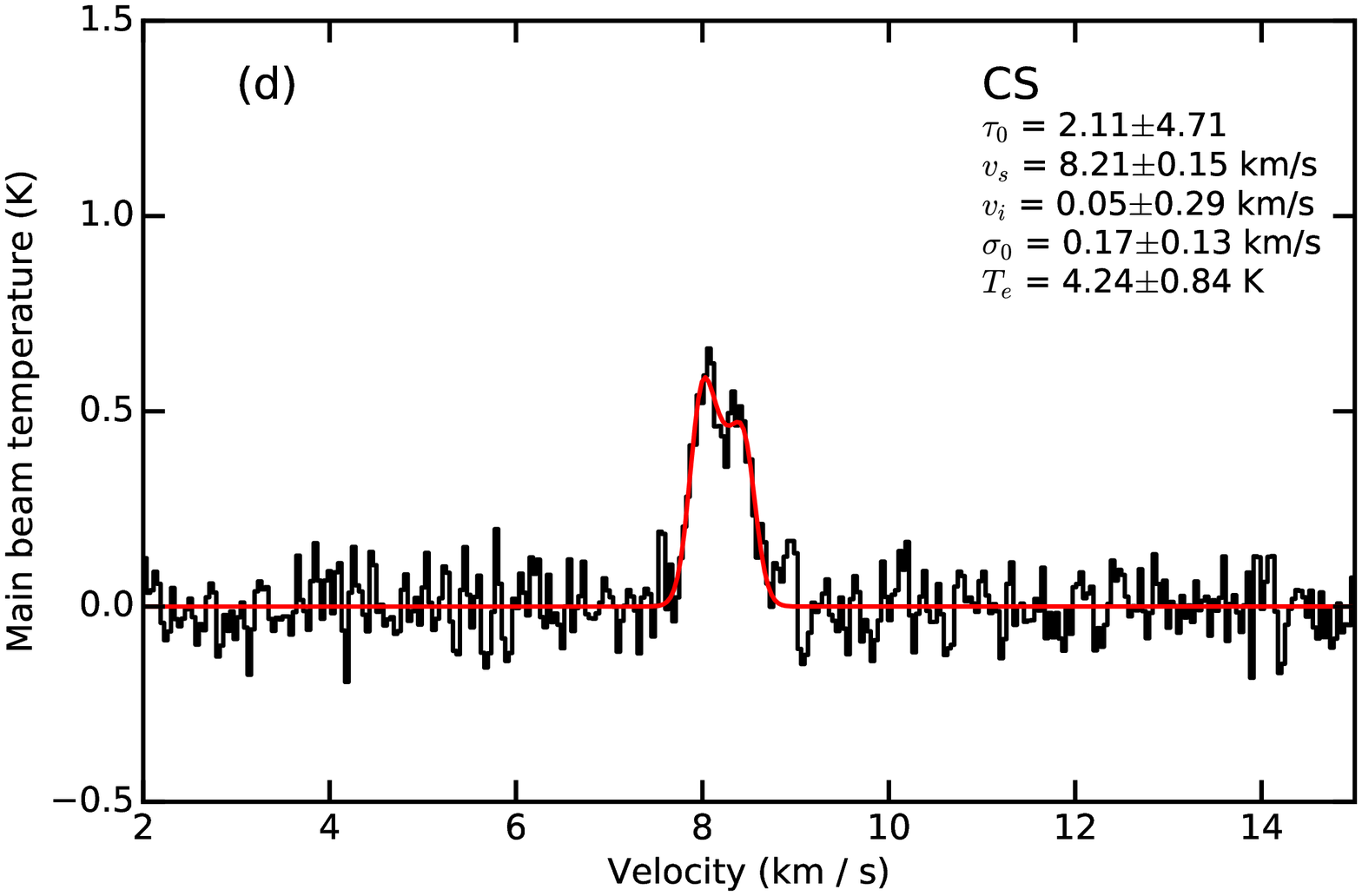}
\caption{{(a) The HNC (1--0) spectrum is averaged over 3$\times$3 pixels centered at an offset of ($-$40\arcsec, 40\arcsec) in Fig.~\ref{Fig:molecules} to increase the signal-to-noise level. (b) Similar to Fig.~\ref{Fig:hill5}a but for an offset of (100\arcsec, $-$120\arcsec). (c) Similar to Fig.~\ref{Fig:hill5}a but for CS (2--1). (d) Similar to Fig.~\ref{Fig:hill5}b but for CS (2--1). All these lines are modeled with the Hill5 model \citep[][; see Sect.~\ref{sec.infall}]{2005ApJ...620..800D}. These offsets are given with respect to $\alpha_{\rm J2000}$=18$^{\rm h}$28$^{\rm m}$49$\rlap{.}^{\rm s}$642, $\delta_{\rm J2000}$=00$^{\circ}$50$^{\prime}$01$\rlap{.}^{\prime \prime}$08. The fitted parameters and their uncertainties are given in the upper right of each panel.}\label{Fig:hill5}}
\end{figure*}

\clearpage
\begin{appendix}
\section{More comments on Source A in the Serpens filament}\label{app.b}
Benefiting from a longer integration on Source A which represents the peak of the F=0--1 line of HCN (1--0) (see Fig.~\ref{Fig:hcn-anam}), our single-point observations have led to the detection of 13 transitions which can be assigned to 8 different molecules and their isotopologues (see Fig.~\ref{Fig:sp-A} and Table~\ref{Tab:a}). As mentioned in Sect.~\ref{sec.mor}, self-absorption is observed in HCO$^{+}$ (1--0), CS (2--1), HCN (1--0), and HNC (1--0), which is analogous to those observed in Serpens South \citep{2013ApJ...766..115K}. The hfs components of H$^{13}$CN (1--0) are detected toward Source A. The fit to its hfs lines gives an optical depth of $\sim$0.70 for the $F=2-1$ line of H$^{13}$CN (1--0). This implies a very high optical depth of $\sim$50 for the corresponding line of HCN (1--0) with a typical isotopic ratio [$^{12}$C/$^{13}$C]=70 in nearby clouds \citep{2015A&A...581A..48G,2016ApJ...824..136L}. Furthermore, our measured H$^{13}$CN (1--0) and HN$^{13}$C (1--0) brightness temperatures are comparable to those in TMC-1 \citep{1998ApJ...503..717H} and higher than those in Serpens South \citep{2013ApJ...766..115K} which were obtained with larger telescopes (Nobeyama 45m and Mopra 22m), neglecting beam dilution effects. The detection of CCS (8$_{7}$--7$_{6}$) also points out that Source A is chemically young, since this molecule tends to be abundant in early phases of star-forming cores \citep{1992ApJ...392..551S}. Furthermore, the peak intensity of HN$^{13}$C (1--0) is about four times that of H$^{13}$CN (1--0), indicative of an extremely cold kinetic temperature \citep{1992A&A...256..595S}. Based on the derived column densities in Table~\ref{Tab:a}, the CH$_{3}$OH A/E abundance ratio is estimated to be 1.7$\pm$0.2, indicating an over-abundance of A type methanol. This could be because CH$_{3}$OH is formed from hydrogenation of solid CO on interstellar grain surfaces and then released into the gas phase via reactive desorption \citep[e.g.,][]{2011A&A...533A..24W,2013ApJ...769...34V}. The frequency of SiO (2--1) has also been observed but the line remains undetected, which suggests that the 3$\sigma$ upper limit of the peak intensity is 0.06~K. 

\section{Position-velocity diagrams of HCN (1--0), N$_{2}$H$^{+}$ (1--0), and the CH$_{3}$OH lines}\label{app.a}
\begin{figure*}[!htbp]
\centering
\includegraphics[width = 0.33 \textwidth]{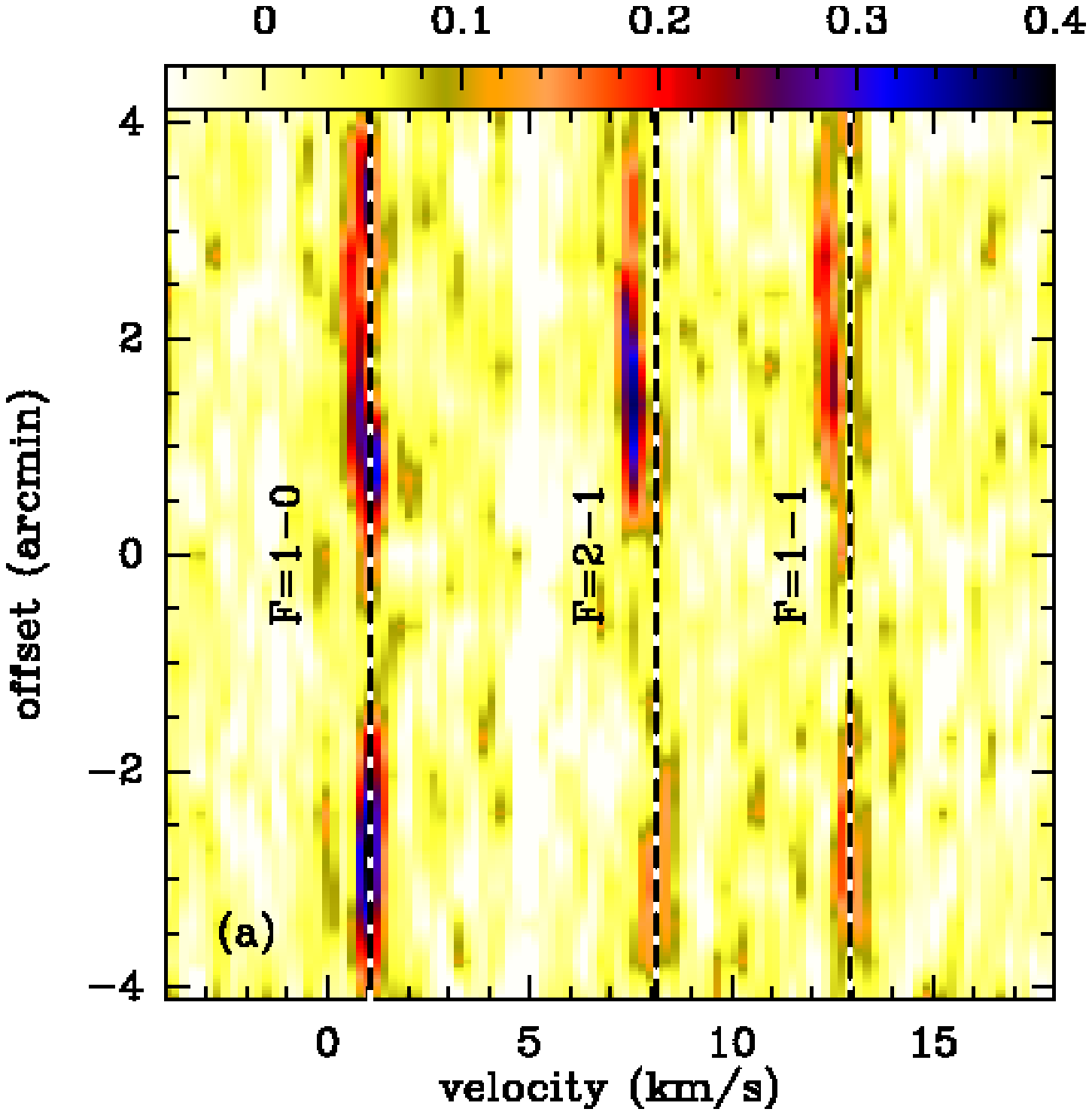}
\includegraphics[width = 0.33 \textwidth]{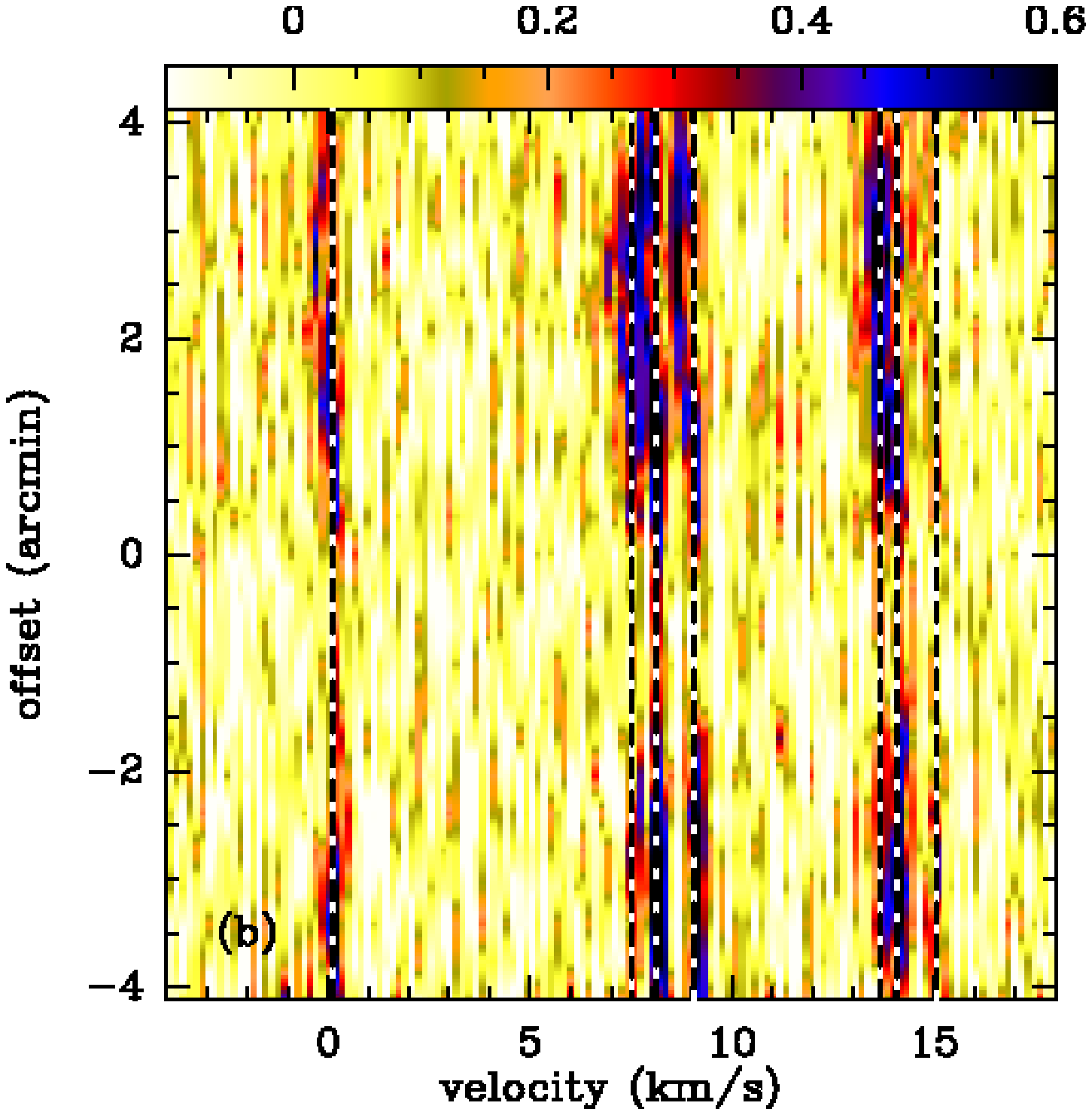}
\includegraphics[width = 0.33 \textwidth]{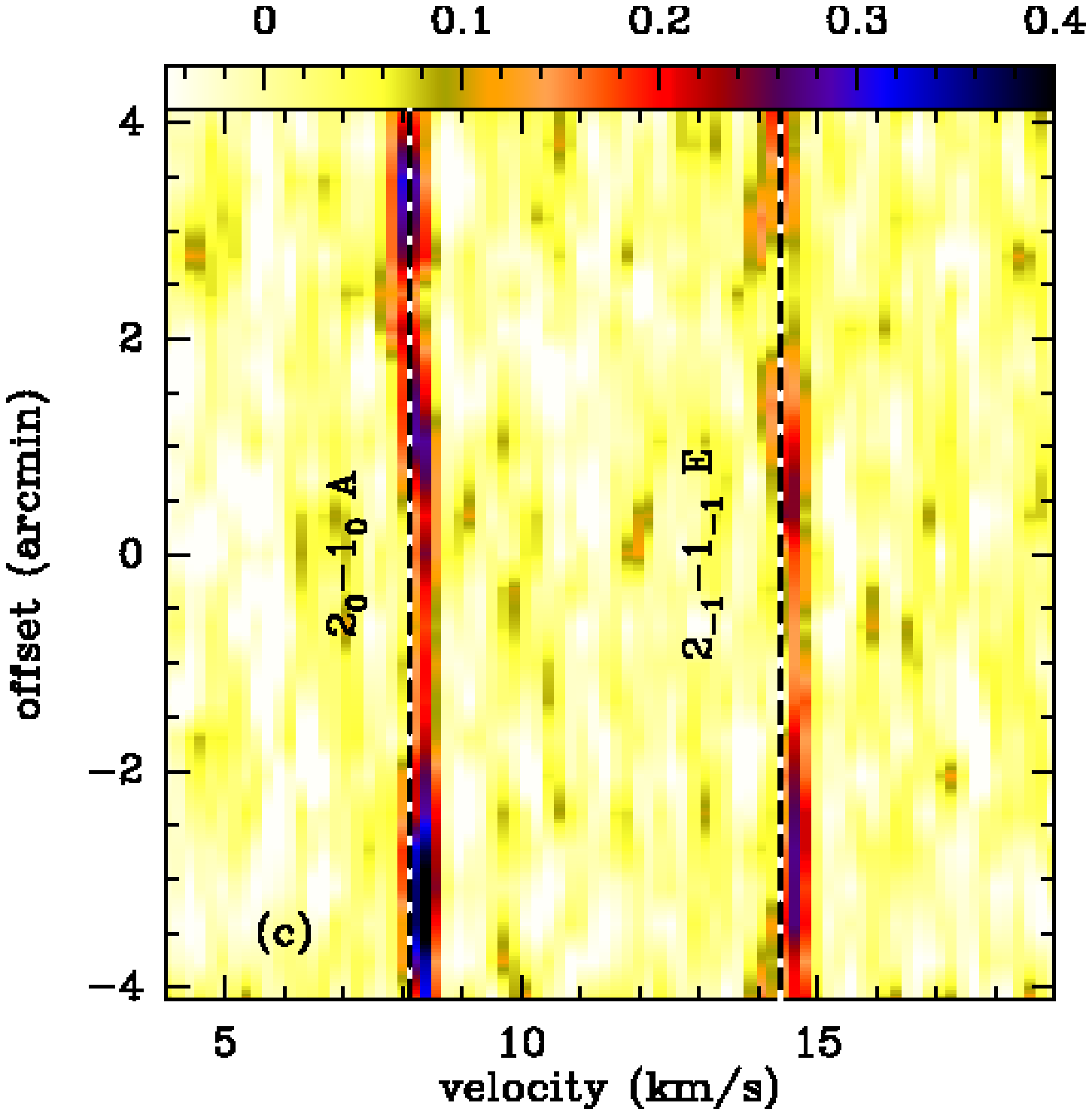}
\caption{{PV diagrams similar to Fig.~\ref{Fig:pv-abs}b but for (a) HCN (1--0), (b) N$_{2}$H$^{+}$ (1--0), and (c) two CH$_{3}$OH lines. In Fig.~\ref{Fig:pv2}a, the three hyperfine structure components are labeled next to their respective emission. In Fig.~\ref{Fig:pv2}b, the channel width of the N$_{2}$H$^{+}$ (1--0) spectra has been binned to 0.16~\kms\,to achieve a higher dynamical range in individual channels. In Fig.~\ref{Fig:pv2}c, the two transitions are labeled next to their respective emission. In all panels, the color bars represent main beam brightness temperatures in units of K. The offsets are given with respect to the center of the PV cut ($\alpha_{\rm J2000}=$18$^{\rm h}$28$^{\rm m}$50$\rlap{.}^{\rm s}865$, $\delta_{\rm J2000}=$00\degr50\arcmin21$\rlap{.}$\arcsec799, also indicated by the white filled circle in Fig.~\ref{Fig:pv-abs}a), and increase from southeast to northwest. In all panels, the vertical dashed lines mark the systemic velocity of $-$8.11~\kms\,for corresponding lines.}\label{Fig:pv2}}
\end{figure*}

In addition to Fig.~\ref{Fig:pv-abs}, Figure~\ref{Fig:pv2} also presents the position-velocity diagrams of HCN (1--0), N$_{2}$H$^{+}$ (1--0), and the CH$_{3}$OH lines. In Fig.~\ref{Fig:pv2}a, we see that the $F=$1--0 component of HCN (1--0) is brighter than the other two components in most of the observed regions, confirming the presence of the HCN hyperfine line anomalies.

\section{Comments on H$_{2}$ column densities}\label{app.c}
H$_{2}$ column densities derived from the Herschel GB team are nearly three times higher than those derived by \citet{2015A&A...584A.119R} in the Serpens filament. The differences may arise from three reasons given below. First, \citet{2015A&A...584A.119R} fitted the SED with data from all five Herschel bands (70--500~$\mu$m), while the Herschel GB team only used observed data points from 160 to 500~$\mu$m \citep{2017MmSAI..88..783F}. Since the 70~$\mu$m emission might trace a hotter dust population, including 70~$\mu$m data in the SED fitting may cause larger uncertainties in deriving the properties of molecular clouds. Second, the two teams used different dust spectral indices, which leads to a difference in the dust mass absorption coefficient by a factor of 1.2-1.5 for the five Herschel bands. Third, the color correction seems different. Although both of them use the same model, the added offset values of the Herschel GB team \citep{2015A&A...584A..91K} are higher than those used in \citet{2015A&A...584A.119R} by a factor of 1.5-2.6, leading to different absolute flux densities.

\begin{figure*}[!htbp]
\centering
\includegraphics[width = 0.48 \textwidth]{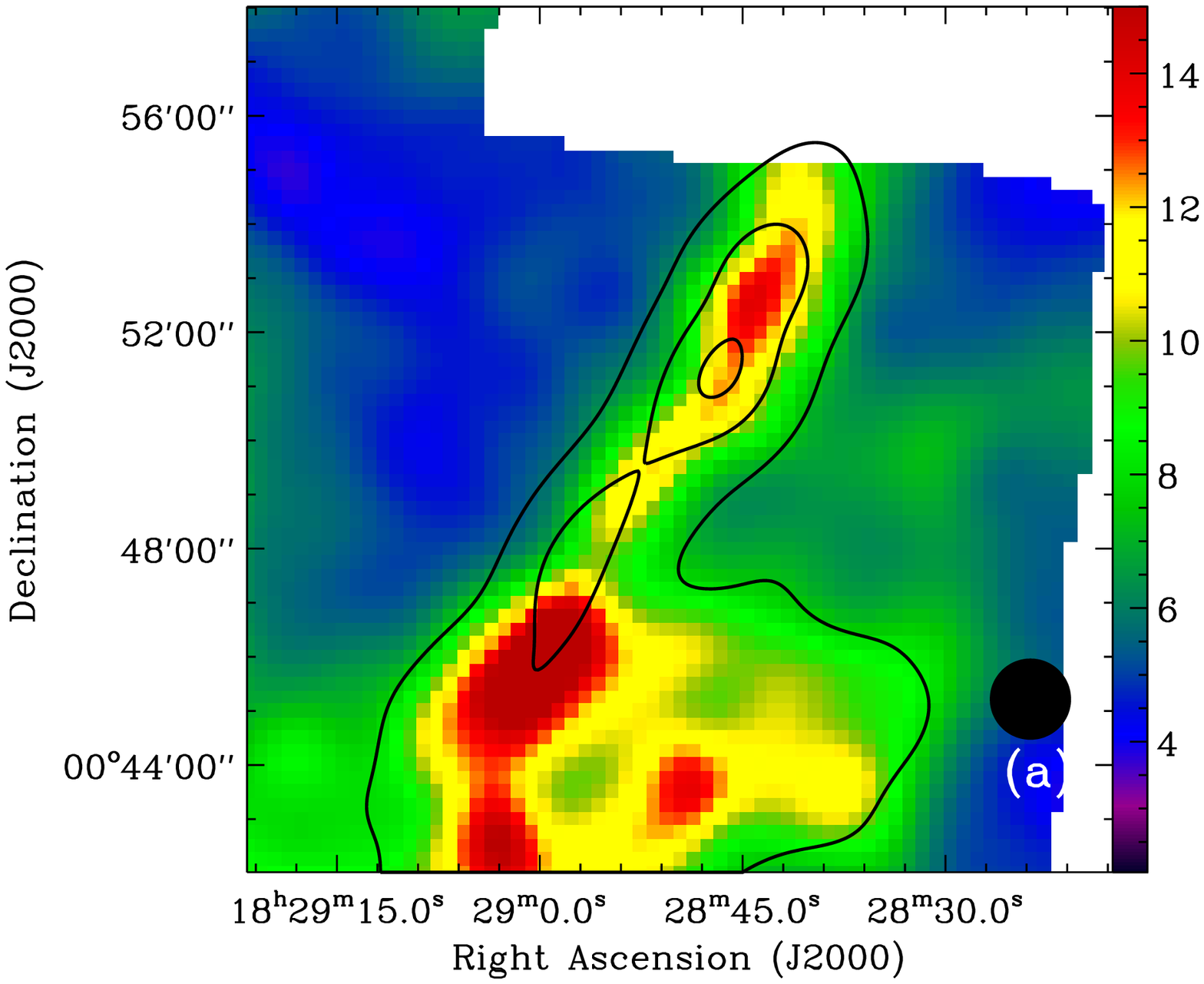}
\includegraphics[width = 0.48 \textwidth]{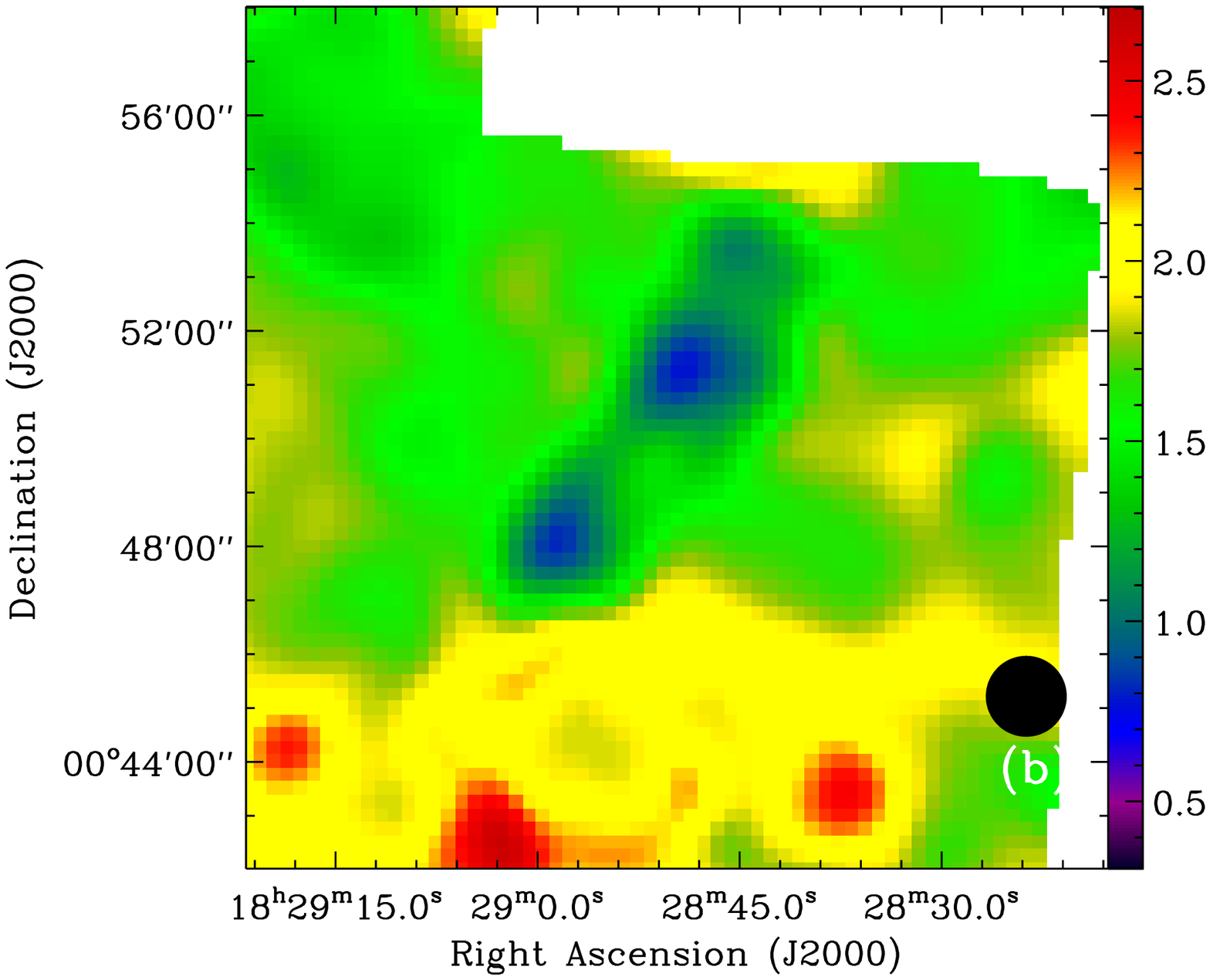}
\caption{{(a) The Spitzer c2d visual extinction map overlaid with the Herschel H$_{2}$ column density contours. The contours increase from 4$\times 10^{21}$~cm$^{-2}$ to 1.2$\times 10^{22}$ cm$^{-2}$ by 4$\times 10^{21}$~cm$^{-2}$. The color bar represents the visual extinction in units of magnitude. (b) The map of the ratios ($N_{\rm H_{2},av}$/$N_{\rm H_{2}}$) between the H$_{2}$ column densities derived from the visual extinction map and those derived from the Herschel SED fitting. The color bar represents the  H$_{2}$ column density ratio. The beam size is shown in the lower right of each panel.}\label{Fig:comparison}}
\end{figure*}

Here, we present a comparison between the Herschel H$_{2}$ column density map used in this work and the visual extinction map that is obtained from the c2d Spitzer Legacy project\footnote{https://irsa.ipac.caltech.edu/data/SPITZER/C2D/images/SER/EXTINCTION\_MAPS/} \citep{2003PASP..115..965E,2009ApJS..181..321E}. This project has provided visual extinction maps with angular resolutions of 90\arcsec--300\arcsec, and we choose the visual extinction map with the highest angular resolution of 90\arcsec\, for the comparison (see Fig.~\ref{Fig:comparison}a). The visual extinction ($A_{\rm V}$) was converted to the H$_{2}$ column density by assuming N$_{\rm H_{2}, av}$ (cm$^{-2}$)=9.4$\times 10^{20}$ $A_{\rm V}$ (mag) \citep{1978ApJ...224..132B}. Meanwhile, the Herschel H$_{2}$ column density map was also convolved to an angular resolution of 90\arcsec\, with a Gaussian kernal size of 82\arcsec, and then was linearly interpolated to the same grid as the visual extinction map. Figure~\ref{Fig:comparison}b presents the ratio ($N_{\rm H_{2},av}$/$N_{\rm H_{2}}$) between the H$_{2}$ column densities derived from the visual extinction and the Herschel SED. We find that the ratios are 0.8--1.5 in NW and SE, and 1.7--2.0 in EX. This suggests that the H$_{2}$ column densities derived from the two methods agree with each other in NW and SE, and the variations may mainly arise from the uncertainties in the Herschel column denisty map and the visual extinction map. In EX, the H$_{2}$ column densities based on the extinction map are nearly twice the H$_{2}$ column densities derived from the Herschel SED. This is probably because the conversion factor from $A_{\rm V}$ to H$_{2}$ column density and the dust spectral index ($\beta$) in EX are different from those in NW and SE. 

\citet{2015A&A...584A.119R} also compared their results with other observations in other regions in Serpens. Their calculated mass of Serpens Core (i.e., Cluster A) is seven times less massive than the mass derived by C$^{18}$O \citep[scaled to the same distance,][]{1995A&A...298..594W}, and their calculated masses of core B and C are also three times less massive than the masses derived with 3 mm continuum emission \citep[scaled to the same distance,][]{1998ApJ...508L..91T,2000ApJ...540L..53T}. Given these facts, the H$_{2}$ column density map derived from the Herschel GB team is used for this work.

\end{appendix}

\bibliographystyle{aa}
\bibliography{references}

\begin{thebibliography}{97}
\expandafter\ifx\csname natexlab\endcsname\relax\def\natexlab#1{#1}\fi

\bibitem[{{Andr{\'e}}(2017)}]{2017CRGeo.349..187A}
{Andr{\'e}}, P. 2017, Comptes Rendus Geoscience, 349, 187

\bibitem[{{Andr{\'e}} {et~al.}(2014){Andr{\'e}}, {Di Francesco},
  {Ward-Thompson}, {Inutsuka}, {Pudritz}, \& {Pineda}}]{2014prpl.conf...27A}
{Andr{\'e}}, P., {Di Francesco}, J., {Ward-Thompson}, D., {et~al.} 2014,
  Protostars and Planets VI, 27

\bibitem[{{Andr{\'e}} {et~al.}(2010){Andr{\'e}}, {Men'shchikov}, {Bontemps},
  {K{\"o}nyves}, {Motte}, {Schneider}, {Didelon}, {Minier}, {Saraceno},
  {Ward-Thompson}, {di Francesco}, {White}, {Molinari}, {Testi}, {Abergel},
  {Griffin}, {Henning}, {Royer}, {Mer{\'{\i}}n}, {Vavrek}, {Attard},
  {Arzoumanian}, {Wilson}, {Ade}, {Aussel}, {Baluteau}, {Benedettini},
  {Bernard}, {Blommaert}, {Cambr{\'e}sy}, {Cox}, {di Giorgio}, {Hargrave},
  {Hennemann}, {Huang}, {Kirk}, {Krause}, {Launhardt}, {Leeks}, {Le Pennec},
  {Li}, {Martin}, {Maury}, {Olofsson}, {Omont}, {Peretto}, {Pezzuto}, {Prusti},
  {Roussel}, {Russeil}, {Sauvage}, {Sibthorpe}, {Sicilia-Aguilar}, {Spinoglio},
  {Waelkens}, {Woodcraft}, \& {Zavagno}}]{2010A&A...518L.102A}
{Andr{\'e}}, P., {Men'shchikov}, A., {Bontemps}, S., {et~al.} 2010, \aap, 518,
  L102

\bibitem[{{Arzoumanian} {et~al.}(2011{\natexlab{a}}){Arzoumanian}, {Andr{\'e}},
  {Didelon}, {K{\"o}nyves}, {Schneider}, {Men'shchikov}, {Sousbie}, {Zavagno},
  {Bontemps}, {di Francesco}, {Griffin}, {Hennemann}, {Hill}, {Kirk}, {Martin},
  {Minier}, {Molinari}, {Motte}, {Peretto}, {Pezzuto}, {Spinoglio},
  {Ward-Thompson}, {White}, \& {Wilson}}]{2011A&A...529L...6A}
{Arzoumanian}, D., {Andr{\'e}}, P., {Didelon}, P., {et~al.} 2011{\natexlab{a}},
  \aap, 529, L6

\bibitem[{{Arzoumanian} {et~al.}(2011{\natexlab{b}}){Arzoumanian}, {Andr{\'e}},
  {Didelon}, {K{\"o}nyves}, {Schneider}, {Men'shchikov}, {Sousbie}, {Zavagno},
  {Bontemps}, {di Francesco}, {Griffin}, {Hennemann}, {Hill}, {Kirk}, {Martin},
  {Minier}, {Molinari}, {Motte}, {Peretto}, {Pezzuto}, {Spinoglio},
  {Ward-Thompson}, {White}, \& {Wilson}}]{2011AA...529L...6A}
{Arzoumanian}, D., {Andr{\'e}}, P., {Didelon}, P., {et~al.} 2011{\natexlab{b}},
  \aap, 529, L6

\bibitem[{{Arzoumanian} {et~al.}(2013){Arzoumanian}, {Andr{\'e}}, {Peretto}, \&
  {K{\"o}nyves}}]{2013A&A...553A.119A}
{Arzoumanian}, D., {Andr{\'e}}, P., {Peretto}, N., \& {K{\"o}nyves}, V. 2013,
  \aap, 553, A119

\bibitem[{{Astropy Collaboration} {et~al.}(2013){Astropy Collaboration},
  {Robitaille}, {Tollerud}, {Greenfield}, {Droettboom}, {Bray}, {Aldcroft},
  {Davis}, {Ginsburg}, {Price-Whelan}, {Kerzendorf}, {Conley}, {Crighton},
  {Barbary}, {Muna}, {Ferguson}, {Grollier}, {Parikh}, {Nair}, {Unther},
  {Deil}, {Woillez}, {Conseil}, {Kramer}, {Turner}, {Singer}, {Fox}, {Weaver},
  {Zabalza}, {Edwards}, {Azalee Bostroem}, {Burke}, {Casey}, {Crawford},
  {Dencheva}, {Ely}, {Jenness}, {Labrie}, {Lim}, {Pierfederici}, {Pontzen},
  {Ptak}, {Refsdal}, {Servillat}, \& {Streicher}}]{2013A&A...558A..33A}
{Astropy Collaboration}, {Robitaille}, T.~P., {Tollerud}, E.~J., {et~al.} 2013,
  \aap, 558, A33

\bibitem[{{Bally} {et~al.}(1987){Bally}, {Langer}, {Stark}, \&
  {Wilson}}]{1987ApJ...312L..45B}
{Bally}, J., {Langer}, W.~D., {Stark}, A.~A., \& {Wilson}, R.~W. 1987, \apjl,
  312, L45

\bibitem[{{Benson} \& {Myers}(1989)}]{1989ApJS...71...89B}
{Benson}, P.~J. \& {Myers}, P.~C. 1989, \apjs, 71, 89

\bibitem[{{Bergin} {et~al.}(2002){Bergin}, {Alves}, {Huard}, \&
  {Lada}}]{2002ApJ...570L.101B}
{Bergin}, E.~A., {Alves}, J., {Huard}, T., \& {Lada}, C.~J. 2002, \apjl, 570,
  L101

\bibitem[{{Bergin} \& {Tafalla}(2007)}]{2007ARA&A..45..339B}
{Bergin}, E.~A. \& {Tafalla}, M. 2007, \araa, 45, 339

\bibitem[{{Bohlin} {et~al.}(1978){Bohlin}, {Savage}, \&
  {Drake}}]{1978ApJ...224..132B}
{Bohlin}, R.~C., {Savage}, B.~D., \& {Drake}, J.~F. 1978, \apj, 224, 132

\bibitem[{{Burkert} \& {Hartmann}(2004)}]{2004ApJ...616..288B}
{Burkert}, A. \& {Hartmann}, L. 2004, \apj, 616, 288

\bibitem[{{Burleigh} {et~al.}(2013){Burleigh}, {Bieging}, {Chromey}, {Kulesa},
  \& {Peters}}]{2013ApJS..209...39B}
{Burleigh}, K.~J., {Bieging}, J.~H., {Chromey}, A., {Kulesa}, C., \& {Peters},
  W.~L. 2013, \apjs, 209, 39

\bibitem[{{Carey} {et~al.}(1998){Carey}, {Clark}, {Egan}, {Price}, {Shipman},
  \& {Kuchar}}]{1998ApJ...508..721C}
{Carey}, S.~J., {Clark}, F.~O., {Egan}, M.~P., {et~al.} 1998, \apj, 508, 721

\bibitem[{{Chandrasekhar} \& {Fermi}(1953)}]{1953ApJ...118..116C}
{Chandrasekhar}, S. \& {Fermi}, E. 1953, \apj, 118, 116

\bibitem[{{De Vries} \& {Myers}(2005)}]{2005ApJ...620..800D}
{De Vries}, C.~H. \& {Myers}, P.~C. 2005, \apj, 620, 800

\bibitem[{{Dhabal} {et~al.}(2018){Dhabal}, {Mundy}, {Rizzo}, {Storm}, \&
  {Teuben}}]{2018ApJ...853..169D}
{Dhabal}, A., {Mundy}, L.~G., {Rizzo}, M.~J., {Storm}, S., \& {Teuben}, P.
  2018, \apj, 853, 169

\bibitem[{{Du} \& {Yang}(2008)}]{2008ApJ...686..384D}
{Du}, F. \& {Yang}, J. 2008, \apj, 686, 384

\bibitem[{{Enoch} {et~al.}(2009){Enoch}, {Evans}, {Sargent}, \&
  {Glenn}}]{2009ApJ...692..973E}
{Enoch}, M.~L., {Evans}, II, N.~J., {Sargent}, A.~I., \& {Glenn}, J. 2009,
  \apj, 692, 973

\bibitem[{{Enoch} {et~al.}(2007){Enoch}, {Glenn}, {Evans}, {Sargent}, {Young},
  \& {Huard}}]{2007ApJ...666..982E}
{Enoch}, M.~L., {Glenn}, J., {Evans}, II, N.~J., {et~al.} 2007, \apj, 666, 982

\bibitem[{{Evans} {et~al.}(2003){Evans}, {Allen}, {Blake}, {Boogert}, {Bourke},
  {Harvey}, {Kessler}, {Koerner}, {Lee}, {Mundy}, {Myers}, {Padgett},
  {Pontoppidan}, {Sargent}, {Stapelfeldt}, {van Dishoeck}, {Young}, \&
  {Young}}]{2003PASP..115..965E}
{Evans}, II, N.~J., {Allen}, L.~E., {Blake}, G.~A., {et~al.} 2003, \pasp, 115,
  965

\bibitem[{{Evans} {et~al.}(2009){Evans}, {Dunham}, {J{\o}rgensen}, {Enoch},
  {Mer{\'{\i}}n}, {van Dishoeck}, {Alcal{\'a}}, {Myers}, {Stapelfeldt},
  {Huard}, {Allen}, {Harvey}, {van Kempen}, {Blake}, {Koerner}, {Mundy},
  {Padgett}, \& {Sargent}}]{2009ApJS..181..321E}
{Evans}, II, N.~J., {Dunham}, M.~M., {J{\o}rgensen}, J.~K., {et~al.} 2009,
  \apjs, 181, 321

\bibitem[{{Fiorellino} {et~al.}(2017){Fiorellino}, {Pezzuto}, {Liu},
  {Benedettini}, {Schisano}, {Elia}, {Andr{\'e}}, {K{\"o}nyves}, {Ladjelate},
  \& {Herschel Gould Belt Survey Consortium}}]{2017MmSAI..88..783F}
{Fiorellino}, E., {Pezzuto}, S., {Liu}, S.~J., {et~al.} 2017, \memsai, 88, 783

\bibitem[{{Friberg} {et~al.}(1988){Friberg}, {Hjalmarson}, {Madden}, \&
  {Irvine}}]{1988A&A...195..281F}
{Friberg}, P., {Hjalmarson}, A., {Madden}, S.~C., \& {Irvine}, W.~M. 1988,
  \aap, 195, 281

\bibitem[{{Ginsburg} \& {Mirocha}(2011)}]{2011ascl.soft09001G}
{Ginsburg}, A. \& {Mirocha}, J. 2011, {PySpecKit: Python Spectroscopic
  Toolkit}, Astrophysics Source Code Library

\bibitem[{{Gong} {et~al.}(2017){Gong}, {Fang}, {Mao}, {Zhang}, {Wang}, {Su},
  {Chen}, {Yang}, {Wang}, \& {Lu}}]{2017ApJ...835L..14G}
{Gong}, Y., {Fang}, M., {Mao}, R., {et~al.} 2017, \apjl, 835, L14

\bibitem[{{Gong} {et~al.}(2015){Gong}, {Henkel}, {Thorwirth}, {Spezzano},
  {Menten}, {Walmsley}, {Wyrowski}, {Mao}, \& {Klein}}]{2015A&A...581A..48G}
{Gong}, Y., {Henkel}, C., {Thorwirth}, S., {et~al.} 2015, \aap, 581, A48

\bibitem[{{Griffin} {et~al.}(2010){Griffin}, {Abergel}, {Abreu}, {Ade},
  {Andr{\'e}}, {Augueres}, {Babbedge}, {Bae}, {Baillie}, {Baluteau}, {Barlow},
  {Bendo}, {Benielli}, {Bock}, {Bonhomme}, {Brisbin}, {Brockley-Blatt},
  {Caldwell}, {Cara}, {Castro-Rodriguez}, {Cerulli}, {Chanial}, {Chen},
  {Clark}, {Clements}, {Clerc}, {Coker}, {Communal}, {Conversi}, {Cox},
  {Crumb}, {Cunningham}, {Daly}, {Davis}, {de Antoni}, {Delderfield}, {Devin},
  {di Giorgio}, {Didschuns}, {Dohlen}, {Donati}, {Dowell}, {Dowell}, {Duband},
  {Dumaye}, {Emery}, {Ferlet}, {Ferrand}, {Fontignie}, {Fox}, {Franceschini},
  {Frerking}, {Fulton}, {Garcia}, {Gastaud}, {Gear}, {Glenn}, {Goizel},
  {Griffin}, {Grundy}, {Guest}, {Guillemet}, {Hargrave}, {Harwit}, {Hastings},
  {Hatziminaoglou}, {Herman}, {Hinde}, {Hristov}, {Huang}, {Imhof}, {Isaak},
  {Israelsson}, {Ivison}, {Jennings}, {Kiernan}, {King}, {Lange}, {Latter},
  {Laurent}, {Laurent}, {Leeks}, {Lellouch}, {Levenson}, {Li}, {Li},
  {Lilienthal}, {Lim}, {Liu}, {Lu}, {Madden}, {Mainetti}, {Marliani}, {McKay},
  {Mercier}, {Molinari}, {Morris}, {Moseley}, {Mulder}, {Mur}, {Naylor},
  {Nguyen}, {O'Halloran}, {Oliver}, {Olofsson}, {Olofsson}, {Orfei}, {Page},
  {Pain}, {Panuzzo}, {Papageorgiou}, {Parks}, {Parr-Burman}, {Pearce},
  {Pearson}, {P{\'e}rez-Fournon}, {Pinsard}, {Pisano}, {Podosek}, {Pohlen},
  {Polehampton}, {Pouliquen}, {Rigopoulou}, {Rizzo}, {Roseboom}, {Roussel},
  {Rowan-Robinson}, {Rownd}, {Saraceno}, {Sauvage}, {Savage}, {Savini},
  {Sawyer}, {Scharmberg}, {Schmitt}, {Schneider}, {Schulz}, {Schwartz},
  {Shafer}, {Shupe}, {Sibthorpe}, {Sidher}, {Smith}, {Smith}, {Smith},
  {Spencer}, {Stobie}, {Sudiwala}, {Sukhatme}, {Surace}, {Stevens}, {Swinyard},
  {Trichas}, {Tourette}, {Triou}, {Tseng}, {Tucker}, {Turner}, {Vaccari},
  {Valtchanov}, {Vigroux}, {Virique}, {Voellmer}, {Walker}, {Ward}, {Waskett},
  {Weilert}, {Wesson}, {White}, {Whitehouse}, {Wilson}, {Winter}, {Woodcraft},
  {Wright}, {Xu}, {Zavagno}, {Zemcov}, {Zhang}, \&
  {Zonca}}]{2010A&A...518L...3G}
{Griffin}, M.~J., {Abergel}, A., {Abreu}, A., {et~al.} 2010, \aap, 518, L3

\bibitem[{{Gutermuth} {et~al.}(2008){Gutermuth}, {Bourke}, {Allen}, {Myers},
  {Megeath}, {Matthews}, {J{\o}rgensen}, {Di Francesco}, {Ward-Thompson},
  {Huard}, {Brooke}, {Dunham}, {Cieza}, {Harvey}, \&
  {Chapman}}]{2008ApJ...673L.151G}
{Gutermuth}, R.~A., {Bourke}, T.~L., {Allen}, L.~E., {et~al.} 2008, \apjl, 673,
  L151

\bibitem[{{Hacar} {et~al.}(2016{\natexlab{a}}){Hacar}, {Alves}, {Burkert}, \&
  {Goldsmith}}]{2016A&A...591A.104H}
{Hacar}, A., {Alves}, J., {Burkert}, A., \& {Goldsmith}, P. 2016{\natexlab{a}},
  \aap, 591, A104

\bibitem[{{Hacar} {et~al.}(2016{\natexlab{b}}){Hacar}, {Kainulainen},
  {Tafalla}, {Beuther}, \& {Alves}}]{2016A&A...587A..97H}
{Hacar}, A., {Kainulainen}, J., {Tafalla}, M., {Beuther}, H., \& {Alves}, J.
  2016{\natexlab{b}}, \aap, 587, A97

\bibitem[{{Hacar} {et~al.}(2017){Hacar}, {Tafalla}, \&
  {Alves}}]{2017A&A...606A.123H}
{Hacar}, A., {Tafalla}, M., \& {Alves}, J. 2017, \aap, 606, A123

\bibitem[{{Hacar} {et~al.}(2018){Hacar}, {Tafalla}, {Forbrich}, {Alves},
  {Meingast}, {Grossschedl}, \& {Teixeira}}]{2018A&A...610A..77H}
{Hacar}, A., {Tafalla}, M., {Forbrich}, J., {et~al.} 2018, \aap, 610, A77

\bibitem[{{Hacar} {et~al.}(2013){Hacar}, {Tafalla}, {Kauffmann}, \&
  {Kov{\'a}cs}}]{2013A&A...554A..55H}
{Hacar}, A., {Tafalla}, M., {Kauffmann}, J., \& {Kov{\'a}cs}, A. 2013, \aap,
  554, A55

\bibitem[{{Hirota} {et~al.}(1998){Hirota}, {Yamamoto}, {Mikami}, \&
  {Ohishi}}]{1998ApJ...503..717H}
{Hirota}, T., {Yamamoto}, S., {Mikami}, H., \& {Ohishi}, M. 1998, \apj, 503,
  717

\bibitem[{{Inutsuka} \& {Miyama}(1992)}]{1992ApJ...388..392I}
{Inutsuka}, S.-I. \& {Miyama}, S.~M. 1992, \apj, 388, 392

\bibitem[{{Jeans}(1902)}]{1902RSPTA.199....1J}
{Jeans}, J.~H. 1902, Philosophical Transactions of the Royal Society of London
  Series A, 199, 1

\bibitem[{{J{\o}rgensen} {et~al.}(2004){J{\o}rgensen}, {Sch{\"o}ier}, \& {van
  Dishoeck}}]{2004A&A...416..603J}
{J{\o}rgensen}, J.~K., {Sch{\"o}ier}, F.~L., \& {van Dishoeck}, E.~F. 2004,
  \aap, 416, 603

\bibitem[{{Kainulainen} {et~al.}(2016){Kainulainen}, {Hacar}, {Alves},
  {Beuther}, {Bouy}, \& {Tafalla}}]{2016AA...586A..27K}
{Kainulainen}, J., {Hacar}, A., {Alves}, J., {et~al.} 2016, \aap, 586, A27

\bibitem[{{Kauffmann} {et~al.}(2008){Kauffmann}, {Bertoldi}, {Bourke}, {Evans},
  \& {Lee}}]{2008A&A...487..993K}
{Kauffmann}, J., {Bertoldi}, F., {Bourke}, T.~L., {Evans}, II, N.~J., \& {Lee},
  C.~W. 2008, \aap, 487, 993

\bibitem[{{Kirk} {et~al.}(2013){Kirk}, {Myers}, {Bourke}, {Gutermuth},
  {Hedden}, \& {Wilson}}]{2013ApJ...766..115K}
{Kirk}, H., {Myers}, P.~C., {Bourke}, T.~L., {et~al.} 2013, \apj, 766, 115

\bibitem[{{K{\"o}nyves} {et~al.}(2015){K{\"o}nyves}, {Andr{\'e}},
  {Men'shchikov}, {Palmeirim}, {Arzoumanian}, {Schneider}, {Roy}, {Didelon},
  {Maury}, {Shimajiri}, {Di Francesco}, {Bontemps}, {Peretto}, {Benedettini},
  {Bernard}, {Elia}, {Griffin}, {Hill}, {Kirk}, {Ladjelate}, {Marsh}, {Martin},
  {Motte}, {Nguy{\^e}n Luong}, {Pezzuto}, {Roussel}, {Rygl}, {Sadavoy},
  {Schisano}, {Spinoglio}, {Ward-Thompson}, \& {White}}]{2015A&A...584A..91K}
{K{\"o}nyves}, V., {Andr{\'e}}, P., {Men'shchikov}, A., {et~al.} 2015, \aap,
  584, A91

\bibitem[{{Larson}(1985)}]{1985MNRAS.214..379L}
{Larson}, R.~B. 1985, \mnras, 214, 379

\bibitem[{{Li} {et~al.}(2016{\natexlab{a}}){Li}, {Burkert}, {Megeath}, \&
  {Wyrowski}}]{2016arXiv160305720L}
{Li}, G.-X., {Burkert}, A., {Megeath}, T., \& {Wyrowski}, F.
  2016{\natexlab{a}}, ArXiv e-prints [\eprint[arXiv]{1603.05720}]

\bibitem[{{Li} {et~al.}(2016{\natexlab{b}}){Li}, {Urquhart}, {Leurini},
  {Csengeri}, {Wyrowski}, {Menten}, \& {Schuller}}]{2016A&A...591A...5L}
{Li}, G.-X., {Urquhart}, J.~S., {Leurini}, S., {et~al.} 2016{\natexlab{b}},
  \aap, 591, A5

\bibitem[{{Li} {et~al.}(2016{\natexlab{c}}){Li}, {Shen}, {Wang}, {Chen}, {Wu},
  {Zhao}, {Wang}, {Zuo}, {Fan}, {Hong}, {Jiang}, {Li}, {Liang}, {Ling}, {Liu},
  {Qian}, {Zhang}, {Zhong}, \& {Ye}}]{2016ApJ...824..136L}
{Li}, J., {Shen}, Z.-Q., {Wang}, J., {et~al.} 2016{\natexlab{c}}, \apj, 824,
  136

\bibitem[{{Liu} {et~al.}(2015){Liu}, {Galv{\'a}n-Madrid}, {Jim{\'e}nez-Serra},
  {Rom{\'a}n-Z{\'u}{\~n}iga}, {Zhang}, {Li}, \& {Chen}}]{2015ApJ...804...37L}
{Liu}, H.~B., {Galv{\'a}n-Madrid}, R., {Jim{\'e}nez-Serra}, I., {et~al.} 2015,
  \apj, 804, 37

\bibitem[{{Loren}(1989)}]{1989ApJ...338..925L}
{Loren}, R.~B. 1989, \apj, 338, 925

\bibitem[{{Loughnane} {et~al.}(2012){Loughnane}, {Redman}, {Thompson}, {Lo},
  {O'Dwyer}, \& {Cunningham}}]{2012MNRAS.420.1367L}
{Loughnane}, R.~M., {Redman}, M.~P., {Thompson}, M.~A., {et~al.} 2012, \mnras,
  420, 1367

\bibitem[{{Lu} {et~al.}(2018){Lu}, {Zhang}, {Liu}, {Sanhueza}, {Tatematsu},
  {Feng}, {Smith}, {Myers}, {Sridharan}, \& {Gu}}]{2018ApJ...855....9L}
{Lu}, X., {Zhang}, Q., {Liu}, H.~B., {et~al.} 2018, \apj, 855, 9

\bibitem[{{Malinen} {et~al.}(2012){Malinen}, {Juvela}, {Rawlings},
  {Ward-Thompson}, {Palmeirim}, \& {Andr{\'e}}}]{2012AA...544A..50M}
{Malinen}, J., {Juvela}, M., {Rawlings}, M.~G., {et~al.} 2012, \aap, 544, A50

\bibitem[{{Mangum} \& {Shirley}(2015)}]{2015PASP..127..266M}
{Mangum}, J.~G. \& {Shirley}, Y.~L. 2015, \pasp, 127, 266

\bibitem[{{Mattern} {et~al.}(2018){Mattern}, {Kauffmann}, {Csengeri},
  {Urquhart}, {Leurini}, {Wyrowski}, {Giannetti}, {Barnes}, {Beuther},
  {Bronfman}, {Duarte-Cabral}, {Henning}, {Kainulainen}, {Menten}, {Schisano},
  \& {Schuller}}]{2018arXiv180807499M}
{Mattern}, M., {Kauffmann}, J., {Csengeri}, T., {et~al.} 2018, ArXiv e-prints
  [\eprint[arXiv]{1808.07499}]

\bibitem[{{Mauersberger} \& {Henkel}(1991)}]{1991A&A...245..457M}
{Mauersberger}, R. \& {Henkel}, C. 1991, \aap, 245, 457

\bibitem[{{Menten} {et~al.}(2005){Menten}, {Pillai}, \&
  {Wyrowski}}]{2005IAUS..227...23M}
{Menten}, K.~M., {Pillai}, T., \& {Wyrowski}, F. 2005, in IAU Symposium, Vol.
  227, Massive Star Birth: A Crossroads of Astrophysics, ed. R.~{Cesaroni},
  M.~{Felli}, E.~{Churchwell}, \& M.~{Walmsley}, 23--34

\bibitem[{{Miville-Desch{\^e}nes} {et~al.}(2010){Miville-Desch{\^e}nes},
  {Martin}, {Abergel}, {Bernard}, {Boulanger}, {Lagache}, {Anderson},
  {Andr{\'e}}, {Arab}, {Baluteau}, {Blagrave}, {Bontemps}, {Cohen},
  {Compiegne}, {Cox}, {Dartois}, {Davis}, {Emery}, {Fulton}, {Gry}, {Habart},
  {Huang}, {Joblin}, {Jones}, {Kirk}, {Lim}, {Madden}, {Makiwa}, {Menshchikov},
  {Molinari}, {Moseley}, {Motte}, {Naylor}, {Okumura}, {Pinheiro Gon{\c
  c}alves}, {Polehampton}, {Rod{\'o}n}, {Russeil}, {Saraceno}, {Schneider},
  {Sidher}, {Spencer}, {Swinyard}, {Ward-Thompson}, {White}, \&
  {Zavagno}}]{2010A&A...518L.104M}
{Miville-Desch{\^e}nes}, M.-A., {Martin}, P.~G., {Abergel}, A., {et~al.} 2010,
  \aap, 518, L104

\bibitem[{{Monsch} {et~al.}(2018){Monsch}, {Pineda}, {Liu}, {Zucker}, {How-Huan
  Chen}, {Pattle}, {Offner}, {Di Francesco}, {Ginsburg}, {Ercolano}, {Arce},
  {Friesen}, {Kirk}, {Caselli}, \& {Goodman}}]{2018ApJ...861...77M}
{Monsch}, K., {Pineda}, J.~E., {Liu}, H.~B., {et~al.} 2018, \apj, 861, 77

\bibitem[{{Mullins} {et~al.}(2016){Mullins}, {Loughnane}, {Redman}, {Wiles},
  {Guegan}, {Barrett}, \& {Keto}}]{2016MNRAS.459.2882M}
{Mullins}, A.~M., {Loughnane}, R.~M., {Redman}, M.~P., {et~al.} 2016, \mnras,
  459, 2882

\bibitem[{{Myers}(2005)}]{2005ApJ...623..280M}
{Myers}, P.~C. 2005, \apj, 623, 280

\bibitem[{{Myers}(2009)}]{2009ApJ...700.1609M}
{Myers}, P.~C. 2009, \apj, 700, 1609

\bibitem[{{Myers} {et~al.}(1996){Myers}, {Mardones}, {Tafalla}, {Williams}, \&
  {Wilner}}]{1996ApJ...465L.133M}
{Myers}, P.~C., {Mardones}, D., {Tafalla}, M., {Williams}, J.~P., \& {Wilner},
  D.~J. 1996, \apjl, 465, L133

\bibitem[{{Ortiz-Le{\'o}n} {et~al.}(2017){Ortiz-Le{\'o}n}, {Dzib}, {Kounkel},
  {Loinard}, {Mioduszewski}, {Rodr{\'{\i}}guez}, {Torres}, {Pech}, {Rivera},
  {Hartmann}, {Boden}, {Evans}, {Brice{\~n}o}, {Tobin}, \&
  {Galli}}]{2017ApJ...834..143O}
{Ortiz-Le{\'o}n}, G.~N., {Dzib}, S.~A., {Kounkel}, M.~A., {et~al.} 2017, \apj,
  834, 143

\bibitem[{{Ostriker}(1964)}]{1964ApJ...140.1056O}
{Ostriker}, J. 1964, \apj, 140, 1056

\bibitem[{{Peretto} {et~al.}(2014){Peretto}, {Fuller}, {Andr{\'e}},
  {Arzoumanian}, {Rivilla}, {Bardeau}, {Duarte Puertas}, {Guzman Fernandez},
  {Lenfestey}, {Li}, {Olguin}, {R{\"o}ck}, {de Villiers}, \&
  {Williams}}]{2014A&A...561A..83P}
{Peretto}, N., {Fuller}, G.~A., {Andr{\'e}}, P., {et~al.} 2014, \aap, 561, A83

\bibitem[{{Peretto} {et~al.}(2007){Peretto}, {Hennebelle}, \&
  {Andr{\'e}}}]{2007A&A...464..983P}
{Peretto}, N., {Hennebelle}, P., \& {Andr{\'e}}, P. 2007, \aap, 464, 983

\bibitem[{{Pety}(2005)}]{2005sf2a.conf..721P}
{Pety}, J. 2005, in SF2A-2005: Semaine de l'Astrophysique Francaise, ed.
  F.~{Casoli}, T.~{Contini}, J.~M. {Hameury}, \& L.~{Pagani}, 721

\bibitem[{{Pillai} {et~al.}(2006){Pillai}, {Wyrowski}, {Carey}, \&
  {Menten}}]{2006A&A...450..569P}
{Pillai}, T., {Wyrowski}, F., {Carey}, S.~J., \& {Menten}, K.~M. 2006, \aap,
  450, 569

\bibitem[{{Pineda} {et~al.}(2011){Pineda}, {Goodman}, {Arce}, {Caselli},
  {Longmore}, \& {Corder}}]{2011ApJ...739L...2P}
{Pineda}, J.~E., {Goodman}, A.~A., {Arce}, H.~G., {et~al.} 2011, \apjl, 739, L2

\bibitem[{{Poglitsch} {et~al.}(2010){Poglitsch}, {Waelkens}, {Geis},
  {Feuchtgruber}, {Vandenbussche}, {Rodriguez}, {Krause}, {Renotte}, {van
  Hoof}, {Saraceno}, {Cepa}, {Kerschbaum}, {Agn{\`e}se}, {Ali}, {Altieri},
  {Andreani}, {Augueres}, {Balog}, {Barl}, {Bauer}, {Belbachir}, {Benedettini},
  {Billot}, {Boulade}, {Bischof}, {Blommaert}, {Callut}, {Cara}, {Cerulli},
  {Cesarsky}, {Contursi}, {Creten}, {De Meester}, {Doublier}, {Doumayrou},
  {Duband}, {Exter}, {Genzel}, {Gillis}, {Gr{\"o}zinger}, {Henning},
  {Herreros}, {Huygen}, {Inguscio}, {Jakob}, {Jamar}, {Jean}, {de Jong},
  {Katterloher}, {Kiss}, {Klaas}, {Lemke}, {Lutz}, {Madden}, {Marquet},
  {Martignac}, {Mazy}, {Merken}, {Montfort}, {Morbidelli}, {M{\"u}ller},
  {Nielbock}, {Okumura}, {Orfei}, {Ottensamer}, {Pezzuto}, {Popesso},
  {Putzeys}, {Regibo}, {Reveret}, {Royer}, {Sauvage}, {Schreiber}, {Stegmaier},
  {Schmitt}, {Schubert}, {Sturm}, {Thiel}, {Tofani}, {Vavrek}, {Wetzstein},
  {Wieprecht}, \& {Wiezorrek}}]{2010A&A...518L...2P}
{Poglitsch}, A., {Waelkens}, C., {Geis}, N., {et~al.} 2010, \aap, 518, L2

\bibitem[{{Pon} {et~al.}(2011){Pon}, {Johnstone}, \&
  {Heitsch}}]{2011ApJ...740...88P}
{Pon}, A., {Johnstone}, D., \& {Heitsch}, F. 2011, \apj, 740, 88

\bibitem[{{Roccatagliata} {et~al.}(2015){Roccatagliata}, {Dale}, {Ratzka},
  {Testi}, {Burkert}, {Koepferl}, {Sicilia-Aguilar}, {Eiroa}, \&
  {Gaczkowski}}]{2015A&A...584A.119R}
{Roccatagliata}, V., {Dale}, J.~E., {Ratzka}, T., {et~al.} 2015, \aap, 584,
  A119

\bibitem[{{Schilke} {et~al.}(1992){Schilke}, {Walmsley}, {Pineau Des Forets},
  {Roueff}, {Flower}, \& {Guilloteau}}]{1992A&A...256..595S}
{Schilke}, P., {Walmsley}, C.~M., {Pineau Des Forets}, G., {et~al.} 1992, \aap,
  256, 595

\bibitem[{{Schneider} {et~al.}(2010){Schneider}, {Csengeri}, {Bontemps},
  {Motte}, {Simon}, {Hennebelle}, {Federrath}, \&
  {Klessen}}]{2010A&A...520A..49S}
{Schneider}, N., {Csengeri}, T., {Bontemps}, S., {et~al.} 2010, \aap, 520, A49

\bibitem[{{Schneider} \& {Elmegreen}(1979)}]{1979ApJS...41...87S}
{Schneider}, S. \& {Elmegreen}, B.~G. 1979, \apjs, 41, 87

\bibitem[{{Shan} {et~al.}(2012){Shan}, {Yang}, {Shi}, {Yao}, {Zuo}, {Lin},
  {Chen}, {Zhang}, {Duan}, {Cao}, {Li}, {Li}, {Liu}, \&
  {Zhong}}]{2012ITTST...2..593S}
{Shan}, W., {Yang}, J., {Shi}, S., {et~al.} 2012, IEEE Transactions on
  Terahertz Science and Technology, 2, 593

\bibitem[{{Shirley}(2015)}]{2015PASP..127..299S}
{Shirley}, Y.~L. 2015, \pasp, 127, 299

\bibitem[{{Sokolov} {et~al.}(2017){Sokolov}, {Wang}, {Pineda}, {Caselli},
  {Henshaw}, {Tan}, {Fontani}, {Jim{\'e}nez-Serra}, \&
  {Lim}}]{2017A&A...606A.133S}
{Sokolov}, V., {Wang}, K., {Pineda}, J.~E., {et~al.} 2017, \aap, 606, A133

\bibitem[{{Sousbie}(2011)}]{2011MNRAS.414..350S}
{Sousbie}, T. 2011, \mnras, 414, 350

\bibitem[{{Sun} {et~al.}(2018){Sun}, {Lu}, {Yang}, {Su}, {Zhang}, {Zhou}, \&
  {Lin}}]{2018AcASn..59....3S}
{Sun}, J.~X., {Lu}, D.~R., {Yang}, J., {et~al.} 2018, Acta Astronomica Sinica,
  59, 3

\bibitem[{{Suzuki} {et~al.}(1992){Suzuki}, {Yamamoto}, {Ohishi}, {Kaifu},
  {Ishikawa}, {Hirahara}, \& {Takano}}]{1992ApJ...392..551S}
{Suzuki}, H., {Yamamoto}, S., {Ohishi}, M., {et~al.} 1992, \apj, 392, 551

\bibitem[{{Tan} {et~al.}(2014){Tan}, {Beltr{\'a}n}, {Caselli}, {Fontani},
  {Fuente}, {Krumholz}, {McKee}, \& {Stolte}}]{2014prpl.conf..149T}
{Tan}, J.~C., {Beltr{\'a}n}, M.~T., {Caselli}, P., {et~al.} 2014, Protostars
  and Planets VI, 149

\bibitem[{{Testi} \& {Sargent}(1998)}]{1998ApJ...508L..91T}
{Testi}, L. \& {Sargent}, A.~I. 1998, \apjl, 508, L91

\bibitem[{{Testi} {et~al.}(2000){Testi}, {Sargent}, {Olmi}, \&
  {Onello}}]{2000ApJ...540L..53T}
{Testi}, L., {Sargent}, A.~I., {Olmi}, L., \& {Onello}, J.~S. 2000, \apjl, 540,
  L53

\bibitem[{{Ulich} \& {Haas}(1976)}]{1976ApJS...30..247U}
{Ulich}, B.~L. \& {Haas}, R.~W. 1976, \apjs, 30, 247

\bibitem[{{Vastel} {et~al.}(2014){Vastel}, {Ceccarelli}, {Lefloch}, \&
  {Bachiller}}]{2014ApJ...795L...2V}
{Vastel}, C., {Ceccarelli}, C., {Lefloch}, B., \& {Bachiller}, R. 2014, \apjl,
  795, L2

\bibitem[{{Vasyunin} \& {Herbst}(2013)}]{2013ApJ...769...34V}
{Vasyunin}, A.~I. \& {Herbst}, E. 2013, \apj, 769, 34

\bibitem[{{Wang}(2018)}]{2018RNAAS...2b..52W}
{Wang}, K. 2018, Research Notes of the American Astronomical Society, 2, 52

\bibitem[{{Wang} {et~al.}(2015){Wang}, {Testi}, {Ginsburg}, {Walmsley},
  {Molinari}, \& {Schisano}}]{2015MNRAS.450.4043W}
{Wang}, K., {Testi}, L., {Ginsburg}, A., {et~al.} 2015, \mnras, 450, 4043

\bibitem[{{White} {et~al.}(1995){White}, {Casali}, \&
  {Eiroa}}]{1995A&A...298..594W}
{White}, G.~J., {Casali}, M.~M., \& {Eiroa}, C. 1995, \aap, 298, 594

\bibitem[{{Wirstr{\"o}m} {et~al.}(2011){Wirstr{\"o}m}, {Geppert}, {Hjalmarson},
  {Persson}, {Black}, {Bergman}, {Millar}, {Hamberg}, \&
  {Vigren}}]{2011A&A...533A..24W}
{Wirstr{\"o}m}, E.~S., {Geppert}, W.~D., {Hjalmarson}, {\AA}., {et~al.} 2011,
  \aap, 533, A24

\bibitem[{{Wright} {et~al.}(2010){Wright}, {Eisenhardt}, {Mainzer}, {Ressler},
  {Cutri}, {Jarrett}, {Kirkpatrick}, {Padgett}, {McMillan}, {Skrutskie},
  {Stanford}, {Cohen}, {Walker}, {Mather}, {Leisawitz}, {Gautier}, {McLean},
  {Benford}, {Lonsdale}, {Blain}, {Mendez}, {Irace}, {Duval}, {Liu}, {Royer},
  {Heinrichsen}, {Howard}, {Shannon}, {Kendall}, {Walsh}, {Larsen}, {Cardon},
  {Schick}, {Schwalm}, {Abid}, {Fabinsky}, {Naes}, \&
  {Tsai}}]{2010AJ....140.1868W}
{Wright}, E.~L., {Eisenhardt}, P.~R.~M., {Mainzer}, A.~K., {et~al.} 2010, \aj,
  140, 1868

\bibitem[{{Wyrowski} {et~al.}(2016){Wyrowski}, {G{\"u}sten}, {Menten},
  {Wiesemeyer}, {Csengeri}, {Heyminck}, {Klein}, {K{\"o}nig}, \&
  {Urquhart}}]{2016A&A...585A.149W}
{Wyrowski}, F., {G{\"u}sten}, R., {Menten}, K.~M., {et~al.} 2016, \aap, 585,
  A149

\bibitem[{{Xiong} {et~al.}(2017){Xiong}, {Chen}, {Yang}, {Fang}, {Zhang},
  {Zhang}, {Du}, \& {Long}}]{2017ApJ...838...49X}
{Xiong}, F., {Chen}, X., {Yang}, J., {et~al.} 2017, \apj, 838, 49

\bibitem[{{Yuan} {et~al.}(2018){Yuan}, {Li}, {Wu}, {Ellingsen}, {Henkel},
  {Wang}, {Liu}, {Liu}, {Zavagno}, {Ren}, \& {Huang}}]{2018ApJ...852...12Y}
{Yuan}, J., {Li}, J.-Z., {Wu}, Y., {et~al.} 2018, \apj, 852, 12

\bibitem[{{Zhang} {et~al.}(2011){Zhang}, {Yang}, {Xu}, {Pandian}, {Menten}, \&
  {Henkel}}]{2011ApJS..193...10Z}
{Zhang}, S.~B., {Yang}, J., {Xu}, Y., {et~al.} 2011, \apjs, 193, 10

\bibitem[{{Zhou} {et~al.}(1993){Zhou}, {Evans}, {Koempe}, \&
  {Walmsley}}]{1993ApJ...404..232Z}
{Zhou}, S., {Evans}, II, N.~J., {Koempe}, C., \& {Walmsley}, C.~M. 1993, \apj,
  404, 232

\end{thebibliography}

\end{document}